\documentclass[fleqn,usenatbib]{mnras}

\usepackage[T1]{fontenc}

\DeclareRobustCommand{\VAN}[3]{#2}
\let\VANthebibliography\thebibliography
\def\thebibliography{\DeclareRobustCommand{\VAN}[3]{##3}\VANthebibliography}

\usepackage{graphicx}	
\usepackage{amsmath}	
\usepackage{amssymb}	
\usepackage{multirow}
\usepackage{cuted}
\usepackage{hyperref}
\usepackage{url}
\usepackage[dvipsnames]{xcolor}
\usepackage{soul}
\usepackage{todonotes}
\usepackage{longtable}
\usepackage{pdflscape}
\usepackage[hypcap=false]{caption}

\usepackage[export]{adjustbox}
\usepackage{gensymb}

\newcommand{\fxnum}[1]{#1}

\newcommand{\Msun}{\mbox{$\mathrm{M}_{\sun}$}}

\newcommand{\sdss}{SDSS~III~and~IV}
\newcommand{\allsdss}{SDSS~I~to~IV}

\newcommand{\porb}{\mbox{$P_{\mathrm{orb}}$}}
\newcommand{\mdot}{$\dot{M}$}

\newcommand\jaavso{AAVSO} 
\setuptodonotes{inline,noinlinepar,inlinewidth=2cm}
\usepackage{newtxtext,newtxmath}

\title[CVs from SDSS]{A catalogue of cataclysmic variables from 20 years of  the Sloan Digital Sky Survey with new classifications, periods, trends and oddities. }

\author[K. Inight et al.]{
Keith~Inight,$^{1}$\thanks{E-mail: keith.inight@gmail.com }
Boris~T.~G\"ansicke,$^{1}$
Elmé~Breedt,$^{2}$
Henry~T.~Israel,$^{6}$
Stuart~P.~Littlefair,$^{6}$
\newauthor
Christopher~J.~Manser,$^{1,4}$
Tom~R.~Marsh,$^{1}$
Tim~Mulvany,$^{1}$
Anna Francesca~Pala,$^{3}$
\newauthor
John~R.~Thorstensen$^{5}$
\\
$^{1}$Department of Physics, University of Warwick, Coventry, CV4 7AL, UK\\
$^{2}$Institute of Astronomy, University of Cambridge, Madingley Road,
Cambridge CB3 0HA, United Kingdom \\
$^{3}$European Space Agency, European Space Astronomy Centre, Camino Bajo del Castillo s/n, Villanueva de la Cañada, E-28692 Madrid, Spain \\
$^{4}$Astrophysics Group, Department of Physics, Imperial College London, Prince Consort Rd, London, SW7 2AZ, UK \\
$^{5}$Department of Physics and Astronomy, Dartmouth College, Hanover NH 03755, USA \\
$^{6}$Department of Physics and Astronomy, University of Sheffield, Sheffield S3 7RH, UK
}

\date{Accepted XXX. Received YYY; in original form ZZZ}

\pubyear{2020}

\begin{document}
\label{firstpage}
\pagerange{\pageref{firstpage}--\pageref{lastpage}}
\maketitle

\begin{abstract}
We present a catalogue of \fxnum{507} cataclysmic variables (CVs) observed in \allsdss\ including \fxnum{70} new classifications  collated from multiple archival data sets. This represents the largest sample of CVs with high-quality and homogeneous optical spectroscopy. We have used this sample to derive unbiased space densities and period distributions for the major sub-types of CVs. We also report on some peculiar CVs, period bouncers and also CVs exhibiting  large changes in accretion rates. We report \fxnum{70} new CVs, \fxnum{59} new periods, \fxnum{178} unpublished spectra and \fxnum{262} new or updated classifications. From the SDSS spectroscopy,  we also identified \fxnum{18} systems incorrectly identified as CVs in the literature. We discuss the observed properties of 13 peculiar CVS, and we identify a small set of eight CVs that defy the standard classification scheme. We use this sample to investigate the distribution of different CV sub-types, and we estimate their individual space densities, as well as that of the entire CV population. The \allsdss\ sample includes 14 period bounce CVs or candidates.  We discuss the variability of CVs across the Hertzsprung-Russell diagram, highlighting selection biases of variability-based CV detection. Finally, we searched for, and found eight tertiary companions to the SDSS CVs. We anticipate that this catalogue and the extensive material included in the Supplementary Data will be useful for a range of observational population studies of CVs. 
\end{abstract}



\section{Introduction}
Cataclysmic variables (CVs, see \citealt{1995cvs..book.....W} for an exhaustive overview) are an important class of white dwarf binaries within multiple astrophysical contexts, including accretion disc physics  \citep{2014apa..book.....G} and potential gravitational wave sources for \textit{LISA} \citep{1996CQGra..13..247D}. These compact binaries will become increasingly important as the inventory becomes more complete and the emerging population statistics can be used to test evolutionary models. Although $\sim10\,000$ CV candidates are known (see for example \citealt{2017yCat....102027W}) 
that sample contains a fair amount of contaminants, including flaring M-dwarfs, young stellar objects, variable AGN and unexplained transients (see discussion in section 2 of \citealt{2020MNRAS.494.3799P}). To be useful for evolutionary modelling a sample of CVs needs not only to be reasonably complete over a given volume but also possess astrometric data for each CV and, crucially, an identification spectrum. 

The sample of \fxnum{507} CVs spectroscopically observed by the Sloan Digital Sky Survey  (SDSS) \citep{1998AJ....116.3040G,2000AJ....120.1579Y}  presented here is a magnitude-limited sample covering over 25.6 \,per\,cent of the sky (section\,\ref{sec:SDSSfootprint}).

The four phases of SDSS (hereafter \allsdss) were executed between 2000 and 2020 \citep{2003AJ....126.2081A,2008ApJS..175..297A,2011ApJS..193...29A,2019ApJS..240...23A} and addressed many science goals of which the search for quasars \citep{2002AJ....123.2945R} and latterly luminous red galaxies and quasars \citep{2013AJ....145...10D} in the  Baryon Oscillation Spectroscopic Survey (BOSS and eBOSS) are of particular relevance here. Between 2000 and 2009 SDSS obtained photometry, using the five $ugriz$ filter passbands, of over 14\,000 square degrees, predominantly of the northern sky. The imaging camera \citep{1998AJ....116.3040G} was installed on a purpose-built 2.5\,m telescope at Apache Point, New Mexico. This photometry was used to select targets for subsequent spectroscopic surveys using a multi-object, fibre-fed spectrograph capable of taking 640 (SDSS~I and II) and 1000 (SDSS~III and IV) simultaneous spectra spanning $\simeq3800-9200$\,\AA\ (SDSS~I and II) and $\simeq3600-10\,000$\,\AA\ (SDSS~III and IV, \citealt{2013AJ....146...32S}). 

Whereas CVs were targeted within SDSS~I via a multi-dimensional colour selection \citep{2002AJ....123..430S}, the vast majority of CVs were serendipitously identified from the spectroscopy of quasar candidates. SDSS~I \citep{2000AJ....120.1579Y} and II \citep{2005AAS...20718005D,2009AJ....137.4377Y} have proven to be a major source of optical spectroscopy of CVs, both of previously known systems and new discoveries \citep{2002AJ....123..430S,2003AJ....126.1499S,2004AJ....128.1882S,2005AJ....129.2386S,2006AJ....131..973S,2007AJ....134..185S,2009AJ....137.4011S,2011AJ....142..181S}. 

\sdss\, were two subsequent phases spanning from July 2008 until June 2014 \citep{2011AJ....142...72E,2013AJ....145...10D}  and from July 2014 until August 2020 \citep{2017AJ....154...28B,2016AJ....151...44D} respectively.  

SDSS~I and II yielded spectra of 285 CVs \citep{2011AJ....142..181S} and have been particularly successful in identifying CVs with no history of outbursts \citep{2009MNRAS.397.2170G}. Specifically they provided a rich source of short period ($80-86$\,min) CVs that provided the crucial observational evidence for the existence of a ``period minimum spike''. Other groundbreaking achievements include the identification of  low accretion rate, X-ray weak, polars (\citealt{2008PASP..120..160S} and references therein) and CVs with pulsating white dwarfs \citep{2005ASPC..330..325W,2006MNRAS.365..969G,2007ApJ...667..433M}. Beyond these individual discoveries the systematic detection of all types of CVs has enabled statistical studies of population density and CV evolution \citep{2005ASPC..330....3G}. These achievements from SDSS~I and II provided the motivation to identify CVs in SDSS~III and IV. Further motivation came from the opportunity to refine the attributes of the known CVs using new astrometric and photometric survey data. 

We report here the identification of \fxnum{70}  new CVs found in \allsdss (see list in Appendix \ref{section:appendix1}). We have also revisited the remaining  \fxnum{437}  CVs observed during SDSS~I to IV in the light of \textit{Gaia}  \citep{2020yCat.1350....0G} astronomy and photometry from the Catalina Real-Time Transient Survey (CRTS, \citealt{2011arXiv1102.5004D}), Zwicky Transient Facility (ZTF, \citealt{Masci_2018,2019PASP..131a8002B}) and the Transiting Exoplanet Survey Satellite (\textit{TESS}, \citealt{2015JATIS...1a4003R}) which was not previously available. This has resulted in \fxnum{59}  new orbital periods (including some unpublished orbital periods from other sources, details are provided in the supplementary data) and \fxnum{262} new/revised classifications. SDSS objects are identified by their coordinates, e.g. SDSS\,J152212.20+080340.9, which we abbreviate throughout the paper as J1522+0803.

The paper is laid out as follows.  We discuss CV evolution in section \ref{section:evolution} and in section \ref{section:sub_types} we provide an overview describing how the different types evolve in the overall CV life-cycle using examples from \allsdss. In section \ref{section:methods} we describe the methods used to identify new CVs and classify existing ones. Our results are described in section \ref{section:results} and discussed in section \ref{section:discussion}  with an overall summary in section \ref{section:summary}.

\section{CV evolution} \label{section:evolution}
Whilst we do not seek to replicate the extensive literature on CVs (a comprehensive coverage can be found in \citealt{1995cvs..book.....W} and \citealt{2001cvs..book.....H}) we do explain the evolutionary relevance of the different types of CV found by SDSS.

CVs are born out of a common-envelope event \citep{1976IAUS...73...75P,2013A&ARv..21...59I} in which the higher-mass star in a binary system evolves off the main sequence and becomes a giant, engulfing its lower-mass companion. Friction in the common envelope extracts angular momentum and energy, resulting in a dramatic reduction in the orbital separation of the two stars. The result of the common envelope is a close binary consisting of a white dwarf (the core of the former red giant) plus a low-mass companion; this is known as a post common envelope binary (PCEB). Most PCEBs exit the common envelope in a detached configuration (and many are observed in this state, e.g. \citealt{2012MNRAS.419..806R}). However, the stellar wind from the companion causes magnetic braking \citep{1981A&A...100L...7V} and results in a loss of angular momentum. Meanwhile tidal forces on the secondary will eventually result in tidal locking to the orbital period \citep{2019ApJ...881...88F}. Magnetic braking will eventually shrink the orbital separation to the point where the secondary star fills its Roche lobe, and starts mass transfer onto the white dwarf~--~initiating the further evolution of the system as a CV. 

Magnetic braking continues into the CV phase resulting in the system continuing to lose angular momentum and as a consequence the system separation and \porb\ will decrease.

At around $\porb =3$h the donor stars in CVs become fully convective, possibly resulting in a re-configuration of their magnetic field (see observational evidence in \citealt{2009A&A...496..787R} and also fig. 15 in \citealt{2021A&ARv..29....1K}). In the standard scenario of CV evolution \citep{2011ApJS..194...28K}, magnetic braking is strongly reduced, and the donor star, slightly bloated due to prior mass loss, shrinks to its thermal equilibrium radius, detaching from the Roche-lobe and stopping the mass transfer. The observational motivation for this ``disrupted magnetic braking scenario'' \citep{1983ApJ...275..713R} is a paucity of CVs within the range $\mathrm{2.15\,\mathrm{h} \lesssim} \porb \lesssim3.18$\,h; this is the so-called ``period gap'' \citep{1980MNRAS.190..801W,2001ApJ...550..897H,2006MNRAS.373..484K}. Some of this under-representation appears to be due to selection effects (see for example fig.\,9 in \citealt{2021MNRAS.504.2420I} and section\,\ref{section:perioddistribution}). In the case of strongly magnetic white dwarfs, magnetic braking could be reduced, and these systems may not experience a detached phase in their evolution \citep{2002MNRAS.335....1W}.

For periods $\lesssim3$\,h, gravitational radiation \citep{1967AcA....17..287P,1988QJRAS..29....1K} supersedes magnetic braking as the primary cause of angular momentum loss, which will eventually cause the separation and orbital period to decrease sufficiently for the donor to fill its Roche lobe and the system to resume accretion as a CV. 

Throughout the evolution of a CV the mass of the donor, the rate of mass loss from the donor (\mdot), and \porb\ all continue to decrease until the donor mass is too low to sustain hydrogen burning and the system becomes a brown dwarf. This point is the ``period minimum'' of the system \citep{1983ApJ...268..825P}. Following further mass loss from the donor the orbital separation (and hence period) increases and it becomes a ``period bouncer'' (e.g. \citet{2018MNRAS.481.2523P}). 

The previous discussion has assumed that the donor is still on the main sequence when it commences accretion. However the donors in some CVs have undergone substantial nuclear evolution. Because the orbital period at which a donor fills its Roche lobe is related to its density $\rho_\mathrm{d}$ via $\porb \propto \left (\rho_{\mathrm{d}}  \right )^{-0.5}$ (equation 12 in \citealt{2011ApJS..194...28K}) it follows that, for the same donor mass, CVs with nuclear evolved donors have longer orbital periods than their un-evolved counterparts (e.g. \citealt{2009A&A...496..805R}). At periods $\lessapprox 5$\,h, evolved donors have a higher temperature and spectral type than main sequence stars of the same mass. They can appear in the period gap \citep{2014ApJ...790...28R} and evolve to periods below the period minimum \citep{2002ApJ...567L..49T}.

\section{CV sub-types} \label{section:sub_types}
``Cataclysmic Variables'' were named as such because of the large brightness variations (eruptions and outbursts) observed among classical novae and dwarf novae. The modern definition of a CV dates from the 1960's (see for example \citealt{1963AdA&A...2...43K}) and consists of a semi-detached binary in which a white dwarf accretes from a Roche-lobe filling donor star~--~and the current roster of CVs includes many systems that have no recorded outbursts \citep{2009MNRAS.397.2170G}. 
\begin{figure*} 
\centering
\includegraphics[trim={8cm 7.5cm 3.6cm 3cm},clip,width=2.5\columnwidth]{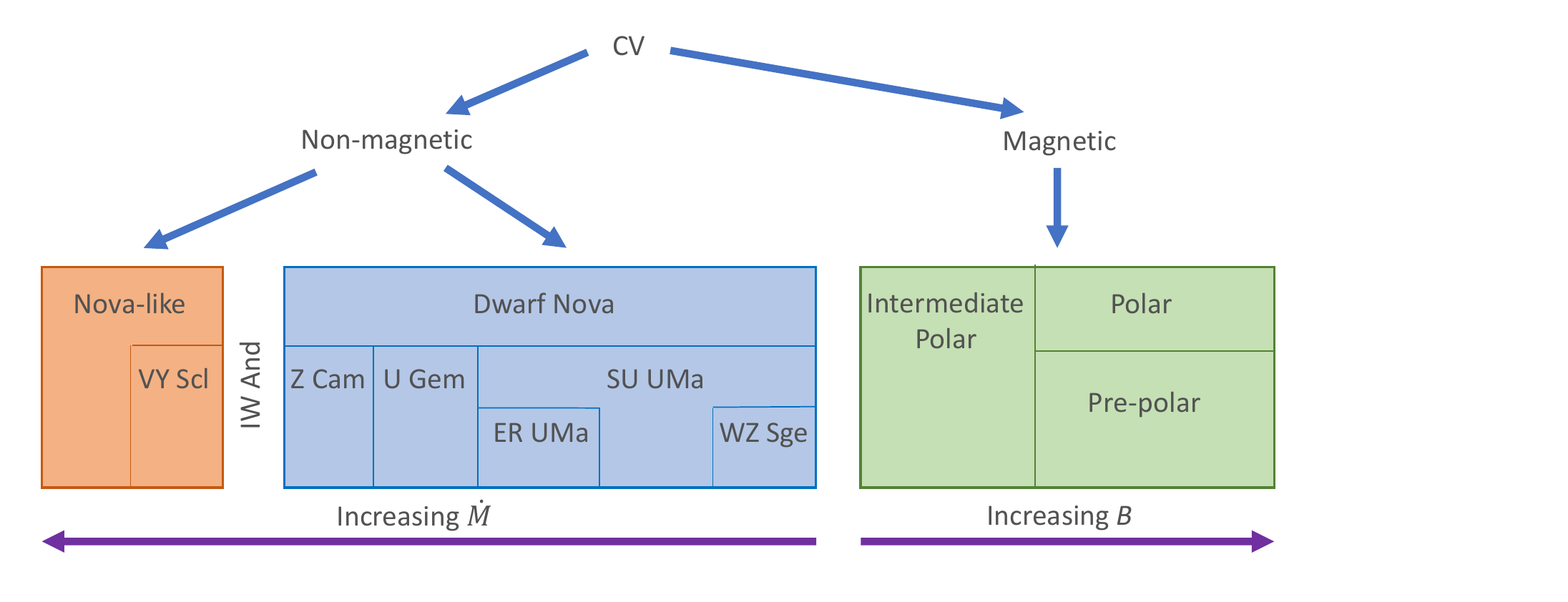}
\caption{\label{fig:diagram2} 
Diagram depicting the CV classification scheme used in this paper. The physical key parameters that define the boundaries between different classes are $\dot M$ and \porb\ (which, together, establish whether or not the accretion disc is subject to thermal instabilities) and $B$ (which determines whether or not a disc forms in the first place). Depending on the exact physical setup, some individual CVs may straddle the boundaries between the sub-types defined above. AM\,CVn systems, with three possible formation channels (Section \ref{section:amcvn}),  follow the same distribution as non-magnetic CVs.}
\end{figure*}

CVs are classified into a number of sub-types. Historically they have been categorised based upon their observable characteristics with categories typically being named after a prototype system, frequently the first discovery. The number of sub-types, their definition, and relationship between different sub-types remain in flux, making their use challenging for  scientists not regularly working in this research area~--~and some care has to be taken with the temptation to define new sub-types for each additional observational wrinkle in a particular CV. Nevertheless, dividing CVs into sub-types is meaningful, as their observational characteristics depend on a small number of physical parameters, which link to the evolutionary state and the accretion geometry of an individual system (see Fig.\,\ref{fig:diagram2}). 

Here we provide an overview of the main CV categories that have been historically used with examples from \allsdss\, and detail their state of evolution and mode of accretion. The key physical parameters are the orbital period, \porb, the rate of mass loss from the donor, \mdot, and the white dwarf magnetic field strength, $B$.   

The strength of the magnetic field of the white dwarf is of particular importance in determining the accretion geometry \citep{2008ApJ...672..524N}. We therefore discuss the classification and evolution of magnetic and non-magnetic systems separately.

\subsection{Non-magnetic CVs}

\subsubsection{Dwarf Novae}

\begin{figure} 
\includegraphics[width=\columnwidth]{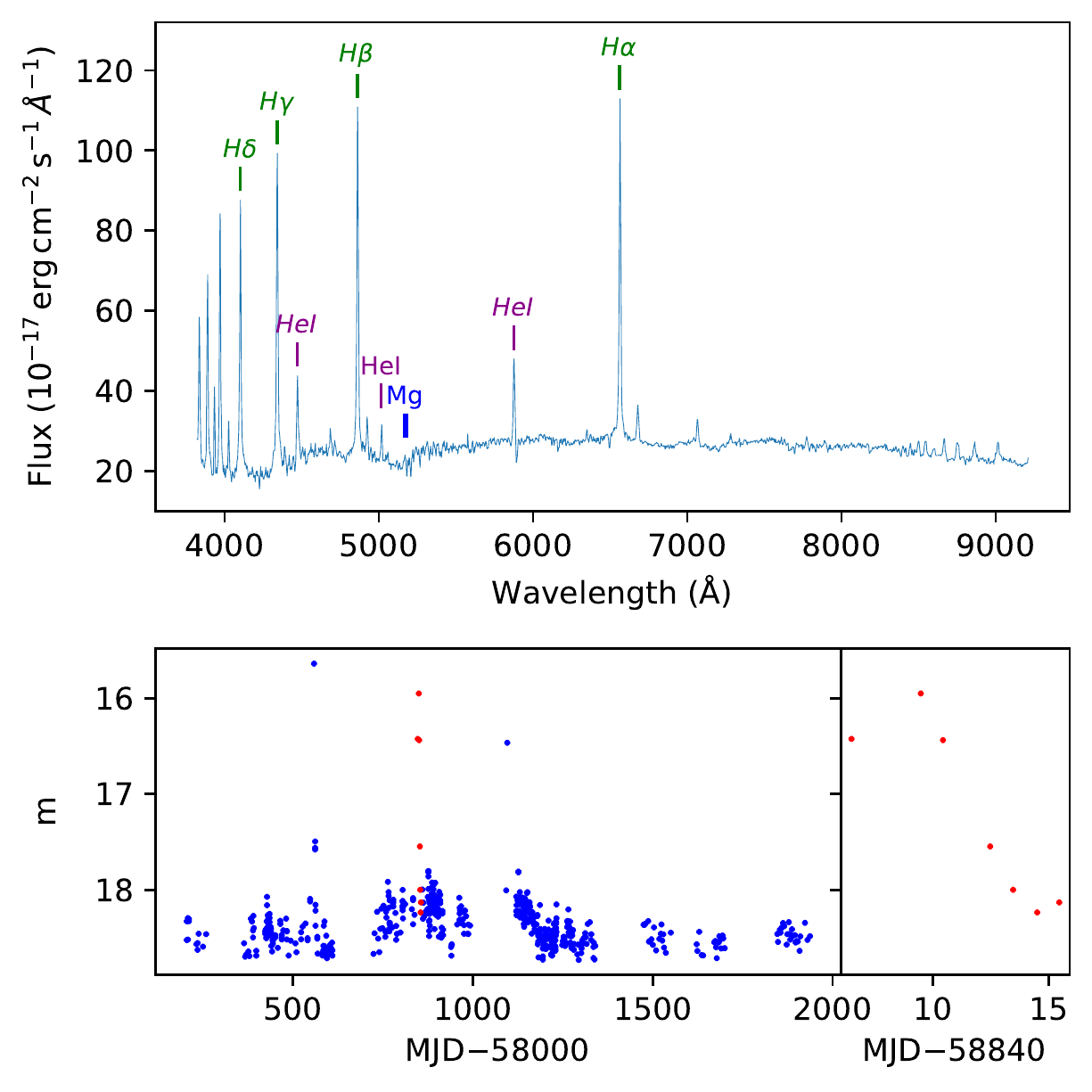}
\caption [.] {\label{fig:UGem} J0813+4528 is an example of a  dwarf nova. Top panel: The quiescent SDSS spectrum (smoothed with a five-point box-car filter) shows strong emission lines. The continuum reveals the spectroscopic signatures of the donor star, indicative of a long orbital period (6.94\,h); this is a U\,Gem type dwarf nova above the period gap.  Bottom left panel: ZTF ($g$-band) light curve showing outbursts. Bottom right panel: Expanded light curve of the outburst highlighted in red in the bottom left panel showing the decay over $\simeq3$\,d. 
}
\end{figure}

The dwarf nova thermal limit cycle starts with increasing mass being retained in the disc which continues to grow until it becomes thermally unstable and a dwarf nova occurs. The magnitude of a dwarf nova outburst is typically $\Delta m=3-5$ \citep{1983A&A...121...29M} and typically lasts for a few days (although superoutbursts such as that of GW\,Lib in 2007 can reach $\Delta m \approx9$ and last several weeks) after which the disc cools and the cycle repeats itself.
Four sub-types are discussed below (SU\,UMa, WZ\,Sge, ER\,UMa and Z\,Cam) with the remaining dwarf nova CVs being classified as U\,Gem. In quiescence dwarf nova spectra (Fig.\,\ref{fig:UGem}) exhibit hydrogen and also sometimes helium emission lines whilst in outburst their spectra resemble those of novalike variables (see below).

\begin{figure} 
\includegraphics[width=\columnwidth]{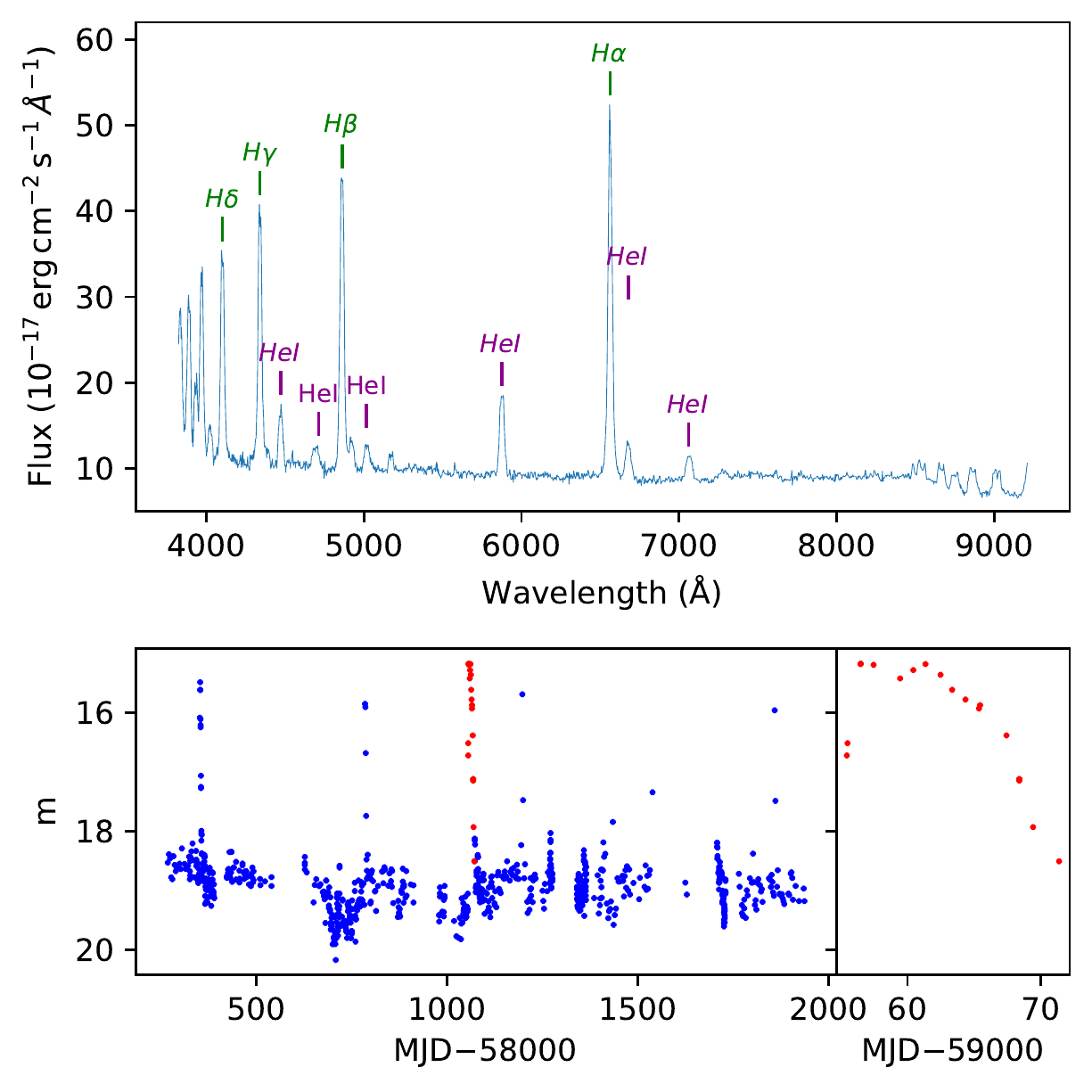}
\caption [.] {\label{fig:SUUMa} J0032+3145 is an example of an SU\,UMa CV. Top panel: Smoothed SDSS spectrum in quiescence showing the typical, slightly double-peaked,  Balmer emission lines. The Helium emission lines are not always present in SU\,UMas. Bottom left panel: ZTF ($g$-band) light curve showing outbursts and a superoutburst in red. Bottom right panel: Expanded light curve showing the decay of the superoutburst and characteristic superhumps. 
}
\end{figure}

\paragraph{SU\,UMa} dwarf novae are characterised by occasional ``superoutbursts'' (Fig.\,\ref{fig:SUUMa}) typically lasting several weeks and recurring with a supercycle of a few hundred days \citep{1989PASJ...41.1005O,2002A&A...386..891B,2005PASP..117.1204P}. These superoutbursts are significantly brighter than the regular dwarf novae which occur in between the superoutbursts.  \citet{1990PASJ...42..135H} showed that as the separation of CVs decreases tidal forces eventually cause the disc to become eccentric and non-axisymmetric. The material in the outer edge of the disc rotates more rapidly than the orbital period and will stabilise at a resonant ratio of 3:1. The gradual accumulation of mass in the disc eventually results in a superoutburst; 
\citet{1990PASJ...42..135H} showed analytically that superoutbursts only occur when $q=M_{\mathrm{d}}/M_\mathrm{WD}\lesssim 0.3$, with $M_\mathrm{WD}$ and $M_\mathrm{d}$ the masses of the white dwarf and of the donor star. Assuming a typical value of $M_\mathrm{WD} \simeq 0.8\,\Msun$ and using the relation $M_\mathrm{d}\approx 0.065P_{\mathrm{orb}}^{5/4}$ (\citealt{1995cvs..book.....W} section 2.8) implies $\porb\lesssim2.8$\,h.  
After reaching the peak luminosity, the superoutburst light curves are modulated by periodic oscillations called superhumps. The superhump period is a few per cent longer than $P_{\mathrm{orb}}$ and the relation $P_{\mathrm{orb}}=0.9162 \times P_{\mathrm{sh}}+5.39/60$\,h  can be used to estimate $P_{\mathrm{orb}}$ \citep{2009MNRAS.397.2170G}. In reverse, if $P_\mathrm{sh}$ and $P_\mathrm{orb}$ are both measured, the superhump period excess $\epsilon=(P_{\mathrm{sh}}-P_{\mathrm{orb}})/P_\mathrm{orb}$ can be used to estimate $q$ \citep{2005PASP..117.1204P,2013PASJ...65..115K} (note that the latter authors distinguish between the slightly different superhump periods observed as the superoutbursts evolve).

\begin{figure} 
\includegraphics[width=\columnwidth]{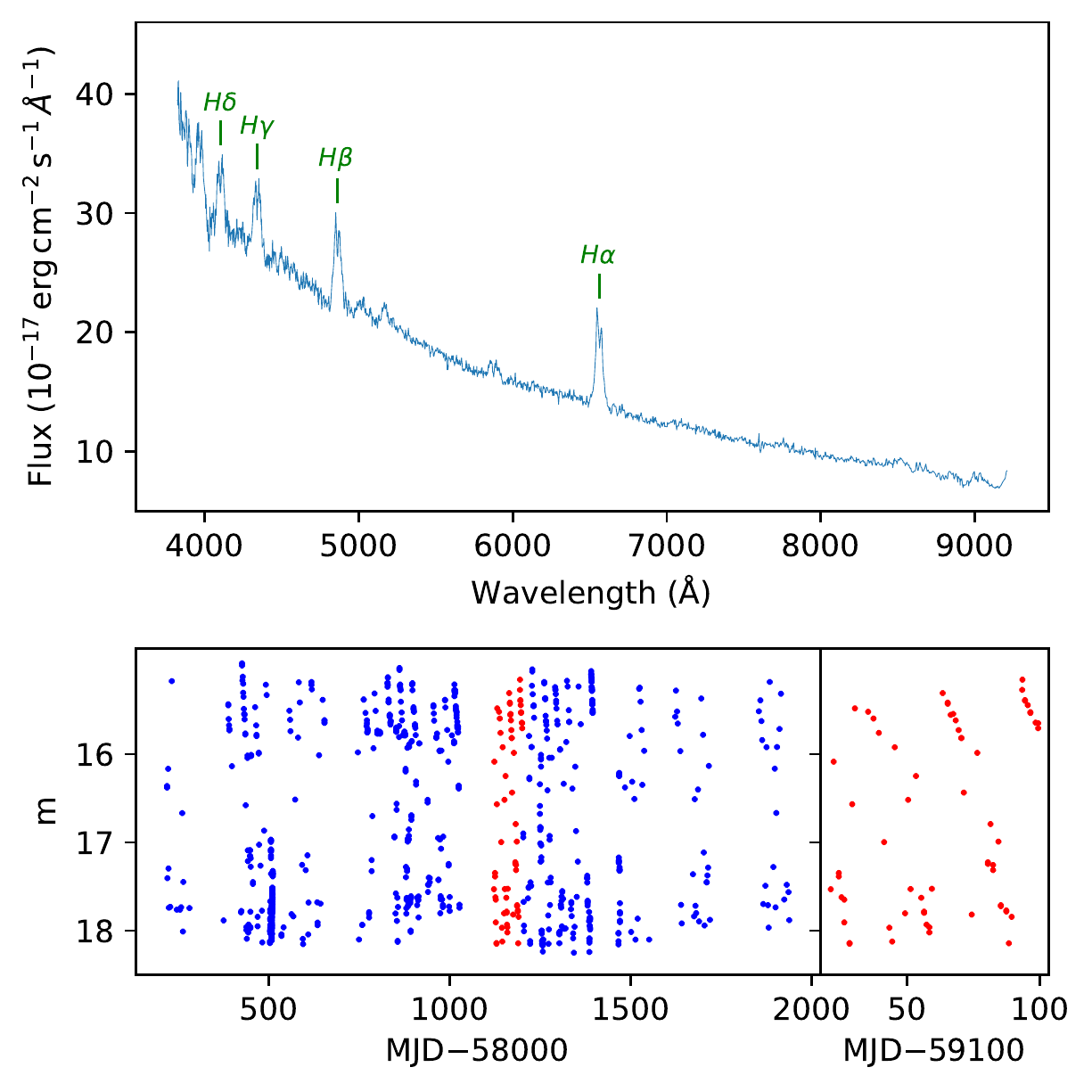}
\caption [.] {\label{fig:ERUMa} J0912+5053 (DI\,UMa) is an example of an  ER\,UMa CV. It has a high \mdot\ which is inconsistent with its short orbital period of 78\,min at the period minimum \citep{1996PASJ...48L..93K,1999PASP..111.1275F}, resulting in extremely frequent superoutbursts. The supercycle is 25\,d whilst normal outbursts occur every four days. Top panel: Smoothed SDSS spectrum in quiescence showing Balmer emission lines. Bottom left panel: ZTF ($g$-band) light curve showing frequent superoutbursts. Bottom right panel: Expanded light curve showing the frequency of the superoutbursts. 
}
\end{figure}

\paragraph{ER\,UMa}
Dwarf novae (Fig.\,\ref{fig:ERUMa}) are a sub-type of SU\,UMa CVs characterised by an unusually short supercycle ($20-50$\,d) so that they spend over a third of the time in superoutburst. The intervening time between superoutbursts is punctuated by normal outbursts every few days  \citep{1995PASP..107..443R,2013arXiv1301.3202K}. The cause for this behaviour is an \mdot\ around four times that of a typical SU\,UMa CV.   

\begin{figure} 
\includegraphics[width=\columnwidth]{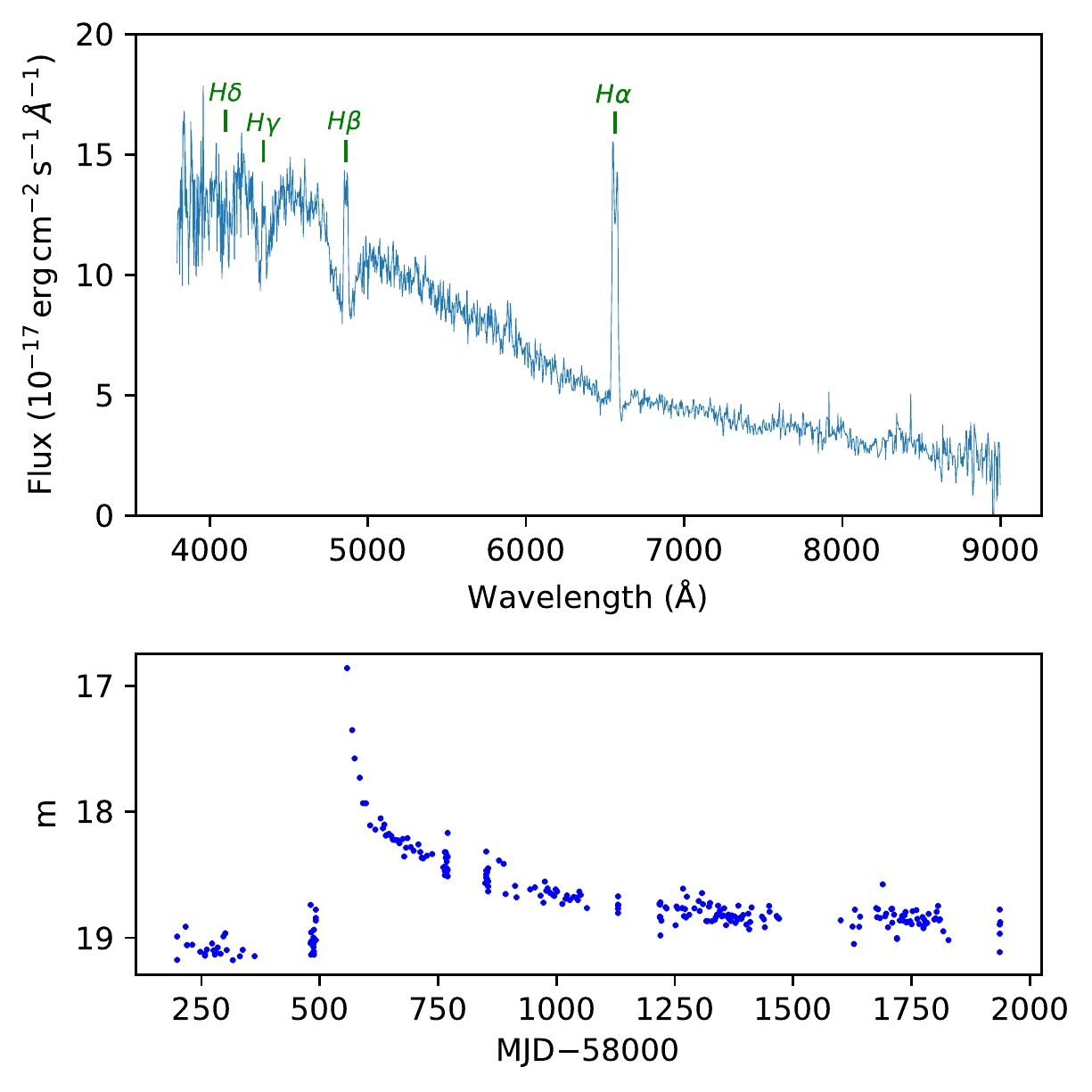}
\caption [.] {\label{fig:WZSge} J1610$-$0102 (V386\,Ser) is an example of a  WZ\,Sge CV. Top panel: Smoothed SDSS spectrum in quiescence showing double-peaked Balmer emission lines from the accretion disc. The spectrum of the white dwarf is visible at the blue end with characteristic wide Balmer absorption lines. Bottom panel: ZTF ($g$-band) light curve showing a single characteristically large superoutburst. Many systems believed to be WZ Sge stars, based on their spectra and periods, have no such recorded outbursts and identification rests on other evidence such as spectra. 
}
\end{figure}

\paragraph{WZ\,Sge}
dwarf novae are also a sub-type \citep{2002PASP..114..721P,2015PASJ...67..108K} of SU\,UMa CVs. Their donors have lost most of their mass and they have low \mdot\ with very infrequent, albeit bright, superoutbursts (Fig.\,\ref{fig:WZSge}). Most WZ\,Sge systems do not exhibit normal outbursts in between their superoutbursts. The low \mdot\ and consequential low disc luminosity, often result in the white dwarf being visible in the optical spectrum. In many cases this is the best evidence for a WZ\,Sge classification as commonly used definitions \citep{2001PASJ...53.1191K,2015PASJ...67..108K} are based upon observation of an outburst, together with the characteristic early superhumps, which may not occur for decades. Note that low \mdot\ is not sufficient in itself for a classification; there also needs to be evidence of the low mass of the donor.

\begin{figure} 
\includegraphics[width=\columnwidth]{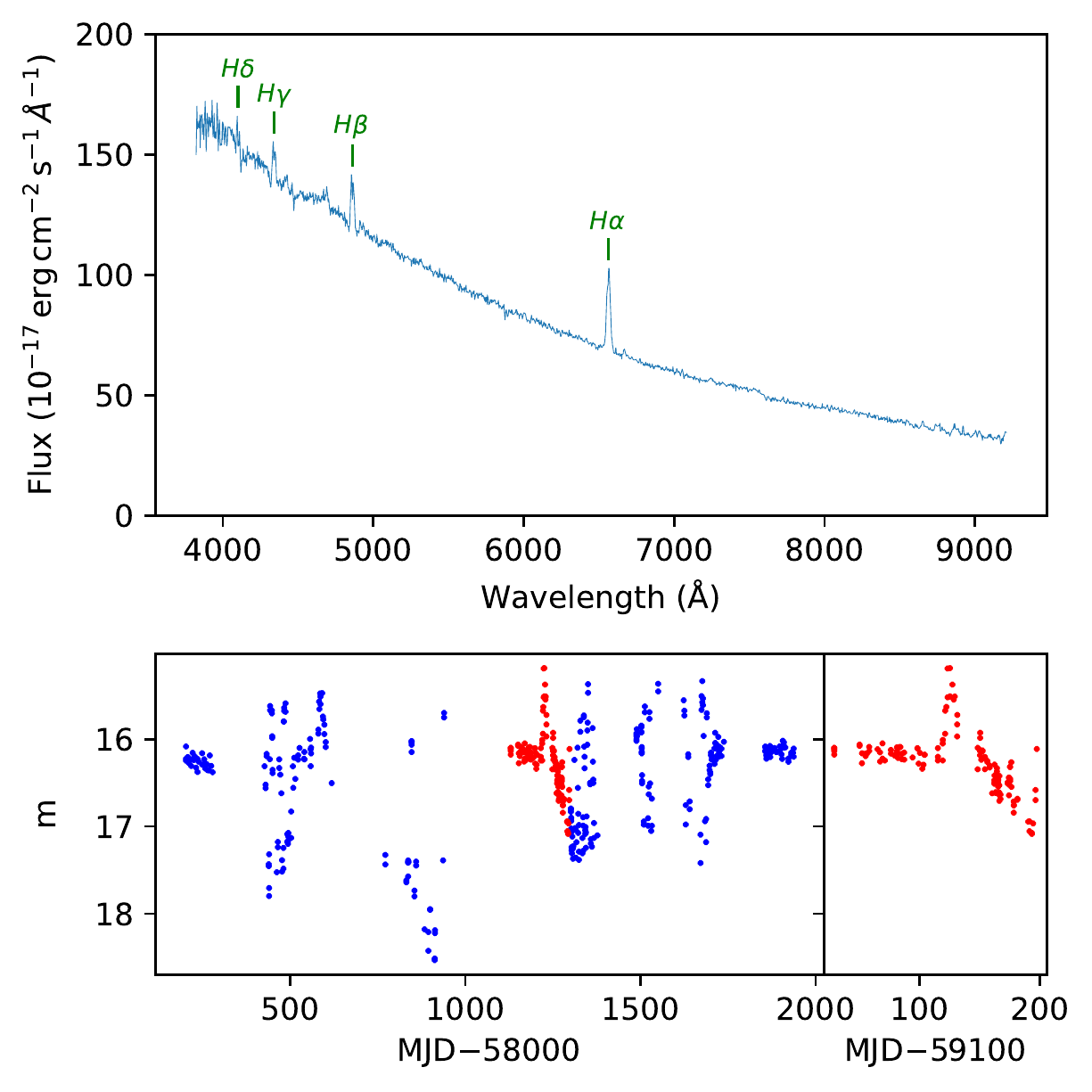}
\caption [.] {\label{fig:ZCam} J1028+2148 (MV\,Leo) is an example of a Z\,Cam CV. Top panel: Smoothed SDSS spectrum in ``standstill'' showing Balmer  emission lines. Bottom left  panel: ZTF (merged and scaled $r-$  and $g-$bands shown in blue) light curve showing standstill periods as well as outbursts and the low, quiescent, level. The outburst is highlighted in red. Bottom right panel: Expanded light curve showing a period of ``standstill'' followed by an outburst and a decline to quiescence.
}
\end{figure}

\paragraph{Z\,Cam} \label{sec:z_cam}
dwarf novae are dwarf novae with the unusual characteristic (Fig.\,\ref{fig:ZCam}) of having ``standstills'' of luminosity midway between the quiescent level and the elevated level associated with a dwarf nova outburst  \citep{2014JAVSO..42..177S}. This is believed to be due to having an \mdot\ that fluctuates around the critical level and is too high to sustain the thermal limit cycle associated with dwarf novae. The IW\,And stars \citep{2014A&A...569A..48H} form a sub-category of Z\,Cam and show continual outbursts (not unlike dwarf novae) during the  ``standstill''. 

\subsubsection{Classical Novae}
Classical novae \citep{bode2012classical} were first observed by ancient Chinese astronomers as eruptions with $\Delta m =8-15$  that decayed over a period of a few to several hundred days \citep{1977QJRAS..18..443C}. They are caused by an accumulation of hydrogen on the surface of the white dwarf which eventually reaches a critical mass and pressure thereby initiating a runaway nuclear reaction \citep{1971MNRAS.155..129S} that expels an expanding shell \citep{1983ApJ...268..689C,2012AJ....143..143S} and may also influence the evolution of the CV \citep{2016ApJ...817...69N,2022ApJ...938...31S} by causing frictional angular momentum loss \citep{2016MNRAS.455L..16S}. Classical novae can therefore occur in all types of CVs including magnetic ones \citep{1988ApJ...332..282S}. The accumulation of hydrogen is influenced by \mdot\ and therefore classical novae are more common in high \mdot\ systems typically with periods $\porb\gtrsim3$\,h. 

\begin{figure} 
\includegraphics[width=\columnwidth]{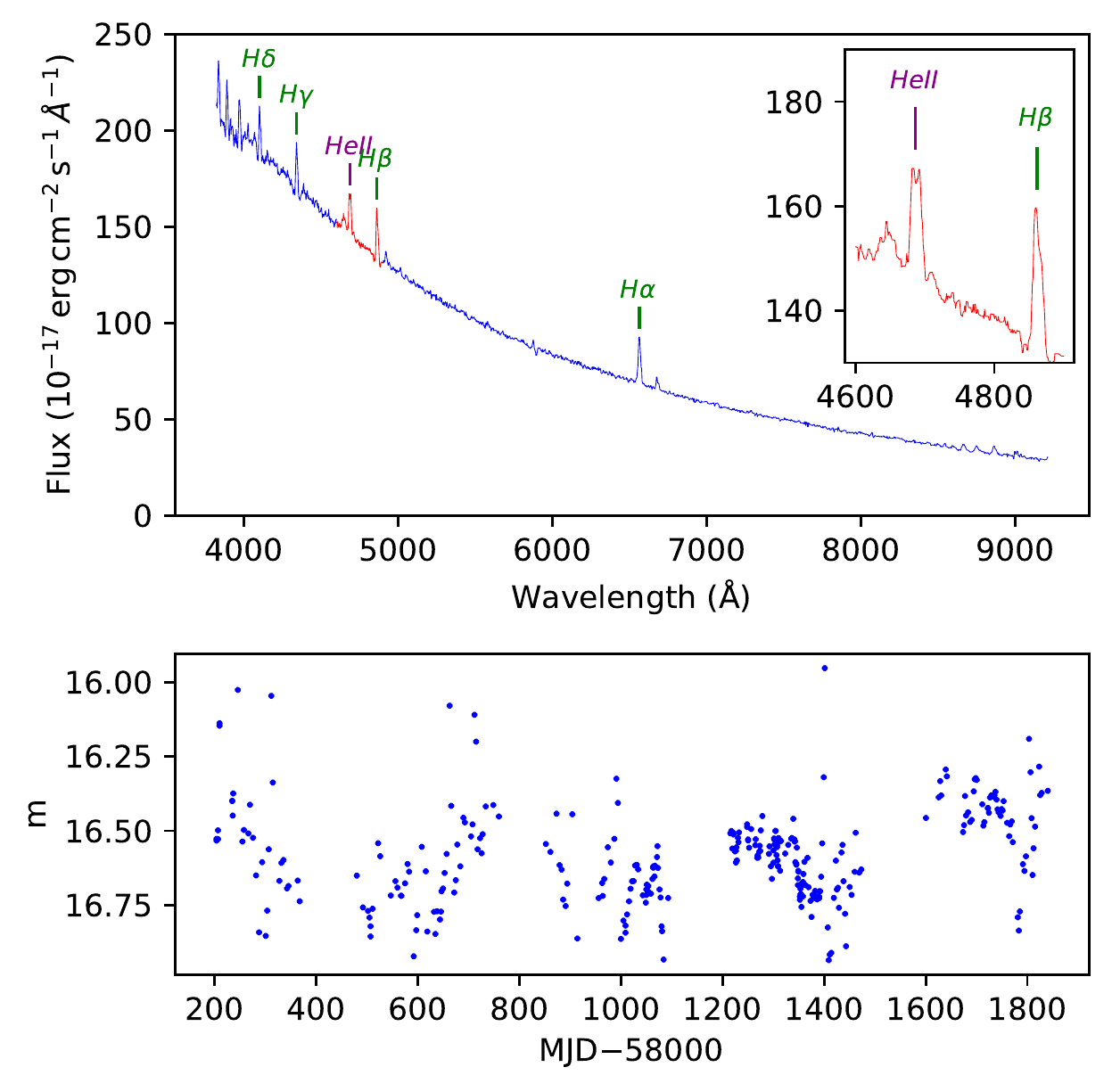}
\caption [.] {\label{fig:NA} J1545+1422 (CT\,Ser) is an example of an  old nova that erupted in 1948, and that now appears as a novalike variable. Top panel: Smoothed SDSS spectrum showing Balmer and \ion{He}{ii} emission lines. Top inset panel: Expanded spectrum showing the relative strength of the H$\beta$ and  \ion{He}{ii} lines. \citep{2014A&A...564A..76M} Bottom panel: ZTF ($g$-band) light curve illustrating the relatively constant emission from the disc in its high state.
}
\end{figure}

By definition an eruption of a classical nova has been recorded at some time. The  spectrum of  a classical nova  varies considerably depending upon the time since this eruption. The rapidly expanding shell will initially dominate the spectrum, with broad emission lines of hydrogen and helium. As it expands further and becomes more diffuse it resembles a planetary nebula and exhibits forbidden lines such as $[\ion{S}{ii}]$~6717\,\AA. The rapid expansion of the shell also results in spectral lines exhibiting  P\,Cygni profiles. Eventually (years to decades) the spectrum becomes that of the pre-eruption CV. Most classical novae revert to being novalike variables (Fig.\,\ref{fig:NA}), which is unsurprising as the nova recurrence time is inversely related to the accretion rate, and hence the nova population is dominated by novalike variables \citep{2005ApJ...628..395T}.

\subsubsection{Novalike variables}
\begin{figure} 
\includegraphics[width=\columnwidth]{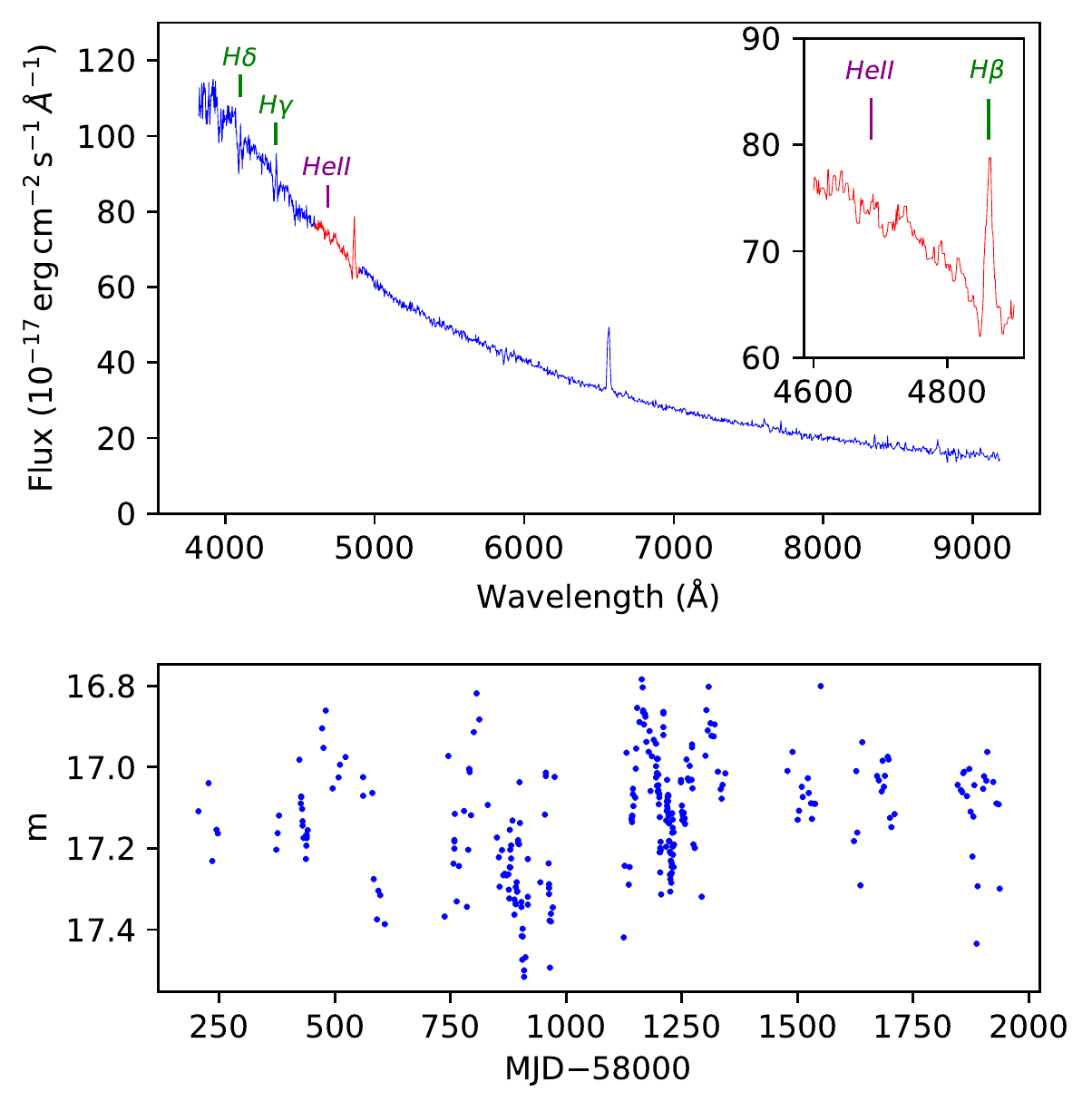}
\caption [.] {\label{fig:NL} J0758+1043 is an example of a  novalike variable. Top panel: Smoothed SDSS spectrum showing Balmer H$\alpha$ and H$\beta$ emission lines together with H$\gamma$ and H$\delta$ absorption lines. Top inset panel: Expanded spectrum showing  the H$\beta$ line and absence of \ion{He}{ii} lines. The strength of   \ion{He}{ii}  4686\,\AA\ emission lines varies with orbital phase far more than Balmer emission lines~$-$~and often exceeding the strength of the H$\beta$ line \citep{2014ApJ...786...68H}. Bottom panel: ZTF ($g$-band) light curve showing the relatively constant emission from the disc in its hot, steady state.
}
\end{figure}

Novalike variables \citep{1996ASSL..208....3D} have bright spectra dominated by the hot and stable disc  (Fig.\,\ref{fig:NL}). The hydrogen in  their discs is almost fully ionised and their spectra therefore exhibit absorption lines, often with inset emission lines. They often have strong disc winds which can appear as P\,Cygni spectral lines \citep{2022MNRAS.510.3605I}. Novalikes sometimes exhibit changes between high and low states and observational phenomena in their light curves have given rise to a number of sub-types. The VY\,Scl sub-type shows relatively deep ($\Delta m =2-5$) low states which can last a few weeks to about a year, during which short brightenings occur \citep{2019MNRAS.489.2961B}, the nature of which is still discussed \citep{2017Natur.552..210S}. This contrasts with the IW\,And class (Section \ref{sec:z_cam}) in which characteristic outbursts occur during the (higher state) standstills \citep{2019PASJ...71...20K}. 

\subsubsection{AM\,CVn} \label{section:amcvn}
\begin{figure} 
\includegraphics[width=\columnwidth]{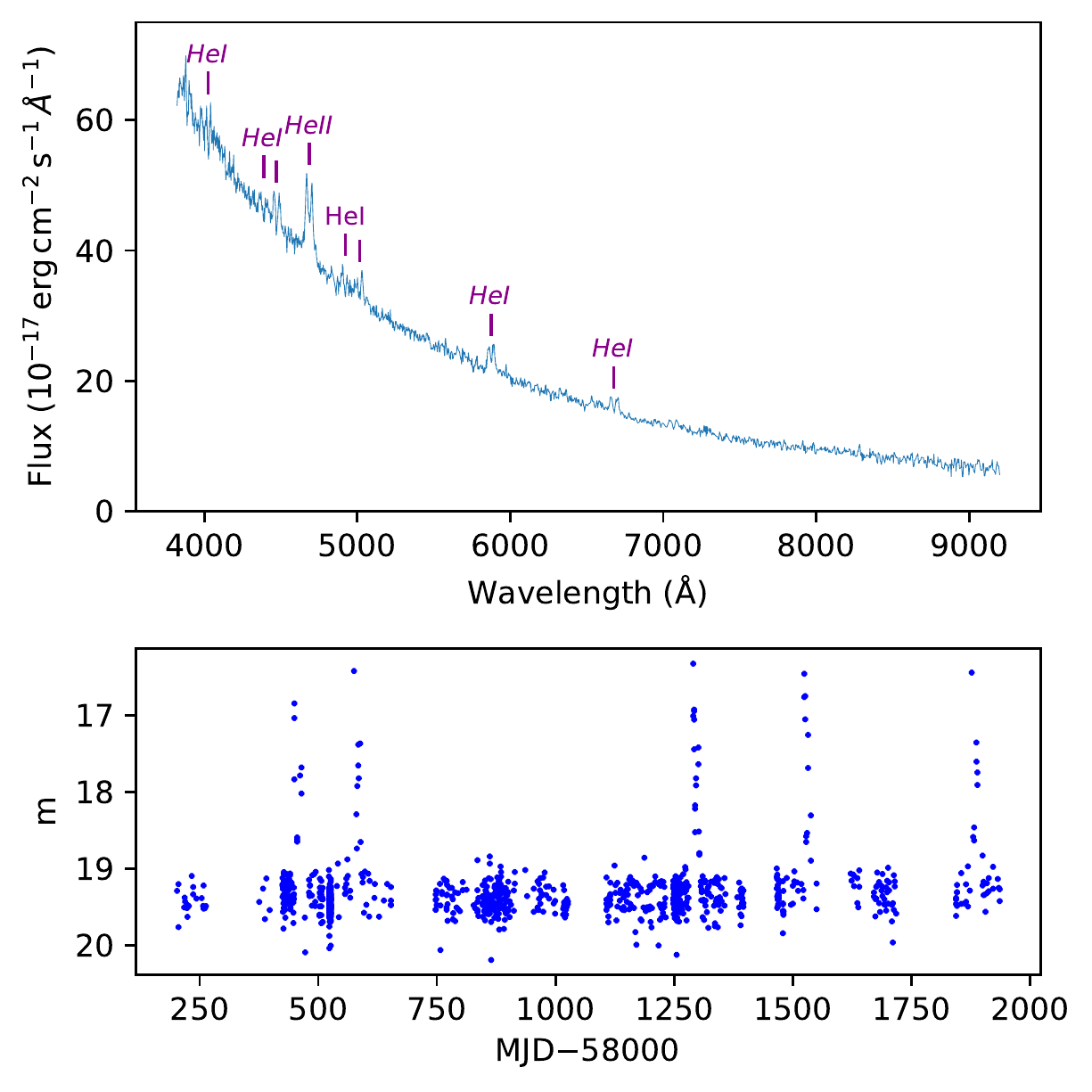}
\caption [.] {\label{fig:AMCVn} J0926+3624 is an example of an AM\,CVn CV. Top panel: The smoothed SDSS spectrum shows \ion{He}{i} and \ion{He}{II}  emission lines but no hydrogen lines. The absence of hydrogen lines is a characteristic of AM\,CVn CVs. Bottom panel: ZTF (merged and scale $r-$ and $g-$bands) light curve showing outbursts.
}
\end{figure}

AM\,CVn systems (Fig.\,\ref{fig:AMCVn}) are identified by helium lines and an absence of hydrogen lines in the optical spectrum \citep{2018A&A...620A.141R}. They are short-period systems   ($\porb\simeq5-68$\,min) but their evolutionary origin is unclear with at least three potential formation channels leading from their progenitors \citep{2010PASP..122.1133S}. AM\,CVn systems are rare, with only $\simeq60$ systems known \citep{2018A&A...620A.141R}. They are good verification binaries for the space-based \textit{LISA} gravitational wave detector \citep{2017arXiv170200786A} given their short periods and separations \citep{2004MNRAS.349..181N,2021ApJ...910...22L}.

\subsection{Magnetic CVs}
Perhaps a third \citep{2020MNRAS.494.3799P} of CVs have a white dwarf with a magnetic field $B\gtrsim1$\,MG and these are termed magnetic. The origin of this strong field is not well understood; it may be formed during the common-envelope event or sometime later \citep{2020MNRAS.492.1523B}. \citet{2021NatAs...5..648S} propose a model whereby the magnetic CV is born without a strong field  and the white dwarf is spun up by accretion. The cooling white dwarf in the CV then forms a dynamo as it crystallises. The strengthening magnetic field causes the system to become detached and the only accretion is from the donor wind which follows magnetic field lines. This pre-polar stage \citep{2022MNRAS.513.3858H} continues until after the donor becomes fully convective and mass transfer recommences at which point the CV is a polar or intermediate polar \citep{2021MNRAS.508..561W}. This model is consistent with observational evidence (fig.\,9 in \citealt{2015SSRv..191..111F}) as  intermediate polars (with $B\lesssim10$\,MG) are, in general,  observed to have orbital periods $\porb\gtrsim3$\,h whilst polars (with $B \gtrsim10$\,MG) are observed to have typical orbital periods $\porb\lesssim2$\,h . 

\subsubsection{Polar}
\begin{figure} 
\includegraphics[width=\columnwidth]{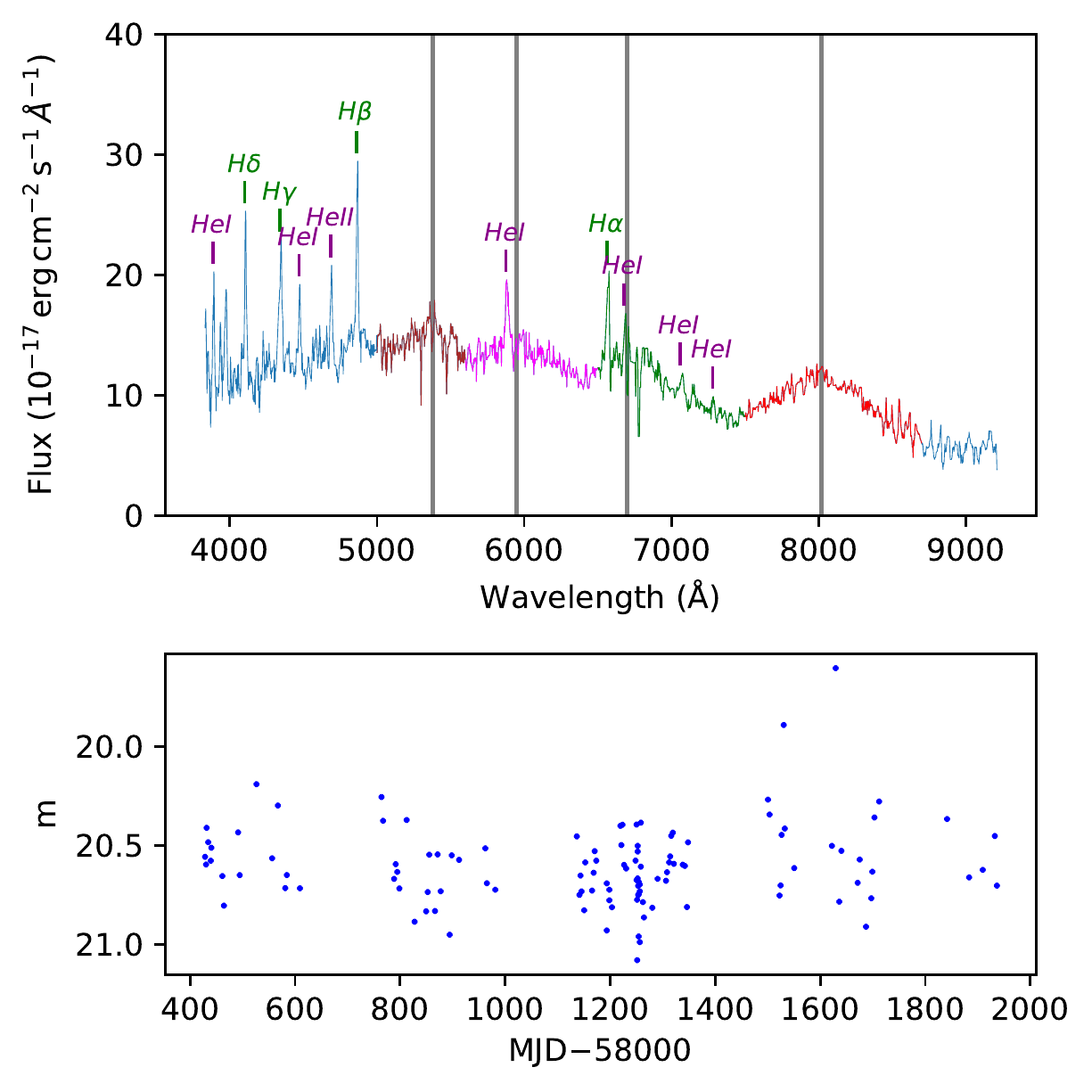}
\caption [.] {\label{fig:polarhump} J0921+2038 is an example of a polar. Top panel: The smoothed SDSS spectrum (blue) shows Balmer, \ion{He}{i} and characteristic \ion{He}{ii} emission lines.  The four cyclotron humps (brown, magenta, green and red) provide  clear evidence that this is a polar.  \citet{2008PASP..120..160S} determined that these are the fifth, sixth, seventh and eighth harmonics (shown in grey) due to a magnetic field of $B=32$\,MG.  Bottom panel: ZTF ($g$-band) light curve showing the variability due to cyclotron beaming. There are no outbursts as polars do not have discs in which to store accreting material and hence they do not undergo a thermal limit cycle.
}
\end{figure}

\begin{figure} 
\includegraphics[width=\columnwidth]{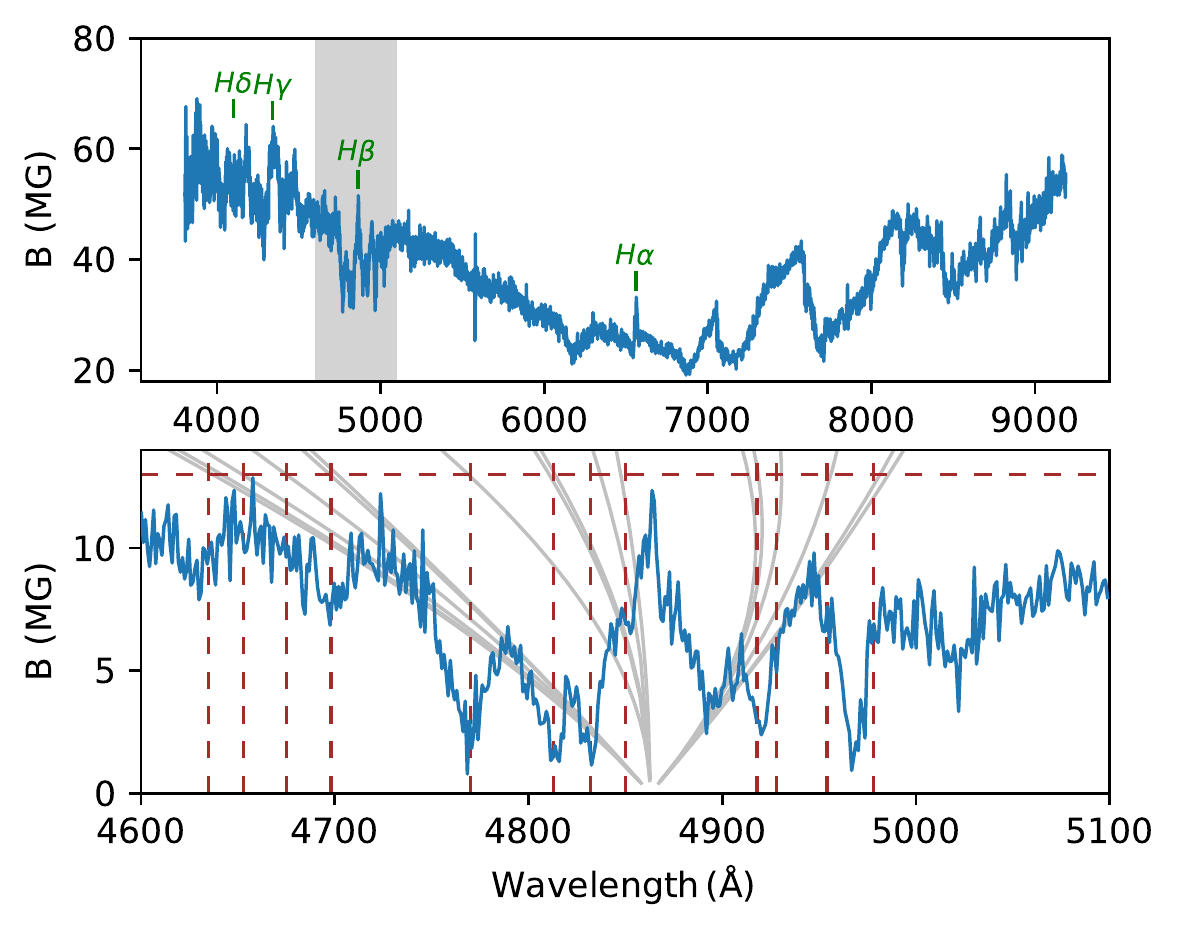}
\caption [.] {\label{fig:zeeman} J1105+2506 (ST\,LMi) is an example of a polar showing Zeeman splitting. The spectrum is shown in the top panel.The lower panel shows a model of the H$\beta$ line evolving vertically (grey lines) with increased magnetic field  \citep{2014ApJS..212...26S}. The spectrum is over-plotted in blue with  flux scaled to roughly match the amplitude of the grey lines. The split lines are identified by vertical dashes; they coincide with the grey lines at $B\simeq 13$\,MG.    
}
\end{figure}

The magnetic field of the white dwarf in polars \citep{1990SSRv...54..195C,2000SSRv...93..611W}, typified by the prototype AM Her, is generally greater than 10\,MG. In this case the Alfv\'en radius is beyond the circularisation radius, preventing the formation of an accretion disc, and the accretion flow follows field lines from the inner Lagrange point to impact on the white dwarf. These shock-heated impact zones ($\approx 10^{8}$\,K) give rise to X-ray radiation and also characteristic \ion{He}{ii} emission lines (Fig.\,\ref{fig:polarhump}). The cyclotron radiation is sometimes observed in spectra as ``humps'' (see for example \citealt{2000A&A...353..646T}). These humps are harmonics of the cyclotron frequency $f=(eB)/(2 \pi m_e)$, with $e$ and $m_e$ the charge and the mass of the electron, due to the magnetic field $B$ and can appear in the infrared (e.g. AM\,Her, $B\simeq15$\,MG \citealt{1991MNRAS.251P..37B}), optical (e.g. MR\,Ser, $B\simeq25$\,MG; \citealt{1993A&A...278..487S}) or ultraviolet (e.g. AR\,UMa, $B\simeq200$\,MG; \citealt{2001ApJ...555..380G})  according to field strength. Cyclotron emission is directional and results in cyclotron beaming which manifests itself as periodic behaviour. Polars can also be identified either by polarimetry, as cyclotron radiation is polarised \citep{1981ApJ...244..569C,2019ASSL..460..247S},  or by the Zeeman splitting of Balmer lines from the white dwarf \citep{1995MNRAS.273...17F} that is sometimes visible in spectra obtained during low states when accretion nearly ceases (Fig.\,\ref{fig:zeeman}) and can be used to estimate the strength of the magnetic field.  A consequence of the strong magnetic field is that the spin period of the white dwarf normally equals \porb. This is thought to be due to the interaction of the strong white dwarf magnetic field with the donor magnetic field generating a torque which over time causes the white dwarf to spin down until it equals \porb; the equilibrium situation \citep{1984MNRAS.211...69C}. Asynchronous polars \citep{1994A&A...283..163P,1999A&A...343..132C} have a white dwarf spin period slightly different from \porb \, which can be detected in periodograms. Asynchronous polars are believed to have been knocked out of exact synchronisation by an event such as a nova \citep{2018MNRAS.479..341P}.

\subsubsection{Pre-polars}

\begin{figure} 

\includegraphics[width=\columnwidth]{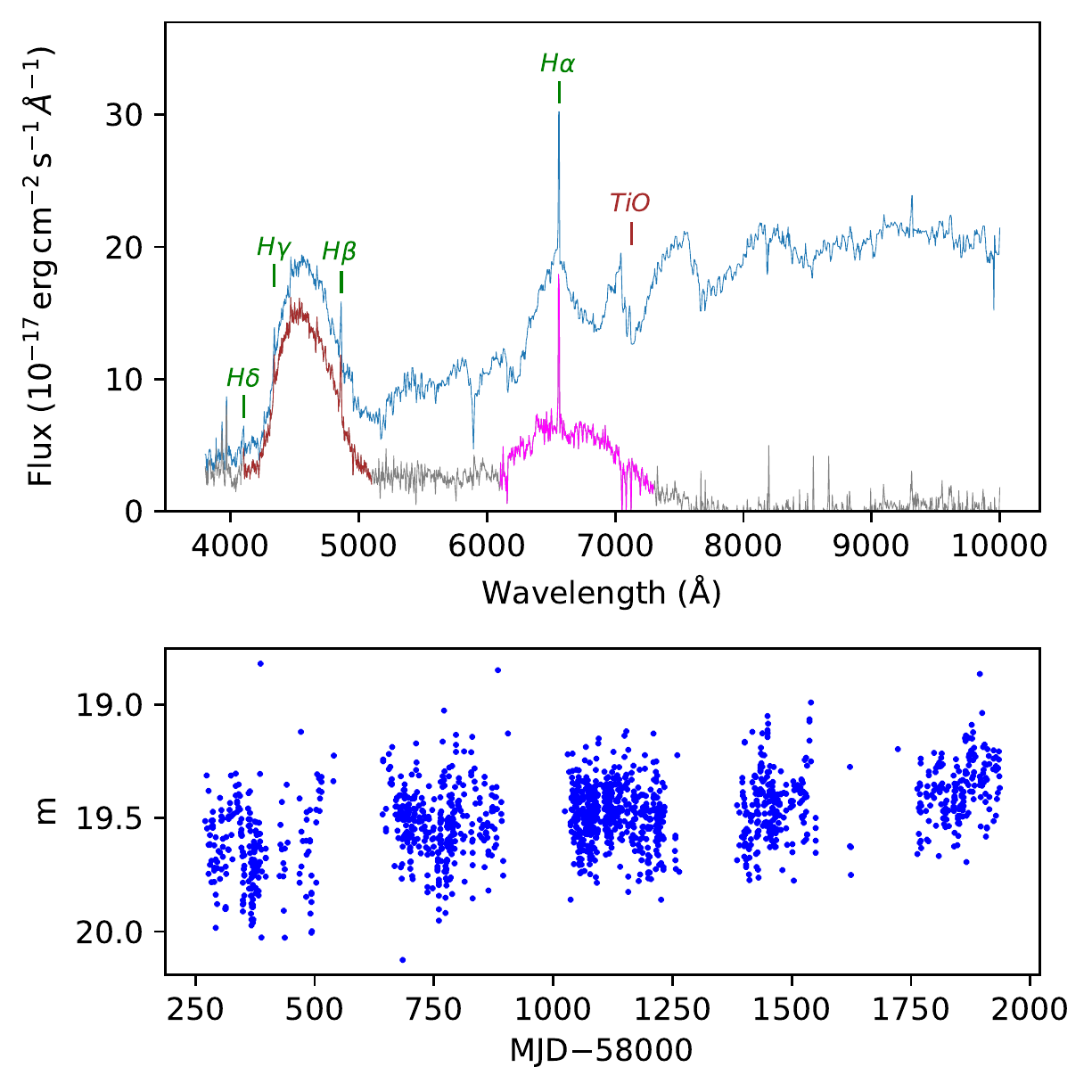}
\caption [.] {\label{fig:prepolarhump} J0049+2226 is an example of a pre-polar. Top panel: The smoothed SDSS spectrum (blue) shows Balmer emission lines. The donor is visible at the red end of the spectrum. The cyclotron hump visible between 4200\,\AA\  and 5000\,\AA\  (brown) is clear evidence that this is a polar or pre-polar. Following \citet{2021MNRAS.502.4305P} we subtracted a model spectrum \citep{2017ApJS..230...16K} of an M3 dwarf (result in grey)  to reveal a further cyclotron hump (magenta) at $6840$\,\AA. Modelling of harmonic frequencies for varying magnetic fields yielded a good fit for a magnetic field strength of $B\simeq78$\,MG with the humps forming the second and third harmonics. The fact that the donor is visible implies that the accretion rate is low and this implies that this is a pre-polar accreting from the donor wind \citep{2021NatAs...5..648S}. Bottom panel: ZTF ($g$-band) light curve showing the variability attributable to cyclotron beaming which revealed the orbital period ($P_{\mathrm{orb}}=5.92$\,h). 
}
\end{figure}

White dwarfs in pre-polars \citep{1999A&A...343..157R,2002ASPC..261..102S,2021MNRAS.502.4305P} (also called low-accretion rate polars) accrete via a stellar wind from the donor rather than through Roche-lobe overflow. These detached systems therefore have very low accretion rates with \mdot\ less than one per cent of that of other CVs \citep{2005ApJ...630.1037S}. Cyclotron humps are the defining characteristic of pre-polars; typically the second and third harmonics are visible in the optical $-$ although the humps can be masked by the donor spectrum (Fig.\,\ref{fig:prepolarhump}). Their cyclotron emission does exhibit circular polarisation \citep{2022MNRAS.513.3858H}, but they have very low X-ray luminosities \citep{2011A&A...530A.117V}. In the evolution scenario of \citet{2021NatAs...5..648S} pre-polars represent a natural phase in the evolution of CVs once the white dwarf develops a strong magnetic field.

\subsubsection{Intermediate polars}
\begin{figure} 
\includegraphics[width=\columnwidth]{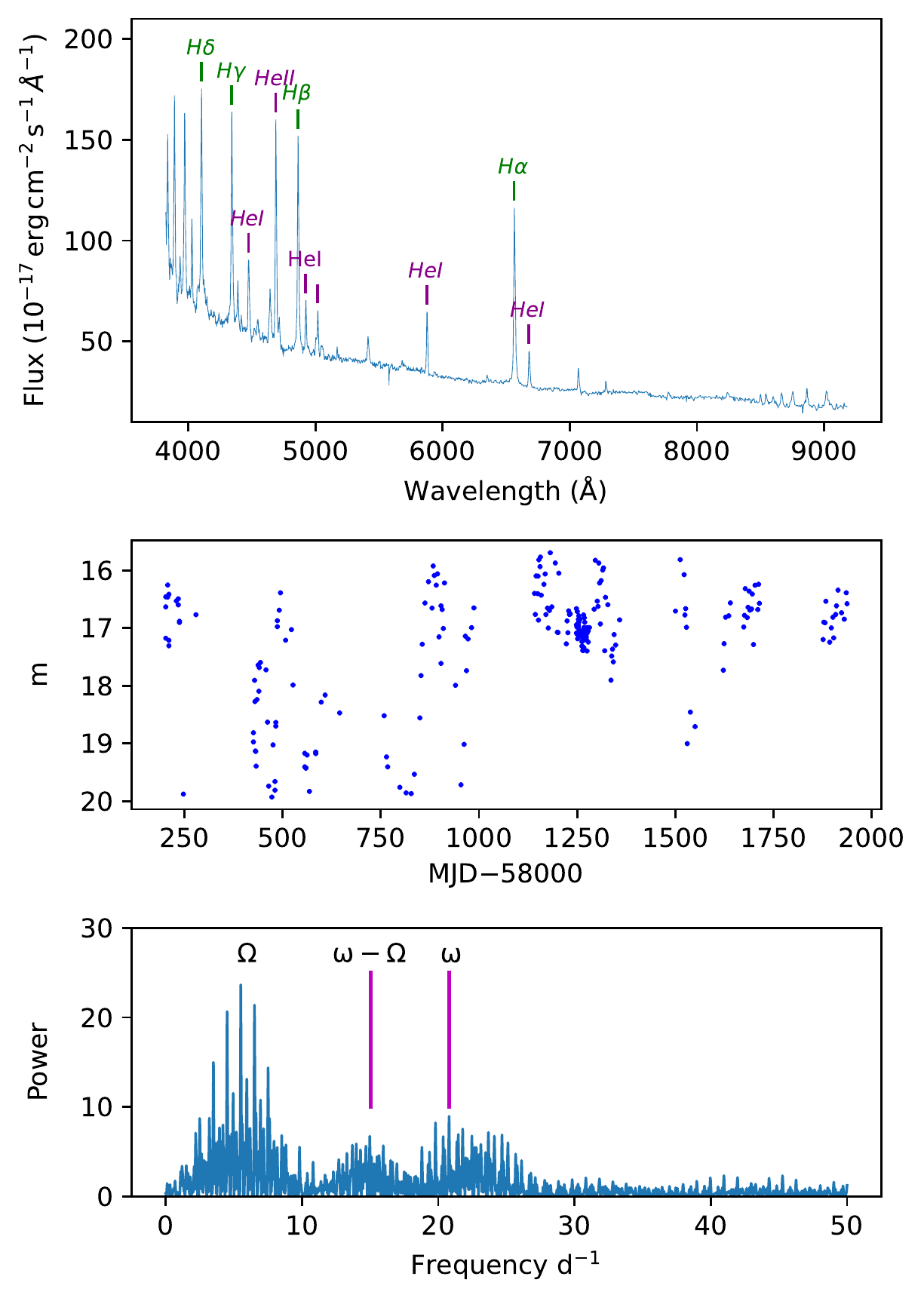}
\caption [.] {\label{fig:IP} J0946+1350 (HY\,Leo, HS\,0943+1404) is an example of an intermediate polar. Top panel: The smoothed SDSS spectrum shows Balmer, \ion{He}{i} and \ion{He}{ii} emission lines. Middle panel:  ZTF ($g$-band) light curve showing the variability due to the spin of the white dwarf and the accretion stream. The alternation between high and low states is also evident. Bottom panel: Periodogram taken from archival Jacobus Kapteyn Telescope data showing the orbital frequency ($\Omega$), spin period ($\omega$) and the beat period together with numerous aliases. For a detailed period analysis, see \citet{2005A&A...440..701R}.
}
\end{figure}

Intermediate polars (\citealt{1994PASP..106..209P}) such as their prototype DQ~Her, have weaker white dwarf magnetic fields than polars.  The magnetic field is too weak to force synchronisation of the spin period of the white dwarf resulting in a spin period significantly shorter than the orbital period
\citep{2004ApJ...614..349N} $-$ the defining characteristic of an intermediate polar  (Fig.\,\ref{fig:IP}). The vast majority of intermediate polars have a truncated accretion disc \citep{2017A&A...602A.102H}. In common with non-magnetic CVs material from the donor passes through the inner Lagrange point and pursues a ballistic path until it merges with the disc. The disc is truncated at the  Alfv\'en radius and material then follows field lines from the innermost part of the disc to one or both magnetic poles.  The presence of the disc can result in outbursts \citep{2017A&A...602A.102H}. Magnetically funneled accretion onto the magnetic poles of the white dwarf results in pulsations in X-ray light curves \citep{1991MNRAS.248..233H}, which are the definitive observational proof of a CV being an intermediate polar .

\section{Methods} \label{section:methods}
In compiling this CV catalogue, we followed a two-step process. In the first step we identified a shortlist of potentially new CVs from the 5\,789\,200 spectra of SDSS~III and IV. In the second step we then undertook a detailed classification of these potential CVs, as well as revisited all the CVs previously identified in SDSS~I and II.

\subsection{Identification}
\allsdss\, focused upon a number of surveys none of which explicitly targeted CVs. Fortuitously the (BOSS and eBOSS) \citep{2013AJ....145...10D,2016AJ....151...44D} targets included  quasars which share a similar colour parameter space to CVs. Since the publication of the first comprehensive orbital period distribution of SDSS \citep{2009MNRAS.397.2170G}, we  progressively built up a list of candidate CVs based upon visual inspection of the flux and wavelength calibrated SDSS spectra \citep{2015PASP..127..397W,2016AJ....151...44D}. The spectra were reviewed multiple times by a range of individuals using a number of pre-selections in terms of colour and filtering on the redshifts determined by the SDSS project. 

This visual process was augmented by a supervised machine learning process. A  data set of known CVs and proven non-CVs was used to train a Random Forest (RF) classifier \citep{rfarticle}. The RF model used features extracted from the equivalent widths of 12 spectral lines and the average gradient of the spectrum measured over five intervals.   The different identification methods resulted in a substantial overlap in the sub-sets of CV candidates that they identified, and whereas we cannot claim to have identified all the CVs that were spectroscopically observed by SDSS, we are confident that our search has picked up the majority of them. CVs that remain overlooked will either have extremely low signal-to-noise ratio spectroscopy, or have colours very close to those of single main-sequence stars, and weak emission lines.

\subsection{Classification}
No single method can reliably classify CVs~--~it is necessary to consider all the available information (SDSS spectroscopy, time-series photometry, \textit{Gaia} astrometry, colours from broad-band photometry, imaging) and seek a classification of CV sub-type that is consistent with all of the data by making use of the diagnostics outlined in section\,\ref{section:sub_types}. Where possible, we attempted to classify the SDSS CVs down to the sub-types illustrated in Fig.\,\ref{fig:diagram2}. However, in $\simeq13$~per cent of  cases only high level classifications such as dwarf nova or magnetic CV are possible due to the limited amount of information that is currently available. In addition, in the $\simeq20$~per cent of cases where a classification is very likely but not certain a colon (:) has been appended.   

We began the classification of each object  with a literature search using Simbad \citep{2000A&AS..143....9W} and a search of the AAVSO VSX database \citep{2006SASS...25...47W}. Where an existing classification exists we checked that it is consistent with the \allsdss\ spectra together with data from \textit{Gaia}, VizieR,  ZTF, CRTS and \textit{TESS}. This process resulted in refining \fxnum{262} classifications. It also revealed \fxnum{18} systems, previously categorised as CVs that are now known not to be CVs (see section \ref{section:misclassifications}). For each CV, the references for the initial discovery, orbital period measurement, and optical spectroscopy are documented in the Supplementary Data. 

Detailed consideration of the available data is discussed in the following sections.  

\subsubsection{Spectral analysis}

We first plotted the SDSS spectrum and then used the spectrum to calculate a synthetic \textit{Gaia} $G$-band magnitude using the \textsc{ python pyphot} \footnote{\url{https://github.com/mfouesneau/pyphot}} package. In addition to the SDSS spectrum we also created a broad-band (ultraviolet to infrared) spectral energy distribution (SED) diagram using data downloaded from VizieR \citep{2014ASPC..485..219A} and over-plotted the spectrum. These plots are available as PNG files in the supplementary data.

We initially compared the SED with the SDSS spectrum. There can be a number of possible explanations if the spectrum is not consistent with the SED. Firstly the flux calibration of all or part of the spectrum may be faulty which we establish by comparing the synthetic magnitude with archive observations. Secondly the spectrum may have been taken at a time when the CV was in a different state to that when the photometry was obtained. Thirdly the SED may have been contaminated by flux from a nearby object~--~which is why we check the PanSTARRS images (see Section \ref{sec:imagery}). In some cases, the photometry from VizieR traces distinct SEDs, suggesting that photometry of the system was obtained in different accretion states. In these cases, the epochs of the SED photometry can be compared with the available light curves (section \ref{section:lightcurves}).

We next tried to identify the donor and/or the white dwarf in the SDSS spectrum. At the blue end of the spectrum broad Balmer absorption lines are indicative of the white dwarf. Conversely if the Balmer jump is visible in emission then the disc or accretion stream is much brighter than the white dwarf, which is typical of SU\,UMa CVs, polars, and intermediate polars. Absorption from the Mg triplet 
at $5167, 5172\, \mathrm{and} \,5183$\,\AA\, (Fig. \ref{fig:UGem}) is indicative of an early M-type or late K-type donor (suggestive of long, $\gtrsim2$\,h, orbital periods) whereas a rising continuum in the red part of the spectrum (Fig.\,\ref{fig:WZSge}), often broken by TiO ``sawtooth'' molecular absorption bands indicates a mid-to-late type M~dwarf (suggestive of shorter orbital periods). In addition, the detection of a blackbody-shaped bump in the ultraviolet or infrared part of the photometric SED can identify a white dwarf or M~dwarf, respectively. Care was taken to distinguish the Balmer absorption lines from a white dwarf from those originating in a disc. Disc absorption lines tend to be narrower and flat-bottomed as they are formed  by Doppler broadening due to Keplerian velocities in the disc whilst the Balmer absorption lines of the white dwarf  are pressure-broadened due to the high surface gravities in the atmosphere. A clear white dwarf signature in the spectrum is indicative of a low level of accretion. If in addition the donor is not visible it suggests that the system is a WZ\,Sge candidate. 

Next, we considered the emission lines detected. In addition to the Balmer and \ion{He}{i} lines, we probed for the presence of $[\ion{S}{ii}]$  6717\,\AA\ which is indicative of contamination by a nebula and is not seen in CV spectra. Broad and strong emission lines at arbitrary wavelengths are associated with highly red-shifted quasars. Finally,  \ion{He}{ii} 4686\,\AA\ emission is indicative of a polar or intermediate polar particularly if the ratio of its equivalent width to that of H$\beta$ is greater than 0.4 \citep{1992PhDT.......119S}

We also considered the shape of emission lines. Balmer emission lines are indicative of an accretion flow, either the disc in a non-magnetic CV, or a stream in polars. In non-magnetic CVs, the width of the Balmer emission from the disc will depend upon the inclination of the system \citep{1986MNRAS.218..761H}, with broad lines and the detection of double-peaks indicating a high inclination. Furthermore deep central absorption dips embedded in  broad double peaked Balmer lines are a good indication that a system is eclipsing. Asymmetric emission lines are associated with the velocity dispersion of the accretion stream of polars whilst very thin lines are typical of background nebulae and detached systems.

Lastly we also superimposed upon the SDSS spectrum the five SDSS photometric magnitudes obtained in the original observations. These can reveal inconsistencies and in particular can reveal cyclotron humps (due to one observation being taken at an orbital phase when the cyclotron emission peaked). 
  
\subsubsection{Light curves}\label{section:lightcurves}

\begin{figure} 
\includegraphics[width=\columnwidth]{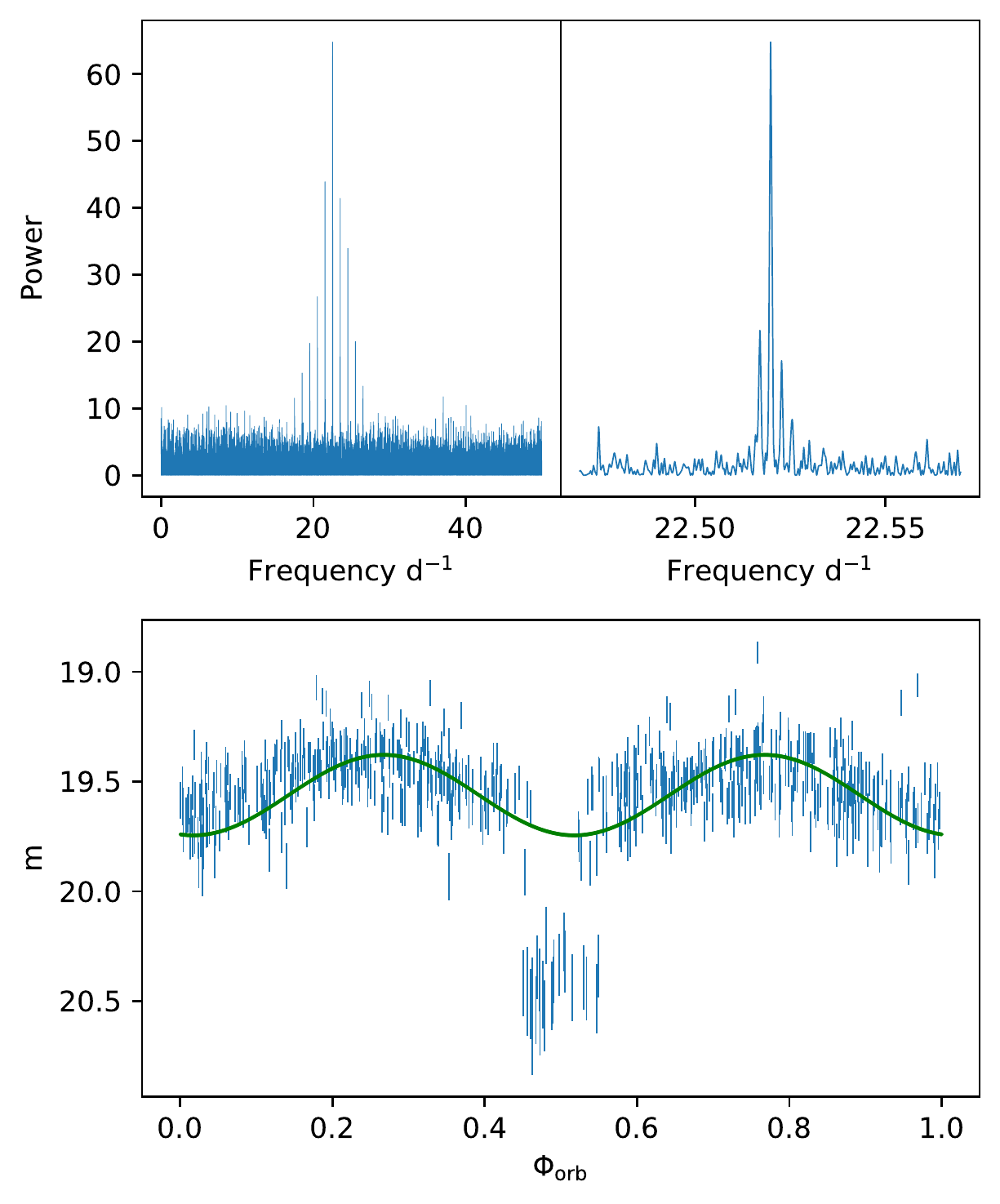}
\caption [.] {\label{fig:LCexample} The Fourier transform of the ZTF light curve of J0158$+$2823, an eclipsing CV, yields a particularly clear periodogram. Top left panel: The aliases around the central frequency are separated by a frequency of once per day~$-$~reflecting the nightly sampling. 
Top right panel:  The central peak provides a good estimate of a period which can then be used to seed a sine fit to the data yielding  $f=22.51981(3)\,\mathrm{d}^{-1}$. Bottom panel: A plot of the data folded at twice the the best-fit period  with a sinusoidal fit superimposed. In this case $P_{\mathrm{orb}}=2/f$ due to ellipsoidal modulation although the corresponding frequency ($11.2599\,\mathrm{d}^{-1}$) is barely visible in the periodogram. This can be seen in the bottom panel with the deep eclipse of the white dwarf by the donor at $\Phi \simeq 0.5$ and the two peaks at $0.25$ in phase either side. In this case there is no doubt which is the correct alias but more generally the period uncertainty has to take account of the possibility of an incorrect alias being selected. 
}
\end{figure}

Archival light curves from the CRTS, ZTF and \textit{TESS} were obtained from online archives;  Caltech, IRSA \citep{2019ASPC..521...36R} and Lightkurve \citep{2018ascl.soft12013L} respectively. We reviewed these looking for outbursts, to provide confirmation of a dwarf nova classification, and also for eclipses to obtain orbital periods. The long span of this data (over 20 years) also reveals many state changes that are indicative of certain CVs (see Figs.\,\ref{fig:polarhump} and \ref{fig:IP}) and novalike variables (Fig.\,\ref{fig:NL}). The light curves for all CVs are shown in the supplementary data. We generated periodograms in an attempt to determine orbital periods from the light curves. Prior to generating the periodogram we cropped any outbursts from the light curve which was then de-trended typically using a low-order polynomial fit. Two periodograms of the light curve were then produced using a Fourier transform and also an orthogonal multi-harmonic analysis of variance \citep{1996ApJ...460L.107S}. Potential periods were identified from peaks in the periodogram that exceeded a false alarm probability of five per cent; the results were then validated by phase-folding the original light curve on the potential period (see Fig.\,\ref{fig:LCexample}). The shape of the phase-folded light curve not only confirms that the period is valid but also helps to identify cases in which the orbital period may be twice that of the strongest signal in the periodogram. The photometric orbital modulation of CVs can have various potential causes. Eclipses may occur once (of the white dwarf) or (rarely) twice (of the white dwarf and the donor, e.g. \citealt{2020MNRAS.499..149A}) per orbital cycle whilst ellipsoidal modulation due to the deformed shape of the donor star will yield two peaks per orbital cycle. Cyclotron beaming is another case where the waveform may show multiple peaks per orbital cycle, as does the emission from two poles in magnetic CVs.

Knowledge of the period assists the classification; SU\,UMa and ER\,UMa CVs typically have  $\porb\lesssim3$\,h \citep{1998A&AS..129...83R}, WZ\,Sge have even shorter periods whilst U\,Gem and Z\,Cam CVs typically have $\porb\gtrsim3$\,h.

Whilst some periodograms show an unambiguous period others leave room for doubt due to aliases and harmonics, we have indicated such periods with a colon (:). 

\subsubsection{Hertzsprung–Russell and Colour-Colour Diagrams}
\begin{figure} 
\includegraphics[width=\columnwidth]{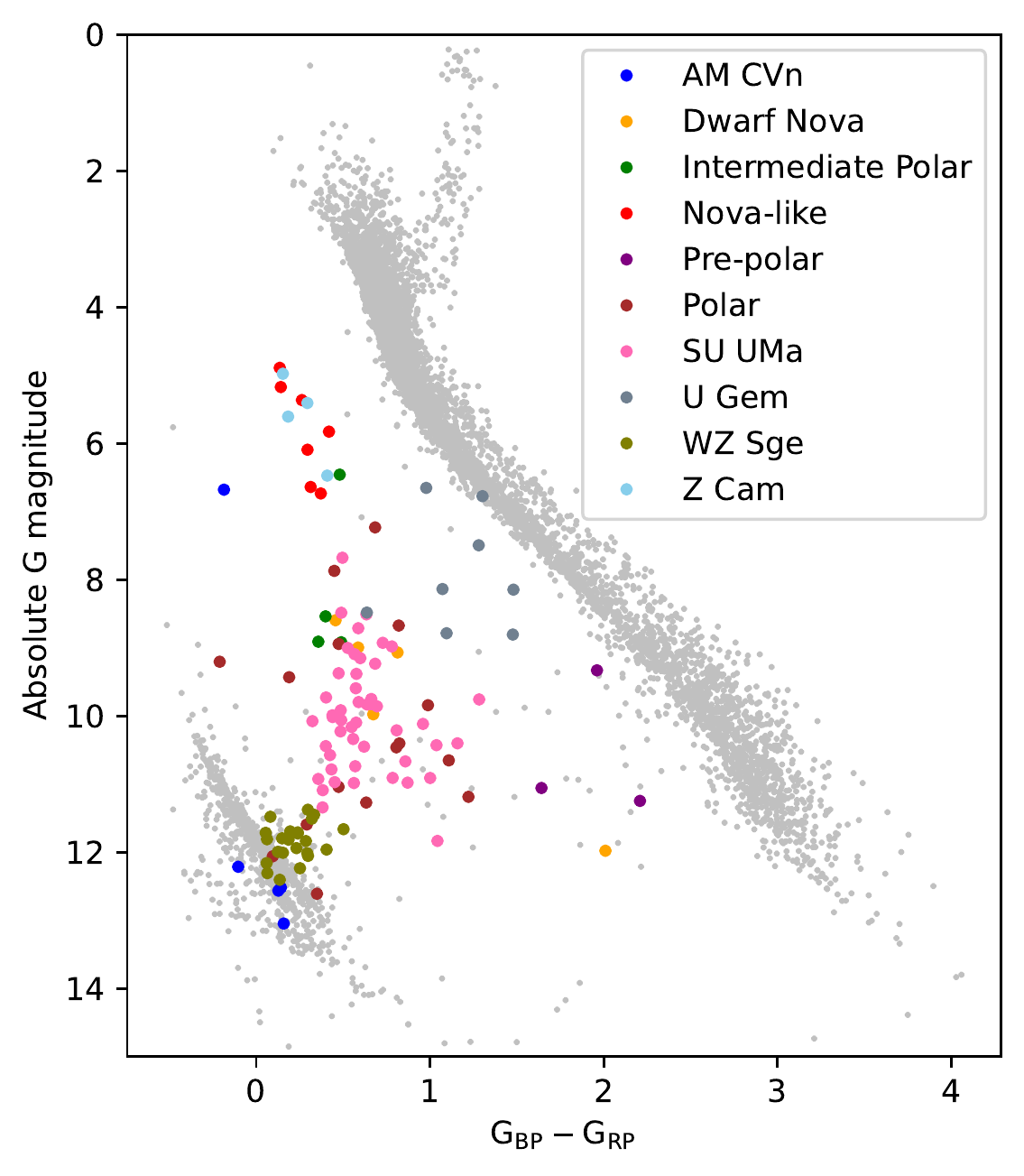}
\caption [.] {\label{fig:CVtypesP} HR diagram showing all the CVs from \allsdss\, where the astrometric and photometric errors are less than 10\,per\,cent. The grey background consists of those \textit{Gaia} objects closer than 100\,pc with \textit{GALEX} observations to increase the proportion of white dwarfs (which are UV-bright).  Note the concentration of AM\,CVn and WZ\,Sge CVs near the white dwarf cooling sequence with SU\,UMa CVs spreading out towards the mains sequence. Novalike variables and Z\,Cam CVs are concentrated around $\mathrm{5<G_{\mathrm{abs}} <7}$ as their high \mdot\ results in large optical luminosity.
}
\end{figure}

The position of a CV on the Hertzsprung–Russell (HR) diagram aids classification as many CV types congregate in specific sections of the diagram (see Fig.\,\ref{fig:CVtypesP} in \citealt{2020MNRAS.492L..40A} and fig.\,8 in \citealt{2021MNRAS.504.2420I}). Absolute magnitudes require knowledge of the distance of a CV which is normally derived from \textit{Gaia} parallaxes. Hence this diagnosis is not available for  CVs without a \textit{Gaia} counterpart. Even when \textit{Gaia} data exists the position of a CV can be misleading if the parallax accuracy is poor or the CV is subject to significant reddening. A further issue is that CVs are intrinsically variable with outbursts, eclipses and orbital variations contributing to the observed spread in their distribution in the HR diagram. Fig.\,\ref{fig:CVtypesP} shows a subset of the CVs from \allsdss\ where systems with inaccurate parallaxes ($\varpi / \sigma_{\varpi} <10$)  or inaccurate  magnitudes ($\texttt{phot\_g\_mean\_flux\_over\_error<10}$, $\texttt{ phot\_bp\_mean\_flux\_over\_error<10}$, or $\texttt{phot\_rp\_mean\_flux\_over\_error<10}$) have been removed in line with \cite{2018A&A...616A...2L,LL:LL-124}. Inspection of Fig.\,\ref{fig:CVtypesP} shows that the location of a CV in the HR diagram can aid its sub-type classification. The HR diagrams, where available, are included as PNG files in the supplementary data.

\begin{figure} 
\includegraphics[width=\columnwidth]{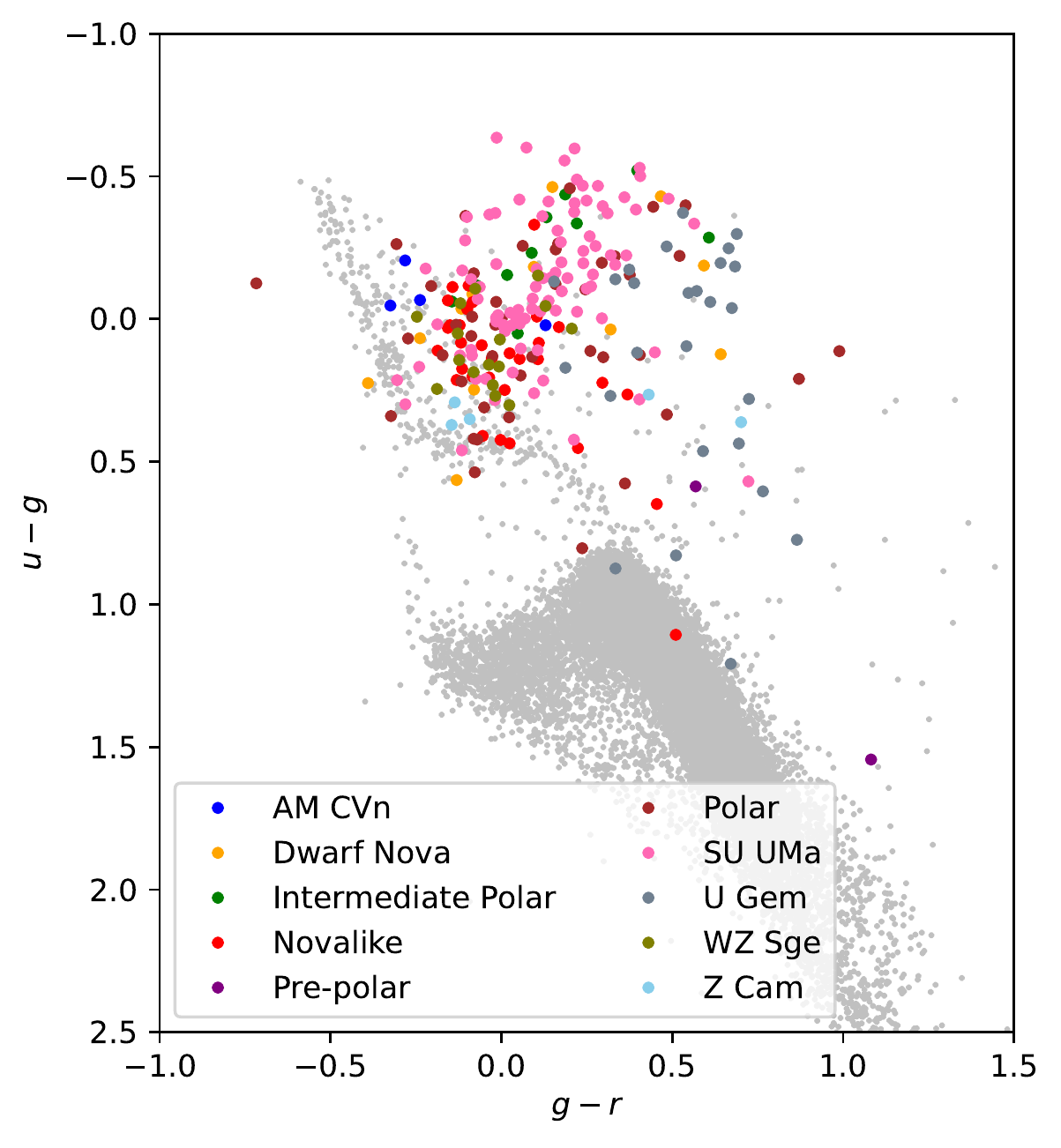}
\caption [.] {\label{fig:CVtypesColCol} Colour-colour diagram showing the CVs from \allsdss\ which have reliable SDSS photometry (where the magnitude of each colour is greater than 15.5 to avoid saturation and the magnitude error is less than 0.05). The grey background shows a collection of spectroscopically observed SDSS stars to guide the eye. The dense lower area forms the main sequence and the upper area is the white dwarf cooling sequence. Note the concentration of SU\,UMa and WZ\,Sge CVs on and above the white dwarf cooling sequence. 
}
\end{figure}

Where a \textit{Gaia} counterpart does not exist, or does not have a parallax, a colour-colour diagram based on SDSS photometry is an alternative (see Fig.\,\ref{fig:CVtypesColCol}). This is typically the case for faint ($m\gtrsim20$) systems. These approaches are somewhat complementary as SDSS photometry saturates at $\mathrm{m\simeq15}$ and is not available for bright CVs.

\subsubsection{Imaging} \label{sec:imagery}
\begin{figure} 
\centerline{\includegraphics[width=0.6\columnwidth]{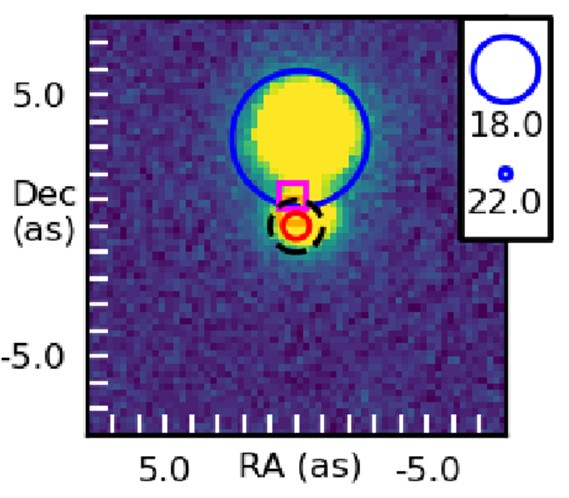}}
\caption [.] {\label{fig:contaminent} Pan-STARRS cutout of J0752+3628 showing the potential for the contamination of fibre spectroscopy. J0751+1444 has an apparent \textit{Gaia} magnitude  $m_{\mathrm{G}}=20.0$ whereas its neighbour, at a distance of 3.35\,arcsec, has $m_\mathrm{G}=16.4$. The red circle denotes the target, the black dotted circle denotes the nominal coverage of an SDSS fibre (2\,arcsec) and the blue circle is centred on the \textit{Gaia} coordinates of the neighbour scaled according to the \textit{Gaia} $G$ magnitude. The \textit{Gaia} coordinates are based upon EDR3 with an epoch of 2016 whilst the Pan-STARRS epoch is around 2012 and so a lack of alignment is a prompt to check proper motions to see if the contaminant was closer when the spectrum was observed. The pink square shows a \textit{GALEX} observation.
}
\end{figure}

Pan-STARRS \citep{2016arXiv161205560C} cutout images for each CV were obtained from the PS1 Image Cutout Server at STScI \citep{2020ApJS..251....7F}. We superimposed circles for each \textit{Gaia} object (with the radius being proportional to the $\textit{Gaia}$ $G$-band magnitude) and for the SDSS fibre (2\,arcsec diameter) (Fig.\,\ref{fig:contaminent}). The resulting image shows whether there is potential contamination of a spectrum from neighbouring objects, nebulae or background galaxies. In addition we superimposed circles for any \textit{GALEX} detections. CVs often have \textit{GALEX} observations but the accuracy of the coordinates of \textit{GALEX} observations can be incorrect by a few arcseconds and the image helps to assess whether a \textit{GALEX} observation could be associated with a different object. These images are available as PNG files in the supplementary data.

\onecolumn
\begin{landscape}
\begingroup
\renewcommand{\arraystretch}{1.2}
\begin{longtable}[l]{lllllrllll}
\caption[width=\columnwidth]{\label{tab:cvtable} CVs from \allsdss. We provide the geometric distances from \citet{2021AJ....161..147B} and CV sub-types according to the definitions in section\, \ref{section:sub_types}.    SDSS $g$-band magnitudes are from data release~7 \citep{2009ApJS..182..543A}. The references for the initial discovery (ID), spectrum (Sp)  and orbital and superhump periods ($P$~--~where known) are also shown. New discoveries, orbital periods and sub-type identifications are shown in blue. Sub-type identifications in black are from VSX \citep{2017yCat....102027W}. : Tentative values, * Superhump periods, \dag low accretion SU\,UMa, \ddag CVs with unusual state changes, \S period bouncers. This is a short extract of the table, the full data is included in the supplementary data. \label{long}}\\

\hline
\multirow{2}{*}{SDSS} & \multirow{2}{*}{Name} & \multirow{2}{*}{\begin{tabular}[c]{@{}l@{}}Gaia EDR3\\ source\_id\end{tabular}} & \multirow{2}{*}{\begin{tabular}[c]{@{}l@{}}$P$ \\ (h)\end{tabular}} & \multirow{2}{*}{\begin{tabular}[c]{@{}l@{}}SDSS $g$\\ (mag)\end{tabular}} & \multirow{2}{*}{\begin{tabular}[c]{@{}l@{}}Distance\\ (pc)\end{tabular}} & \multirow{2}{*}{\begin{tabular}[c]{@{}l@{}}CV Sub-type\\ 
 \end{tabular}}  & \multicolumn{3}{c}{References}                         \\ \cline{8-10} 

 &    &     &       &      &      &         & \multicolumn{1}{l}{ID} & \multicolumn{1}{l}{Sp} & P \\ \hline
 \hline
 \endfirsthead

\hline
\hline
\multirow{2}{*}{SDSS} & \multirow{2}{*}{Name} & \multirow{2}{*}{\begin{tabular}[c]{@{}l@{}}Gaia EDR3\\ source\_id\end{tabular}} & \multirow{2}{*}{\begin{tabular}[c]{@{}l@{}}$P$ \\ (h)\end{tabular}} & \multirow{2}{*}{\begin{tabular}[c]{@{}l@{}}SDSS $g$\\ (mag)\end{tabular}} & \multirow{2}{*}{\begin{tabular}[c]{@{}l@{}}Distance\\ (pc)\end{tabular}} & \multirow{2}{*}{\begin{tabular}[c]{@{}l@{}}CV Sub-type\\ 
 \end{tabular}}  & \multicolumn{3}{c}{References}                         \\ \cline{8-10} 

 &    &     &       &      &      &         & \multicolumn{1}{l}{ID} & \multicolumn{1}{l}{Sp} & P \\ \hline
 \hline
 \endhead
 \hline
 \endfoot
 \hline
 \endlastfoot
\textcolor{blue}{J000014.74+255603.1} &  &  &  & 22.17 &  & \textcolor{blue}{CV:} & 1 & 1 &  \\
J000720.77+200721.6 & CSS 110921:000721+200722 &  &  & 21.48 &  & \textcolor{blue}{WZ Sge:} & 15 & 1 &  \\
J000844.32-014014.6 & Gaia15abi & 2544817841421994624 &  & 17.93 & $1236^ {+220} _{-230}$ & \textcolor{blue}{NL:/Polar} & 15 & 1 &  \\
J001107.26+303235.9 & V0402 And & 2861506395998660096 & 0.063499(3)* & 20.4 & $793^ {+434} _{-225}$ & SU UMa & 11 & 1 & 5 \\
J001153.08-064739.2 & SDSS J001153.08-064739.1 & 2443219535337313280 & 0.10028081(8) & 17.83 & $477^ {+29} _{-19}$ & U Gem & 6 & 6 & 6 \\
J001158.28+315543.7 & CSS 101111:001158+315544 & 2861949606559861120 & \textcolor{blue}{0.0554(1):} & 21.85 &  & \textcolor{blue}{SU UMa} & 15 & 1 & 1 \\
\textcolor{blue}{J001231.54+280011.1} &  & 2859847683924640384 &  & 20.98 &  & \textcolor{blue}{WZ Sge:} & 1 & 1 &  \\
J001538.25+263656.7 & CSS 090918:001538+263657 & 2856493142666940544 & 0.10150(6) & 17.92 & $557^ {+36} _{-34}$ & SU UMa & 15 & 7 & 7 \\
J001856.92+345444.2 & V0479 And & 2876157251000218496 & 0.594093(4)  & 17.46 & $2027^ {+408} _{-259}$ & Polar: & 4 & 4 & 8 \\
J002049.51+280020.2 & MASTER OT J002049.54+280020.5 & 2857063063352792320 & \textcolor{blue}{0.05658(1)} & 19.64 & $347^ {+59} _{-45}$ & \textcolor{blue}{WZ Sge:} & 13 & 1 & 1 \\
J002243.55+061002.7 & MGAB-V295 & 2747622177049417344 & \textcolor{blue}{0.079(1):} & 19.88 & $2113^ {+1596} _{-715}$ & \textcolor{blue}{SU UMa} & 9 & 1 & 1 \\
J002500.17+073349.2 & CSS 081123:002500+073350 & 2748173586426407424 &  & 19.68 & $1484^ {+412} _{-306}$ & \textcolor{blue}{DN/IP:} & 15 & 1 &  \\
J002637.05+242915.6 & CSS 091026:002637+242916 & 2807191999055199232 & \textcolor{blue}{0.08533(1)} & 18.75 & $588^ {+100} _{-87}$ & Polar & 15 & 1 & 1 \\
J002728.00-010828.5 & EN Cet & 2541910801397761152 & 0.05933(5) & 20.78 & $1218^ {+461} _{-547}$ & \textcolor{blue}{WZ Sge} & 10 & 4 & 2 \\
J002842.53+311819.4 &  & 2859593555004235520 & \textcolor{blue}{1.217(1)} & 18.59 & $3642^ {+1152} _{-985}$ & \textcolor{blue}{NL} & 3 & 3 & 1 \\
\textcolor{blue}{J002910.96+324125.8} &  &  &  & 22.12 &  & \textcolor{blue}{DN} & 1 & 1 &  \\
J003151.76+202051.0 & MGAB-V534 & 2796463449923353856 & 0.080871025(2) & 18.06 & $368^ {+19} _{-16}$ & SU UMa\dag  & 9 & 1 & 14 \\
J003153.46+084118.5 & ASASSN-17kg & 2749846901389046656 & 0.0585(3)* & 19.15 & $283^ {+24} _{-17}$ & \textcolor{blue}{WZ Sge} & 12 & 1 & 16 \\
J003203.62+314510.4 & CSS 091220:003204+314510 & 2859621764350006784 &  & 19.1 & $960^ {+335} _{-184}$ & \textcolor{blue}{SU UMa} & 15 & 1 &  \\
J003459.90+273619.0 & CSS 081031:003460+273620 & 2857545615813275136 & \textcolor{blue}{0.07337(1)} & 21.12 & $1349^ {+828} _{-462}$ & \textcolor{blue}{SU UMa} & 15 & 1 & 1
 \end{longtable}
 \endgroup
\noindent
References:\, 1\,\textcolor{blue}{This work}, 2\,\citet{2008MNRAS.386.1568D}, 3\,\citet{2020AJ....159...43H}, 4\,\citet{2005AJ....129.2386S}, 5\,\citet{2012PASJ...64...21K}, 6\,\citet{2014ApJ...790...28R}, 7\,\citet{2016AJ....152..226T}, 8\,\citet{2013A&A...553A..28G}, 9\,\textcolor{blue}{MGAB catalog}, 10\,\citet{1998AcApS..18..214E}, 11\,\citet{1998IBVS.4578....1A}, 12\,\citet{2015MNRAS.446.4078K}, 13\,\citet{2019MNRAS.482.4570G}, 14\,\textcolor{blue}{VSNET23583}, 15\,\textcolor{blue}{CRTS candidate CV}, 16\,\textcolor{blue}{VSNET21317} 
\noindent
\end{landscape}

\onecolumn

\section{Results}\label{section:results}

In the interests of brevity the full  observational data and analysis of the over 500  \allsdss\  CVs has been consigned to the supplementary data. Illustrative examples are given here (Table \ref{tab:cvtable} and Fig.\,\ref{fig:spectraandLCexample}) together with a brief discussion of a small set of the most interesting systems. Spectra have been smoothed by using a  box-car median filter of width five points. A list of the \fxnum{70} newly discovered CVs is included in Table \ref{tab:newCVs}.

\noindent\begin{minipage}{\textwidth}
    \centering 
    \includegraphics[width=0.97\hsize]{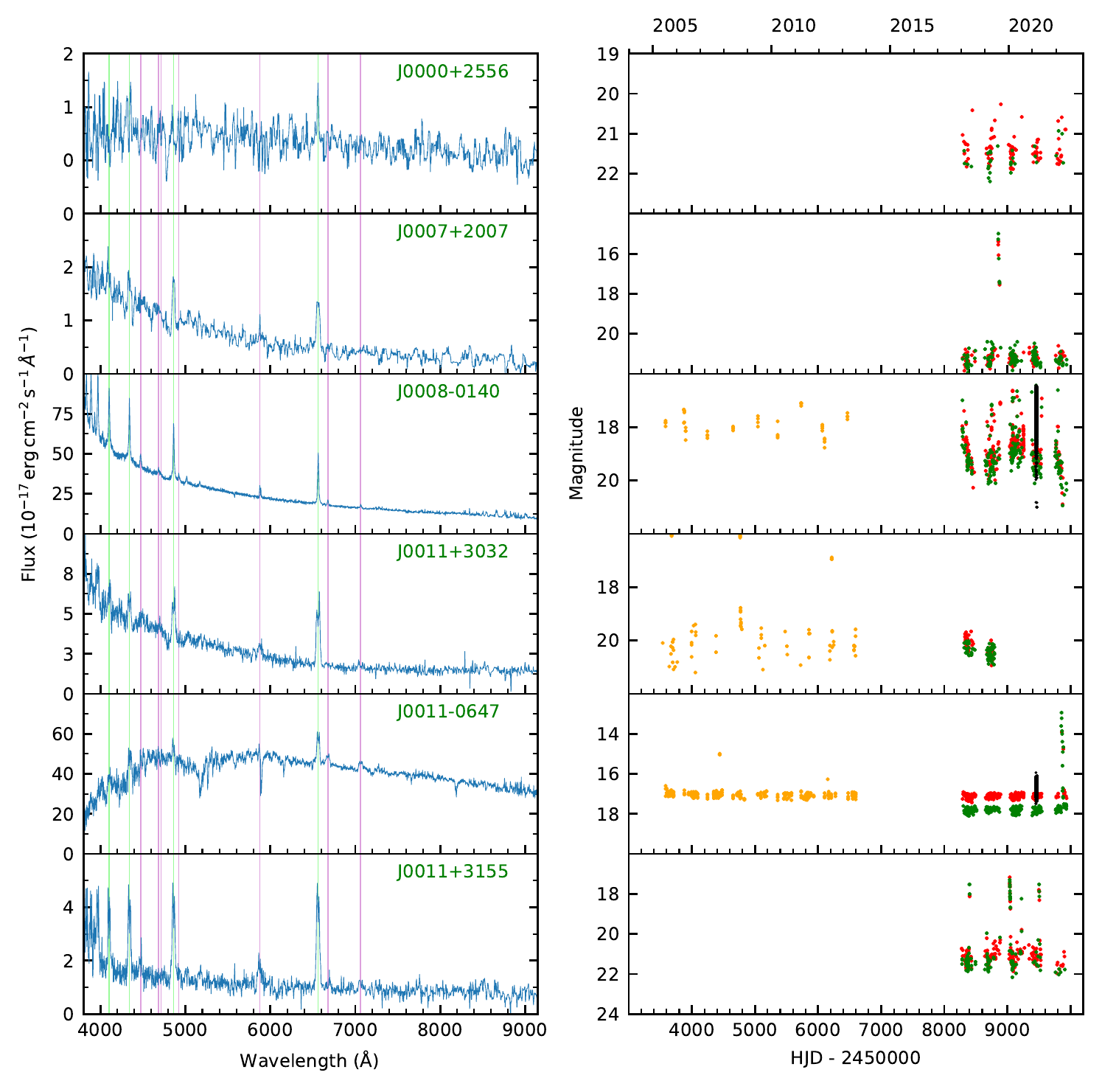} 
    \captionof{figure}{\label{fig:spectraandLCexample}Left panels: Spectra with Balmer lines shown in green and He lines in pink. Right panels: Light curves with CRTS shown in orange, ZTF in red ($r$-band) and green ($g$-band) and \textit{TESS} in black. This is a short extract of the data: the complete set is included in the supplementary data.}
\end{minipage}

\twocolumn

\subsection{Unusual systems}
The following notes discuss a small selection of the systems in \allsdss\, which have unusual features and are worthy of further study. 
\subsubsection{J0028+3118} 
\begin{figure} 
\includegraphics[width=\columnwidth]{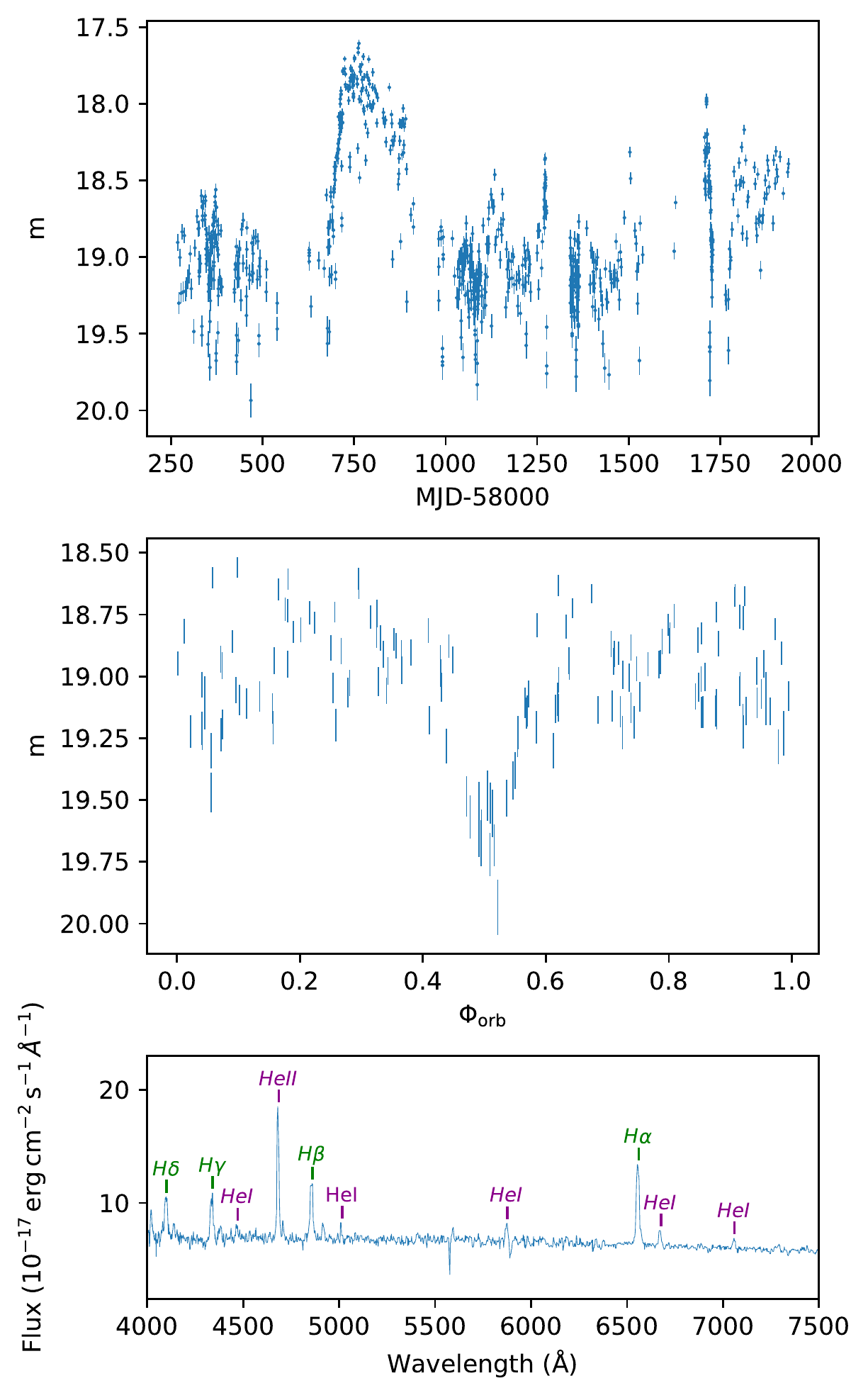}
\caption [.] {\label{fig:interestingJ0028} J0028+3118 is an eclipsing CV with an unusually long orbital period. Top panel: ZTF light curve ($r$-band) showing a brightening event lasting about one year. Middle panel: The section of the light curve with $\mathrm{MJD}<58600$ folded on the 29.2\,h period. 
Bottom panel: SDSS spectrum with a prominent \ion{He}{ii} 4686\,\AA\ emission line.}
\end{figure}

The spectrum displays single-peaked Balmer emission lines above a flat continuum as well as a very strong \ion{He}{ii} 4686\,\AA\ emission line. The system is located close to the main sequence in the HR diagram. There is a ${\simeq}1$\,yr  state change of ${\simeq}2$ magnitudes at $\mathrm{MJD}\simeq58750$ together with three shorter and smaller brightening events in the ZTF light curve. The ZTF light curve (Fig.\,\ref{fig:interestingJ0028}) suggests that $P_{\mathrm{orb}}=29.20(1)$h and also that the system is eclipsing. J0028+3118 is a high-accretion novalike system but the very long period is unusual. The extraordinarily strong \ion{He}{ii} 4686\,\AA\ suggests a hot and helium-rich accretion stream~$-$~possibly the result of a hydrogen depleted donor that is still burning \citep{2018ApJ...861..102H}.  

\subsubsection{J0223$-$0520} 
\begin{figure} 
\includegraphics[width=\columnwidth]{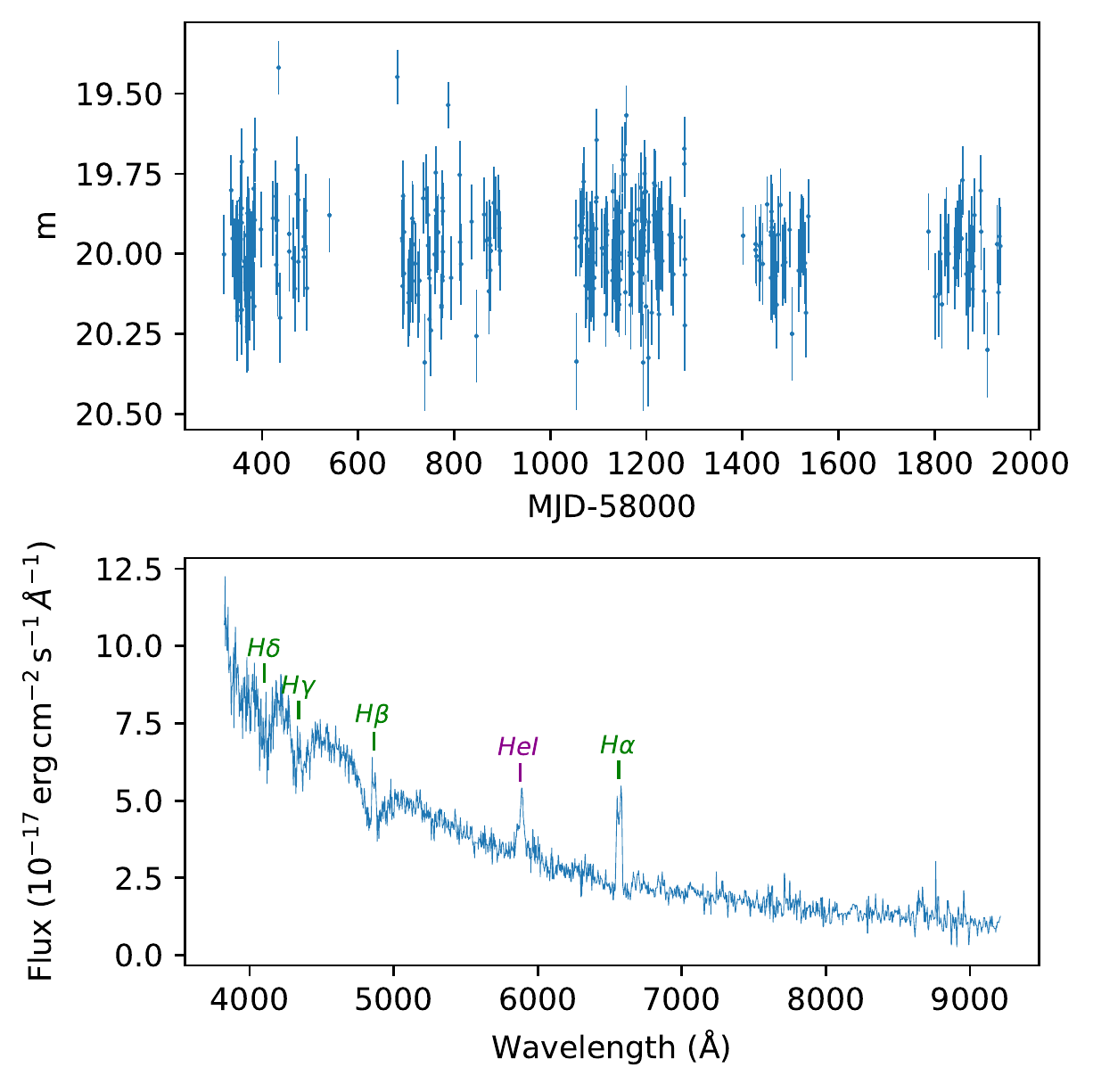}
\caption [.] {\label{fig:interestingJ0223} J0223$-$0520 
is a CV with an unusually hot white dwarf. Top panel: ZTF light curve ($r$-band) showing variability. Bottom panel: SDSS spectrum showing the white dwarf absorption lines.}
\end{figure}

The spectrum shows double-peaked Balmer emission lines with the higher order ones embedded in broad absorption lines from the white dwarf (Fig.\,\ref{fig:interestingJ0223}). The blue slope of the continuum, and the shape of the white dwarf absorption lines suggest a relatively high white dwarf temperature,  which is consistent with the SED being dominated by the white dwarf. The system is brighter in the far-ultraviolet (\textit{GALEX} $m_\mathrm{FUV}=19.5$) than in the optical ($m_\mathrm{g}=19.9$). This is not consistent with its position at the lower end of the white dwarf cooling sequence in the HR diagram with $G_\mathrm{BP}-G_\mathrm{RP}=0.35$, which would imply a temperature $T_{\mathrm{eff}}\simeq8000$\,K. The  ZTF light curve shows variability but does not reveal an orbital  period. There is an \textit{XMM} X-ray detection  which is unusual for a low-\mdot\ system \citep{2020A&A...641A.136W}.   This is probably a WZ\,Sge. It is possible that the raised temperature of the white dwarf is the consequence of a previous eruption.

\subsubsection{J0932+4725 and J0947+4524} 
\begin{figure} 
\includegraphics[width=\columnwidth]{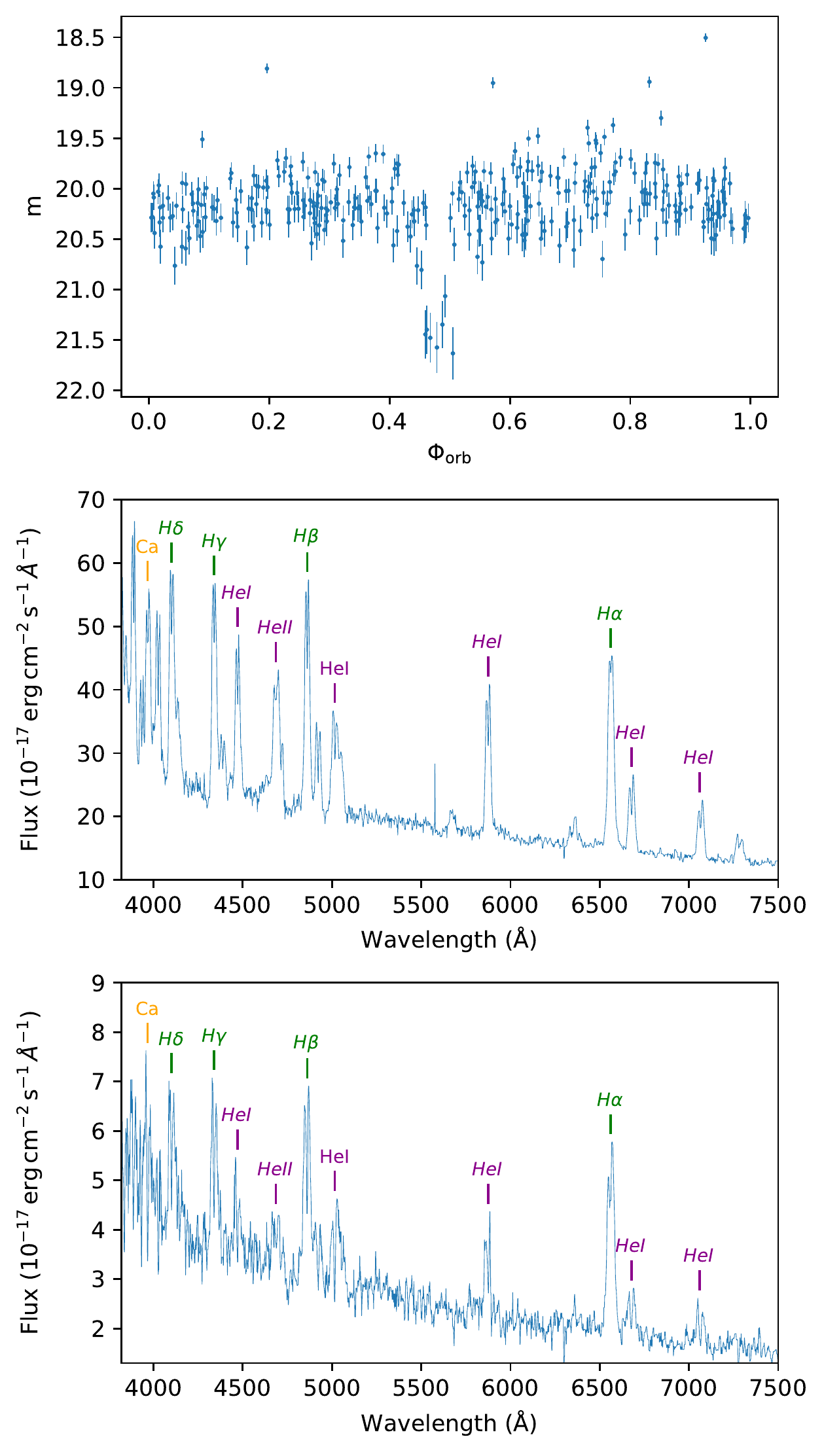}
\caption [.] {\label{fig:interestingJ0947} Two eclipsing CVs  with unusually strong \ion{He}{i} emission lines, which very likely contain stripped nuclear evolved donor stars. Top panel: Folded ZTF light curve of J0947+4525 showing eclipse. Middle panel:  Spectrum of J0932+4725. Bottom panel: Spectrum of J0947+4524.
}
\end{figure}

Double-peaked Balmer, \ion{He}{i}, \ion{He}{ii} and Ca emission lines are visible in the SDSS spectra of these systems as well as the Balmer jump in emission (Fig.\,\ref{fig:interestingJ0947}). There is deep central absorption in the emission lines suggesting that they are eclipsing~--~which is confirmed by \citet{2006AJ....131..562H} for J0932+4725, and by the ZTF light curve for J0947+4524.  Both are closer to the white dwarf cooling sequence than the main sequence  in the HR diagram.  In the CRTS and ZTF light curves J0932+4725 exhibits outbursts and superoutbursts whilst J0947+4524 exhibits two small ($\Delta m\simeq1$) outbursts.  J0932+4725 has a period of 1.588\,h \citep{2006AJ....131..562H} whilst we obtained a period of 1.34963(2)\,h from the ZTF light curve.

The large ratio of the  helium to hydrogen line strengths suggests that the donor is likely to be He-enriched. An \textit{HST}  observation of J0932+4725 by \citet{2017MNRAS.466.2855P} shows a large \ion{N}{v}/\ion{C}{iv}  line flux ratio.  \citet{2003ApJ...594..443G} attribute this type of nitrogen excess to a system having passed through a phase of thermal timescale mass transfer and now accreting CNO processed material from a donor stripped of its external layers. J0932+4725 is therefore most probably such an evolved system.  Whilst there are no \textit{HST} spectra of  J0947+4524 the optical spectrum is very similar to that of J0932+4725 and suggests that this may also have an evolved donor. Systems with evolved donors are interesting as they are potential precursors to AM\,CVn systems \citep{2003MNRAS.340.1214P,2015ApJ...809...80G}

The superoutbursts confirm that J0932+4725 is an SU\,UMa CV. J0947+4524 is a dwarf nova but we cannot classify it more precisely.

\subsubsection{J1146+6759}

\begin{figure} 
\includegraphics[width=\columnwidth]{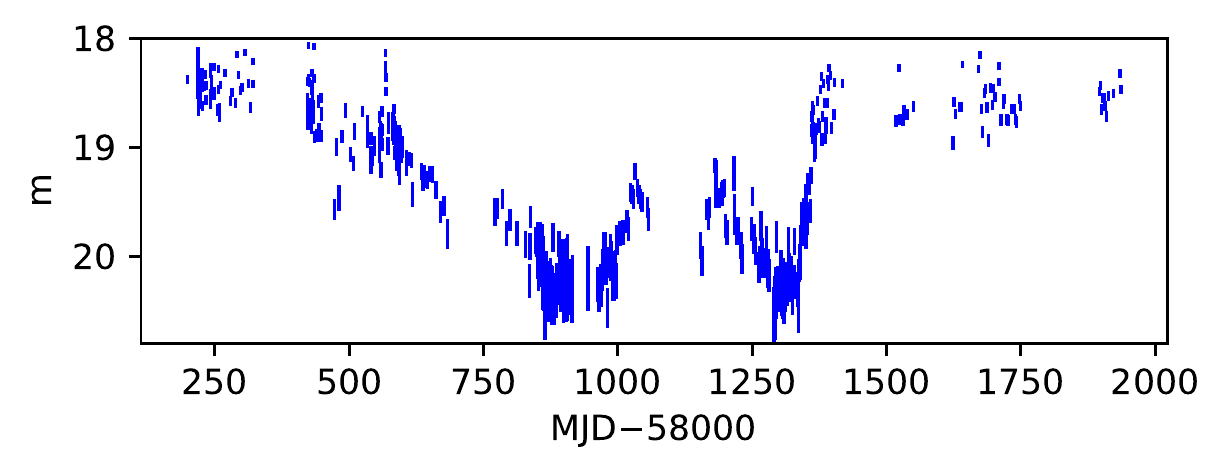}
\caption [.] {\label{fig:interestingJ1146} J1146+6759 has been classified as an SU\,UMa dwarf nova, but has an unusual ZTF light curve ($r$-band) showing a slow decline into a faint state without outbursts.
}
\end{figure}

Double-peaked Balmer and \ion{He}{i}  emission lines are visible in the spectrum as well as the Balmer jump in emission. The donor is not detected. J1146+6759 is located midway between the main sequence and the white dwarf cooling sequence in the HR diagram. The system was observed in a superoutburst on 2011 January 4 and a superhump period of 0.0615(8)\,d was reported (vsnet-alert 12567\footnote{http://ooruri.kusastro.kyoto-u.ac.jp/mailarchive/vsnet-alert/12567}), classifying this system as an SU\,UMa dwarf nova. The ZTF light curve shows very unusual variability (Fig.\,\ref{fig:interestingJ1146}) with a steady decline in amplitude of $\Delta m\simeq1.5$  over three years. This is followed by a rapid return to the original magnitude and then an outburst. It demonstrates that some dwarf novae can undergo state changes (section\,\ref{section:state_changes})

\subsubsection{J1150+4041} 
\begin{figure} 
\includegraphics[width=\columnwidth]{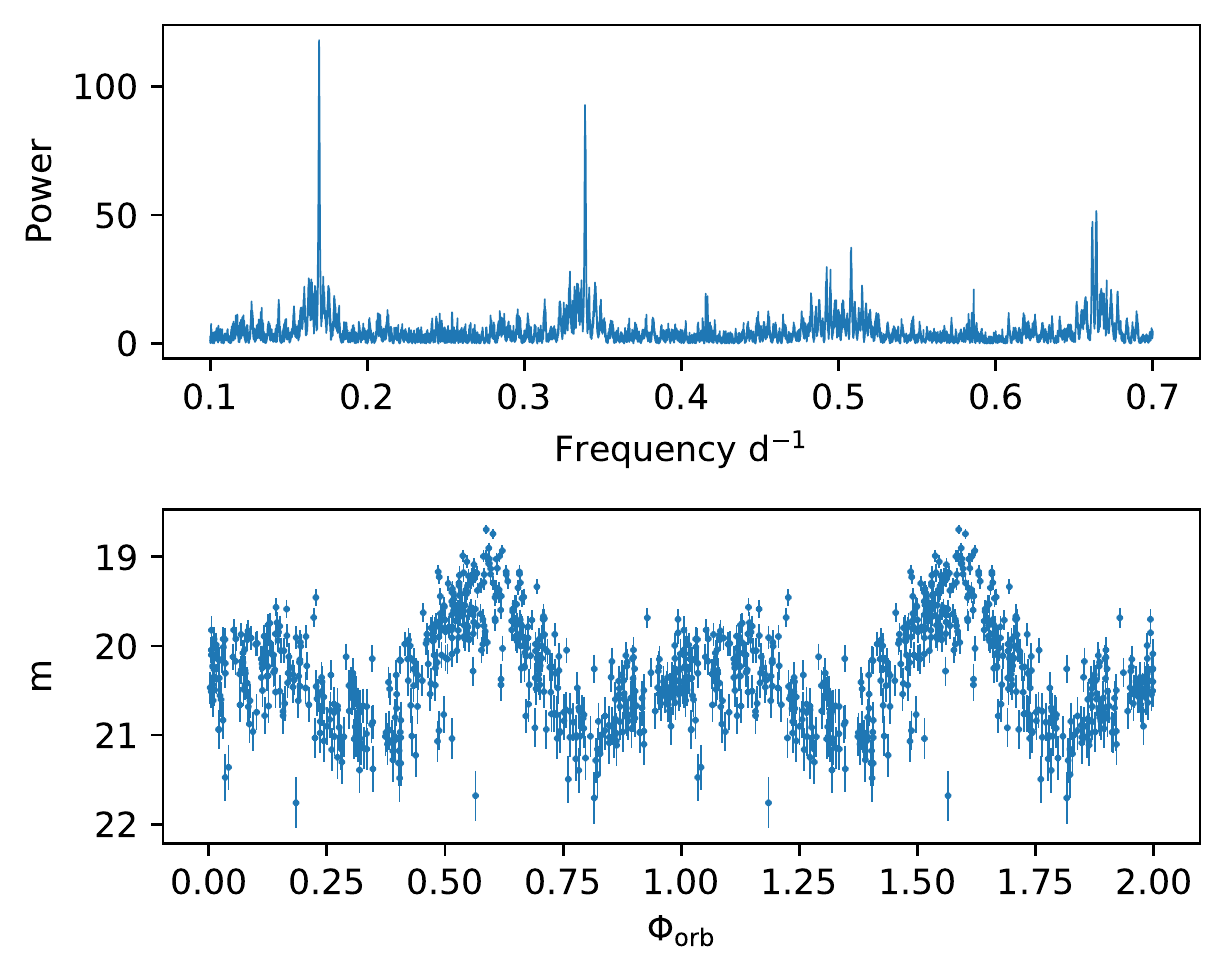}
\caption [.] {\label{fig:interestingJ1150} The ZTF data of J1150+4041 shows a very clear and strong signal at an  unusually long period of 141.65(5)\,h. The highest peak in the periodogram from the ZTF light curve is the second harmonic and the folded light curve shows a double humped profile.
}
\end{figure}

The spectrum shows strong Balmer, \ion{He}{i} and \ion{He}{ii}  asymmetric emission lines, suggesting that this is a magnetic CV. Neither the white dwarf nor the donor are visible. The ZTF light curve shows continuous variability with an amplitude of $\Delta m\simeq1$, but no outbursts. The periodogram computed from the ZTF data shows a clear and strong signal at  an unusually long period of 141.65(5)\,h, or 5.9\,d  (Fig.\,\ref{fig:interestingJ1150}).  If this signal is genuinely the orbital period of the system, it would exceed that of the current record of 5.7\,d (V1017\,Sgr, \citealt{1992Natur.358..563S}). However, in contrast to V1017\,Sgr, the spectrum of J1150+4041 does not reveal any signature of a donor star, normally detected at such a long orbital period. Moreover, the system is located relatively close to the white dwarf cooling sequence in the \textit{Gaia} HR diagram, i.e. intrinsically faint. We have currently no consistent explanation for the observed properties of J1150+4041, but speculate that the long period might be the beat between the white dwarf spin and the orbital period in a slightly asynchronous polar. Radial velocity and fast photometry follow-up is required to unravel the true nature of this system.

\subsubsection{J1224+1841} 
\begin{figure} 
\includegraphics[width=\columnwidth]{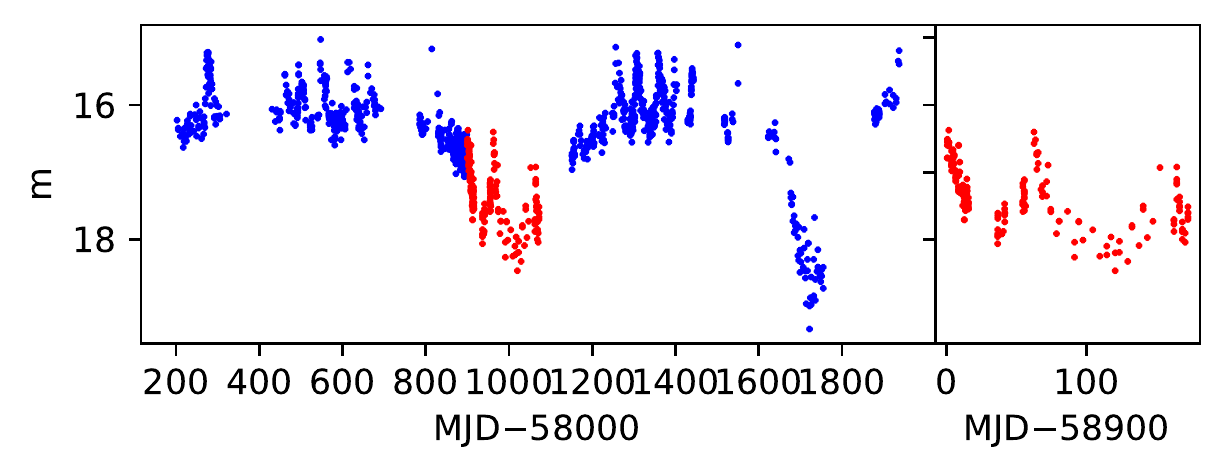}
\caption [.] {\label{fig:interestingJ1224} J1224+1841 is a transitional object between novalike variables and dwarf novae: the ZTF light curve shows both outbursts and a drop-out (red). Outbursts also appear to occur during the faint state.
}
\end{figure}

The spectrum shows a blue continuum with Balmer emission lines, which are embedded in Balmer absorption lines for H$\beta$ and the higher Balmer lines. J1224+1841 is located in the area of the HR diagram occupied by novalike variables. The ZTF light curve shows multiple outbursts and also a drop of $\Delta m\simeq 2$ to a low state lasting several hundred days (Fig.\,\ref{fig:interestingJ1224}). There appear to be outbursts during the low state although it is possible that they are very short reversions to the high state. The CRTS light curve shows two similar low states. J1224+1841 belongs to the growing family of CVs bordering between novalike variables and dwarf novae \citep{2019PASJ...71...20K,2021PASJ...73.1280K,2022MNRAS.510.3605I}. Radial velocity follow-up should be carried out to determine the orbital period of this system.

\subsubsection{J1244+5936} 
\begin{figure} 
\includegraphics[width=\columnwidth]{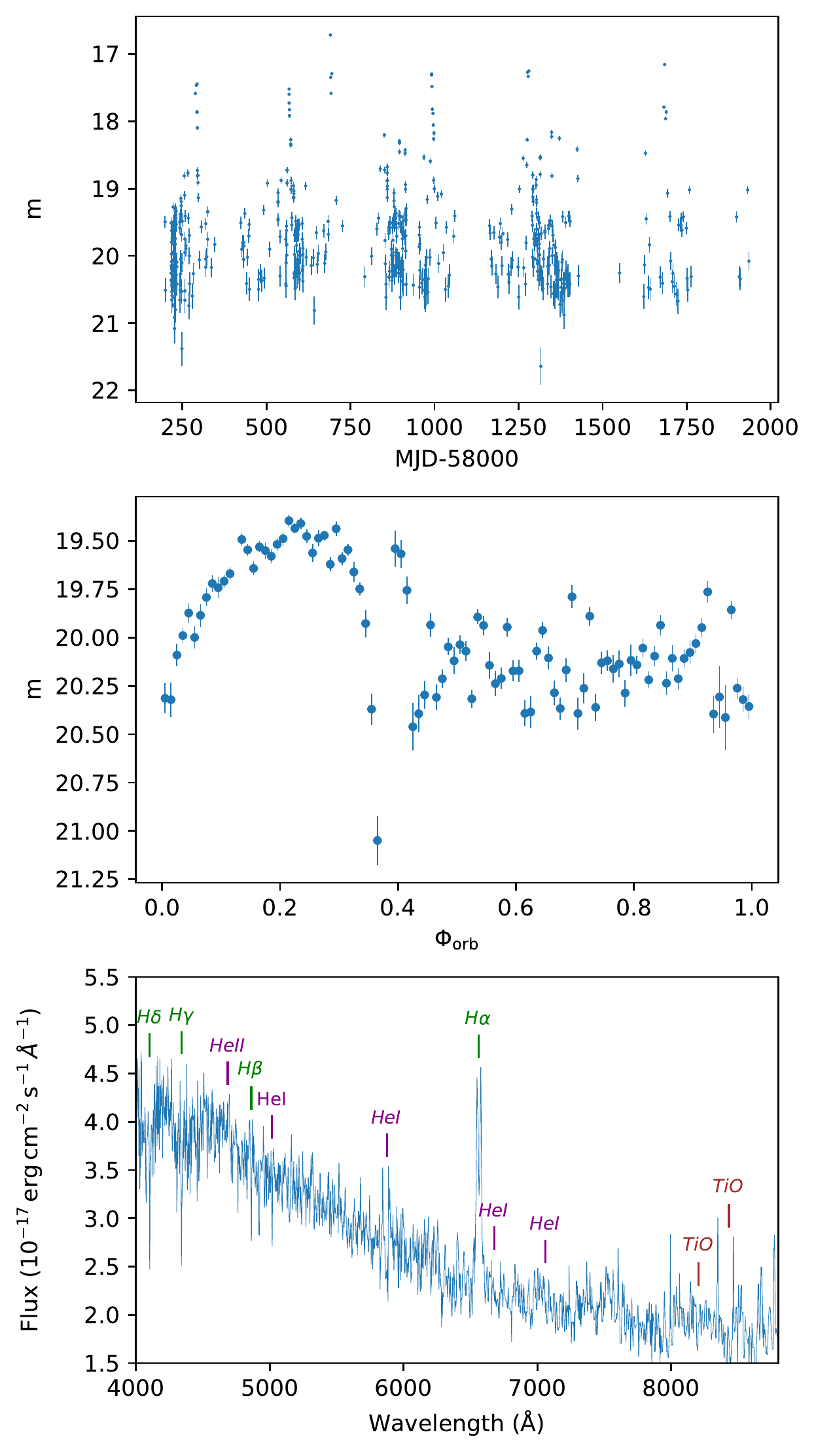}
\caption [.] {\label{fig:interestingJ1244} J1244+5936 (Gaia16adi) is an eclipsing system with a very low accretion rate. Top panel: ZTF light curve ($r$-band). Middle panel: Light curve folded at the orbital period of 1.75\,h, and rebinned into 30 phase bins. Note the eclipse of the disc at $ \Phi_\mathrm{orb}\simeq 0.4$ and the eclipse of the hot spot at  $\Phi_\mathrm{orb}\simeq 0.45$.  Bottom panel: SDSS spectrum showing H$\gamma$ and H$\delta$ absorption lines from the white dwarf and TiO absorption lines from the donor in the red. 
}
\end{figure}

This system is also known as Gaia16adi \citep{2012gfss.conf...21W}. A double-peaked H$\alpha$ emission line is visible in the spectrum as well as other Balmer, \ion{He}{i} and \ion{He}{ii} emission lines (Fig. \ref{fig:interestingJ1244}). The deep central absorption features in the higher Balmer lines, dipping below the continuum flux level, suggest that the system is eclipsing, which is confirmed by the ZTF photometry. Absorption lines from the white dwarf are visible together with TiO molecular bands from the donor. It is closer to the white dwarf cooling sequence than the main sequence in the HR diagram. The ZTF and CRTS light curves show $\simeq10$  outbursts with at least one low amplitude superoutburst accompanied by superhumps. We obtained a reliable period of 1.74954(1)\,h from the ZTF light curve and a consistent period from the CRTS light curve. The folded light curve shows that it is eclipsing with a hot spot just before the eclipse.  J1244+5936 appears to be an SU\,UMa albeit with a very low \mdot. It may have recently started accreting~--~possibly after passing through the period gap; although theory \citep{1989A&A...225..391D} suggests that this transitional period will be short ($\sim 10^6$\,yr).

\subsubsection{J1522+3913}
\begin{figure} 
\includegraphics[width=\columnwidth]{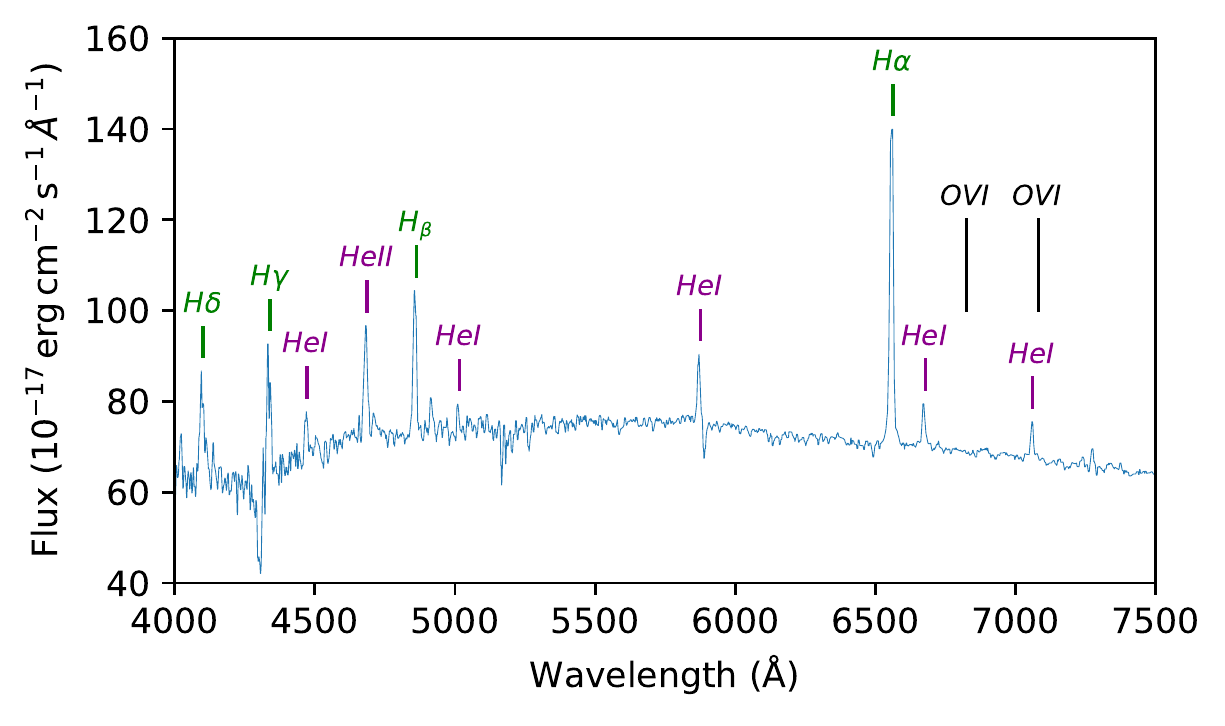}
\caption [.] {\label{fig:interestingJ1522} The spectrum of the long-orbital period (10.8\,d) J1522+3913 contains narrow absorption features from a $\mathrm{G}$-type donor. The helium emission lines suggest that the donor is evolved. There are no molecular and metallic absorption lines or Raman-scattered \ion{O}{VI} lines; two of the conditions in  \citet{2021MNRAS.506.4151M} for a symbiotic star classification.
}
\end{figure}

The spectrum (Fig.\,\ref{fig:interestingJ1522}) shows Balmer emission lines above a relatively red continuum. Narrow absorption lines reveal a $\simeq\mathrm{G}$-type donor star. The SED of the system exhibits a clear ultraviolet excess in the \textit{GALEX} bands, most likely from the accretion disc. The \textit{Gaia} HR diagram shows that J1522+3913 is just evolving off the main sequence into the giant branch. The ZTF light curve shows at least one outburst. We obtained a reliable period of 10.7686\,d (258.4464\,h) from the ZTF light curve confirming the period found by \citet{2014ApJS..213....9D} from the CRTS data. The folded light curve shows a clear double hump, consistent with the detection of ellipsoidal modulation. To our knowledge J1522+3913 is the longest known orbital period for a CV~$-$~twice that of V1017\,Sgr ($\porb=5.714$\,d, \citealt{1992Natur.358..563S}).  We note that a small number of symbiotic stars are located in this region of the HR diagram as well, however, J1522+3913 does not show any of the defining features in its spectrum (see the definition in section 1.1 of \citealt{2021MNRAS.506.4151M}). Instead, we suggest that the very long orbital period, and the location in the HR diagram are consistent with a CV in which the donor has evolved off the main sequence (e.g. for example GK\,Per \citealt{2021MNRAS.507.5805A}).

\subsubsection{J1549+1739} 
\begin{figure} 
\includegraphics[width=\columnwidth]{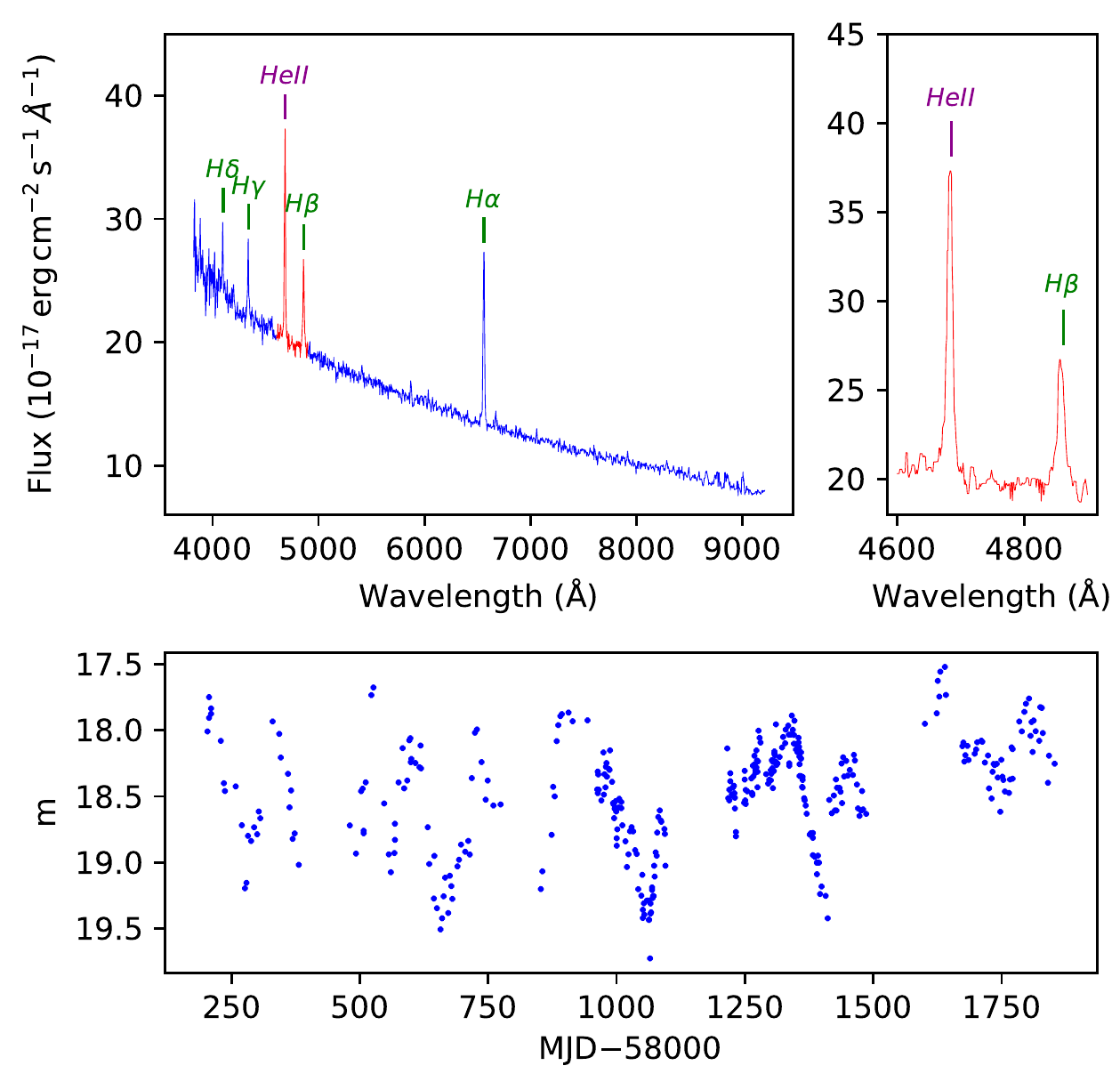}
\caption [.] {\label{fig:interestingJ1549} J1549+1739 is probably a novalike variable. Top panel: The spectrum shows an unusually strong \ion{He}{ii} line. Top inset panel: Expanded plot of the \ion{He}{ii} line. Bottom panel: The ZTF light curve ($r$-band) shows repeated changes of magnitude on timescales of about a year.
}
\end{figure}

The SDSS spectrum shows single-peaked Balmer emission lines and a very strong \ion{He}{ii} emission line on top  of a strong blue continuum (Fig.\,\ref{fig:interestingJ1549}). There is no sign of the white dwarf or the donor. J1549+1739 is located in the novalike variable area of the HR diagram. Whilst there are no outbursts in the CRTS and ZTF light curves, the system meanders on time scales of about a year between $m\simeq17.5-19$. This is probably a novalike variable but the strong \ion{He}{ii} and its light curve are unusual. Spectroscopic follow up is encouraged to explore the spectral changes between the bright and faint states, and to determine the orbital period of the system.

\subsubsection{J1712+3218} 
\begin{figure} 
\includegraphics[width=\columnwidth]{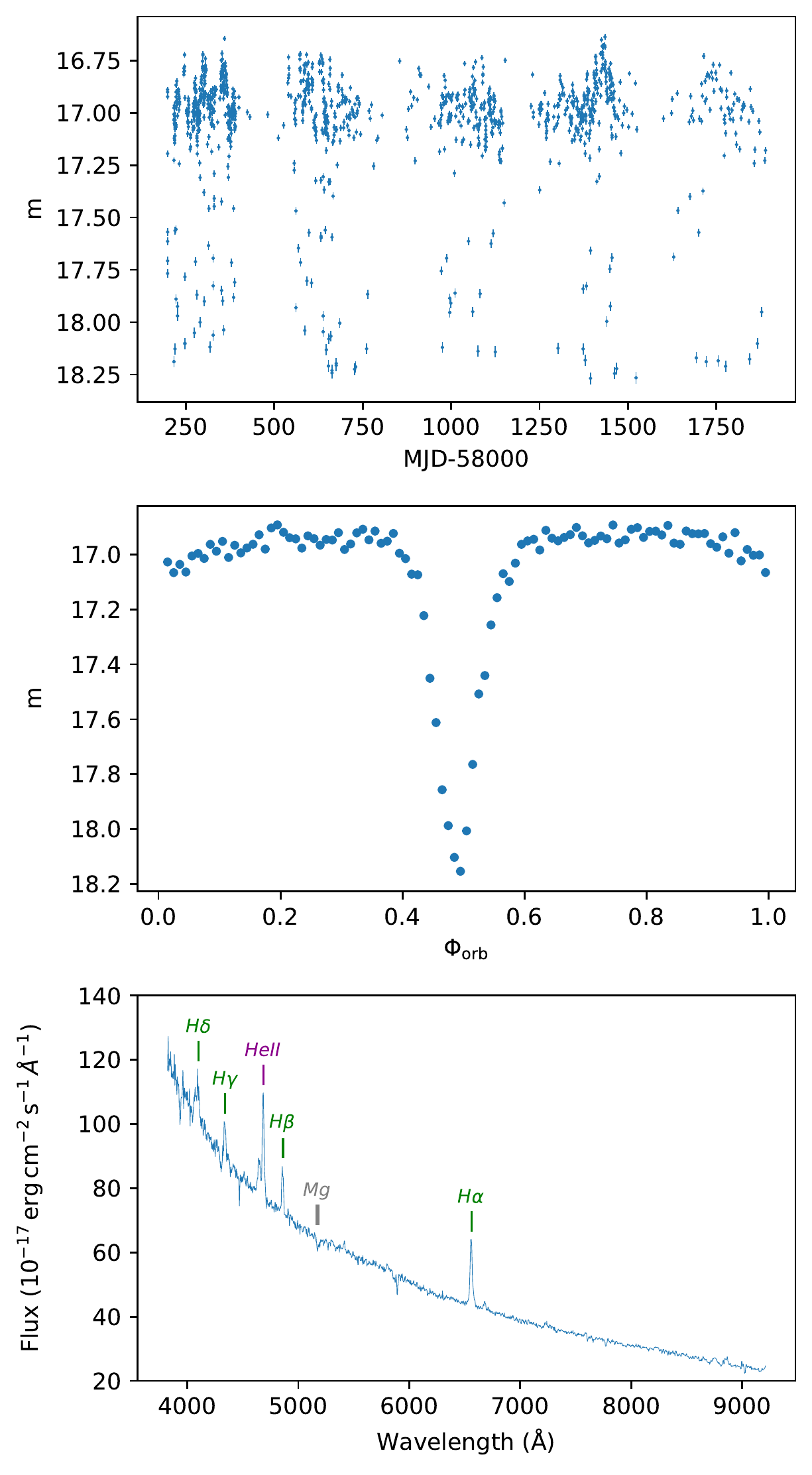}
\caption [.] {\label{fig:interestingJ1712} J1712+3218 is a novalike variable with an unusually long orbital period. Top panel: ZTF light curve ($r$-band) showing eclipses.  Middle panel: ZTF light curve folded on 8.96\,h showing the eclipse. Bottom panel: The spectrum shows absorption lines that may be from an early-type donor (Mg is indicated).}
\end{figure}

The spectrum  (Fig.\,\ref{fig:interestingJ1712}) shows  Balmer and Helium emission lines, including strong \ion{He}{ii}, and the Bowen blend above a blue continuum.  Narrow absorption lines of Mg, Na and Ca are detected, which may be from the donor star. J1712+3218 is located in the novalike variable area of the HR diagram. The ZTF light curve shows variability with clear eclipsing behaviour. \citet{2014ApJS..213....9D} report an orbital period of 17.998\,h and concluded that this is a Beta Persei system. We determine $\porb=8.9599(1)$\,h  from the eclipses in the ZTF light curve. We also have unpublished MDM time series spectroscopy confirming the orbital period. We classify J1712+3218 as a novalike variable. This is an unusually long period for a novalike variable although it would be consistent with an early donor that is visible even though the disc is in a high state.

\subsubsection{J1726+5432} 
\begin{figure} 
\includegraphics[width=\columnwidth]{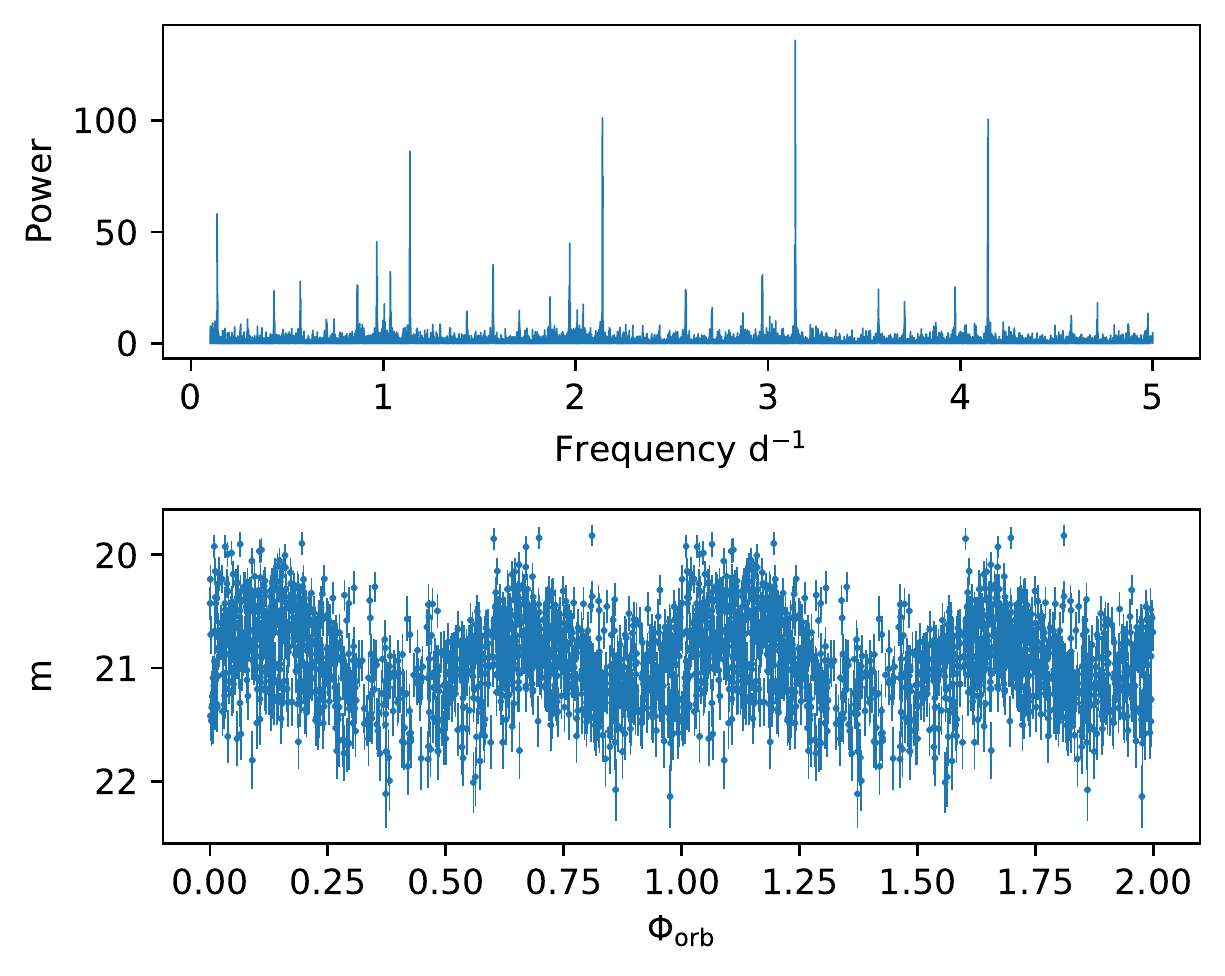}
\caption [.] {\label{fig:interestingJ1726} J1726+5432 has an unusually long  period of $\porb=15.2783(3)$\,h. Top panel: The periodogram peak at 3.14 per day is at twice \porb. Bottom panel: Light curve folded at \porb. 
}
\end{figure}

This system has two SDSS spectra obtained at MJDs 51813 and 51997. Both spectra show single-peaked Balmer, \ion{He}{i} and \ion{He}{ii}  emission lines as well as the Balmer jump in emission  (Fig.\,\ref{fig:interestingJ1726}). The SED shows that it has historically been brighter.  There is no \textit{Gaia} parallax. It is above the white dwarf cooling sequence in the SDSS $u-g, g-r$ diagram, although the SDSS photometry was obtained during a brighter state. The ZTF light curve shows no outbursts but low-level variability. We find an unusually long  period of 15.2783(3)\,h from the ZTF light curve which, together with the helium-rich spectrum may indicate an evolved donor. In the absence of outbursts we cannot assign a sub-type and have classified this generically as a CV.

\subsubsection{J1831+4202} 
\begin{figure} 
\includegraphics[width=\columnwidth]{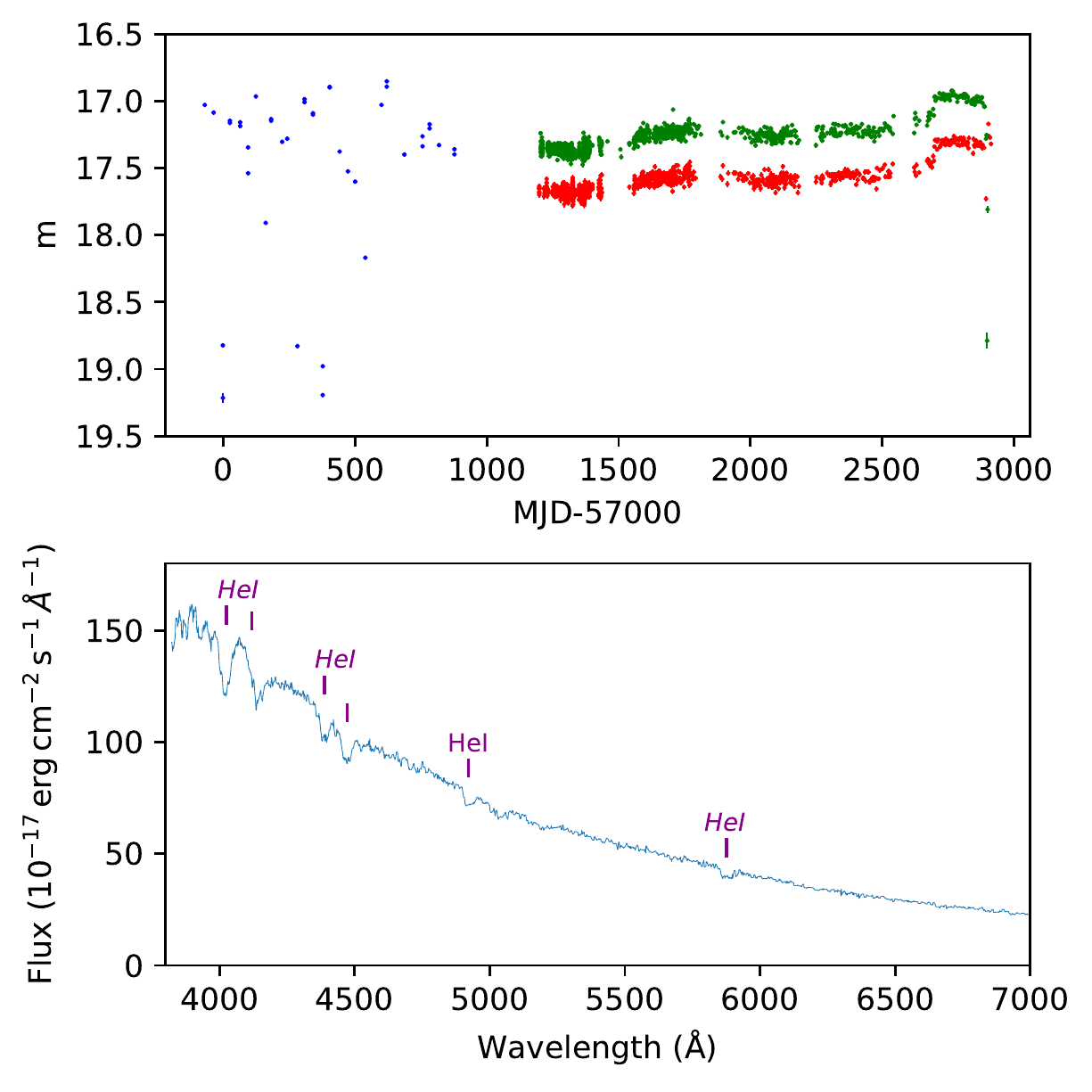}
\caption [.] {\label{fig:interestingJ1831} J1831+4202 is an AM\,CVn with a helium-dominated absorption line spectrum and it exhibits low-amplitude brightness variations on time scales of about a year with occasional drop-outs. Top panel: The ZTF light curve ($r$- and $g$-band data shown in red and green respectively) and the \textit{Gaia} light curve (shown in blue). Bottom panel: Spectrum showing helium absorption lines and an absence of hydrogen lines.
}
\end{figure}

\ion{He}{i} absorption lines are detected in the spectrum superimposed on a blue continuum (Fig.\,\ref{fig:interestingJ1831}). There is no sign of the white dwarf or the donor. J1831+4202 has previously been classified as a  CV \citep{2011MNRAS.417.1210G} and a DB white dwarf \citep{2019ApJ...882..106G,2019A&A...621A..38G} however the DB classification is inconsistent with the variability in the \textit{Gaia} and ZTF light curves which indicates that this is an interacting binary. It is close to AM\,CVn in the HR diagram and we hence classify this system as an accretion-disc dominated AM\,CVn. The drop-outs in the \textit{Gaia} and ZTF light curves are unusual for this type of system.

\subsubsection{J2119+0332}
\begin{figure} 
\includegraphics[width=\columnwidth]{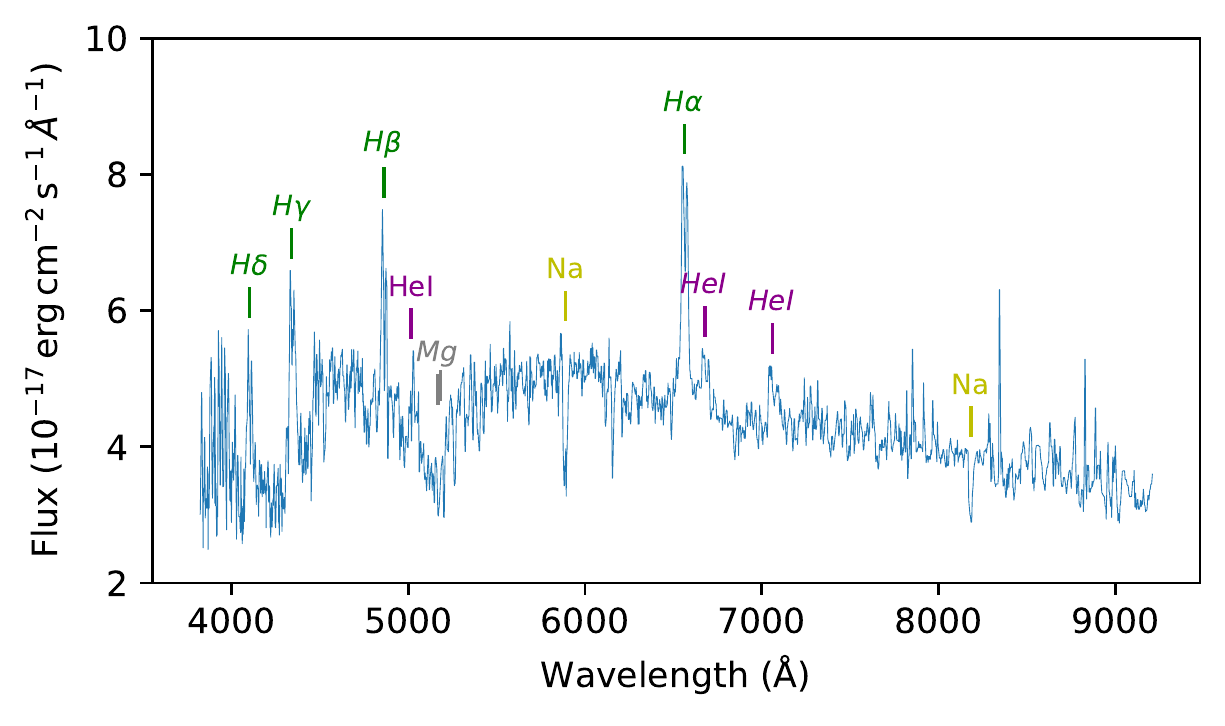}
\caption [.] {\label{fig:interestingJ2119} The spectrum of J2119+0332 shows Na and Mg lines consistent with a K-type star. Whilst bright enough to outshine the disc the spectrum is too dim for a main sequence K-type suggesting a nuclear evolved stripped donor. This hypothesis is corroborated by the short orbital period of 2.4\,h.
}
\end{figure}

Double-peaked Balmer and \ion{He}{i} lines are visible in the spectrum as well as the Balmer jump in emission  (Fig.\,\ref{fig:interestingJ2119}). There are also \ion{Na}{}{} absorption lines at $\simeq8190$\,\AA\ and at $\simeq5890$\,\AA\ together with broad Mg at $\simeq5170$\,\AA\ suggesting a K-type donor.  J2119+0332 is closer to the main sequence than the  white dwarf cooling sequence in the HR diagram. There are no outbursts in either the CRTS or the ZTF light curves. We found ellipsoidal modulation and a reliable orbital period of 2.37223(3)\,h from the ZTF light curve, i.e. within the period gap. The period and relatively low luminosity are inconsistent with a main-sequence K-type donor and reminiscent of SDSS\,J001153.08$-$064739.2 \citep{2014ApJ...790...28R} and QZ\,Ser \citep{2002PASP..114.1117T} where the donor has undergone nuclear evolution on the main sequence before mass transfer to the white dwarf commences. In these cases a period of rapid mass transfer has resulted in a stripped donor and a short period. The strong Na absorption lines are consistent with CNO burning.

\subsection{Non-CVs} 
\label{section:misclassifications}

The classification of a number of systems that were previously classified as CVs have now, with the benefit of new data such as from \textit{Gaia} and ZTF, been revised. Most of the systems previously classified as CVs turned out to be either single white dwarfs or white dwarf plus main sequence binaries, in which emission lines from a nebular background or from the chromosphere of the companion were mistaken for signatures of accretion flows. The new classifications of these non-CV systems are listed in Table \ref{tab:notacv}.


\begin{table*}
\caption{Systems identified as CVs in the literature which we have reclassified as detached binaries or isolated stars. Unless otherwise stated the original classification was taken from \citet{2011AJ....142..181S}. Each system may include a white dwarf~(WD), \nolinebreak{M dwarf (M)} or other type of main sequence star (MS).} \label{tab:notacv}

\begin{tabular}{l|p{100mm}|l}
\hline
SDSS name  & Notes & New type\\
  \hline
    002603.80-093021.0 & \citet{2012AJ....144...81T} determined that the spectrum was not that of a CV. \citet{2013ApJS..204....5K} classified the object as  a (hydrogen) DA white dwarf. We found a ZTF period of 9.3086hh confirming that this is probably a binary. & WD+MS \\
034420.16+093006.8 & \citet{2015MNRAS.449.2194P} found that this is a detached, eclipsing system.  & WD+MS \\
035409.32+100924.4 &  \citet{2013ApJS..204....5K} found that this was an eclipsing system including a  DA white dwarf. This system is within the Eridanus superbubble (see fig.\, 1 in  \citealt{2006ApJ...644L.185R}) and this is likely to be the source of the Balmer emission lines; the remaining spectrum is that of an isolated white dwarf.  & WD \\
053659.12+002215.1 & \citet{2012AJ....144...81T} determined that the spectrum was not that of a CV. \citet{2013ApJS..204....5K} classified the system as a  DA white dwarf. This system is on the edge of the Eridanus superbubble (see fig.\, 1 in  \citealt{2006ApJ...644L.185R}) and this is likely to be the source of the Balmer emission lines; the remaining spectrum is that of an isolated white dwarf. & WD \\
075117.09+144423.5 & The spectrum shows emission lines but these are contamination from a nearby CV (PQ\,Gem) and the object is actually an M-dwarf. & MS Star \\ 
102347.67+003841.2 & \citet{2009ApJ...703.2017W} found that this object is a millisecond radio pulsar - a low mass X-ray binary including a neutron star. & Low mass X-ray binary \\
105443.06+285032.7 & \citet{2010MNRAS.402..620R} concluded that the object is a detached binary including a  DA white dwarf. \citet{2012AJ....144...81T} also determined that the spectrum was not that of a CV. We note that the weak Balmer emission lines appear to vary in strength between multiple SDSS spectra. & WD+M \\
111703.53+182558.1 & \citet{2012AJ....144...81T} determined that the spectrum was not that of a CV. We classify the system as a detached binary containing a hot  white dwarf and a main sequence star. The SDSS and LAMOST \citep{2017ASPC..512..153H} spectra show that the emission lines vary in strength consistent with coming from the irradiated secondary at different phases of the orbit. We find a period of 9.33\,h from the ZTF light curve. & WD+MS \\
121929.46+471522.8 & \citet{2020MNRAS.499.2564G} concluded that this was an isolated magnetic white dwarf with Zeeman-split Balmer line emission. & WD \\
124959.76+035726.6 & \citet{2010MNRAS.402..620R} concluded that the object is a detached binary including a  DA white dwarf. \citet{2012AJ....144...81T} also determined that the spectrum was not that of a CV. We find a period of 17.84\,h from the ZTF light curve. & WD+M \\
130236.21+060148.0 & \citet{2012AJ....144...81T} determined that the spectrum was not that of a CV. \citet{2019AJ....157..231K} found that it was an isolated system~--~either an M or L dwarf. & M/L dwarf \\
131424.68+455528.3 & \citet{2016MNRAS.455.3413K} identified this as a CV from SDSS data release 12. However the spectrum only shows very narrow emission lines and there are no recorded outbursts. It is in a nearby galaxy, UGC 8320  (4.3\,Mpc, {\citealt{2003A&A...398..467K}}), and so it is most likely to be a hot star with the emission lines arising from the background galaxy. & MS Star\\
131709.07+245644.2 & \citet{2012AJ....144...81T} determined that the spectrum was not that of a CV. & WD+M \\
132856.71+310846.0 & \citet{2012AJ....144...81T} determined that the spectrum was not that of a CV. \citet{2019A+A...621A..38G} found that it was a hot luminous subdwarf. & Hot luminous subdwarf? \\
151500.56+191619.6 & \citet{2012AJ....144...81T} determined that the spectrum was not that of a CV.  It appears to be a detached binary with emission lines caused by irradiation. & WD+MS \\
155904.62+035623.4 & \citet{2012AJ....144...81T} determined that the spectrum was not that of a CV. \citet{2013MNRAS.429..256P} concluded that it was a WD+MS. We find a period of 2.5h from the ZTF light curve. & WD+MS \\
202520.13+762222.4 & The spectrum is very noisy with only the H$\alpha$ emission line clearly visible. There are no recorded outbursts and it is on the main sequence. The SED is consistent with this being a single M-dwarf. This is most likely to be a flaring main sequence star rather than a CV.& MS Star \\ 
210241.09-004408.3 & \citet{2012AJ....144...81T} determined that the spectrum was not that of a CV & MS Star or subdwarf?
    \\
  \hline
\end{tabular}
\end{table*}

\section {Discussion} \label{section:discussion}
\subsection{Distribution of periods} \label{section:perioddistribution}

\begin{figure} 
\includegraphics[width=\columnwidth]{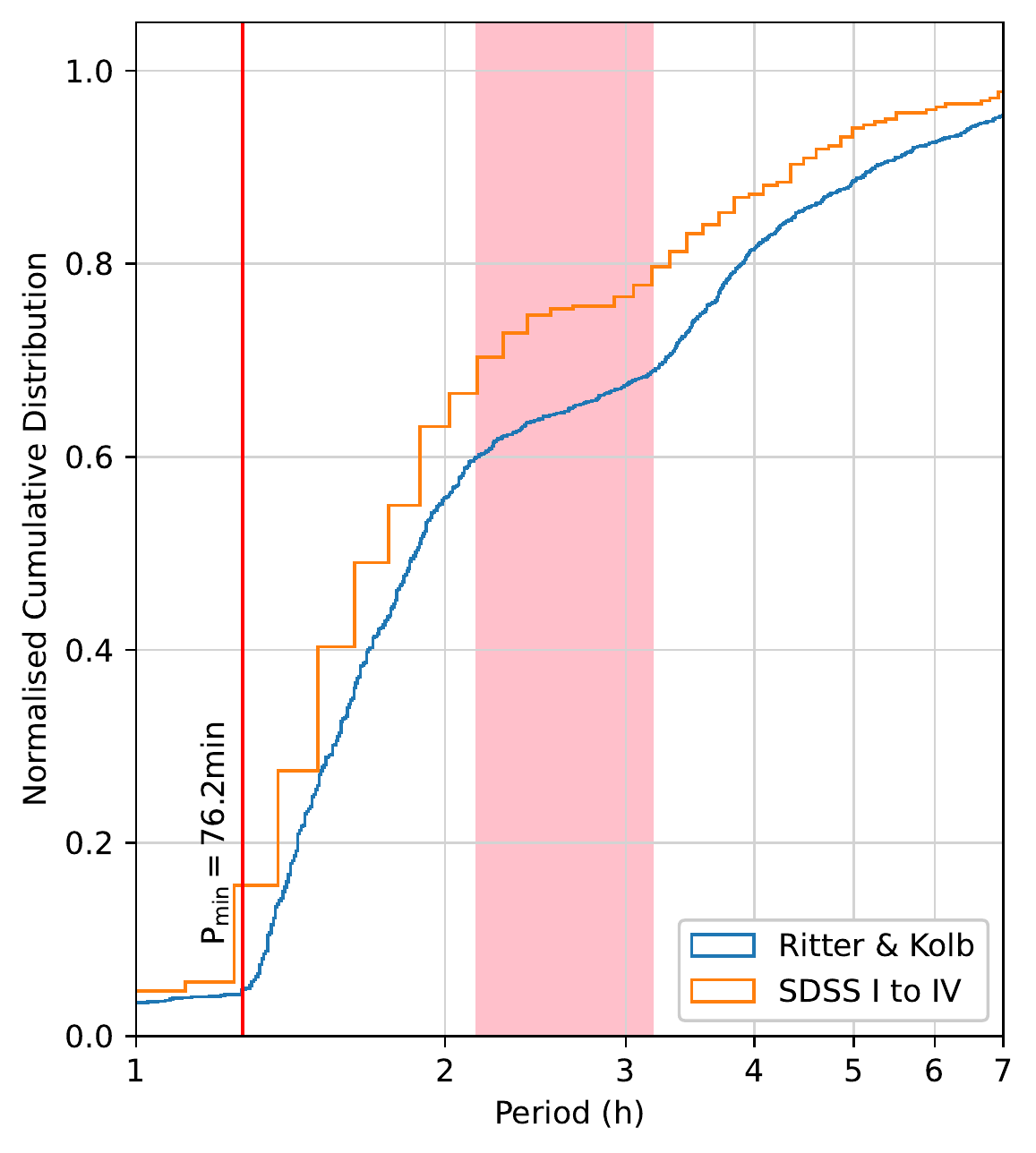}
\caption{\label{fig:PeriodDist}Cumulative distribution of the periods of the CVs in \allsdss. The distribution is compared with that of the Ritter and Kolb catalogue (version 7.24, \citealt{2003A&A...404..301R}). The period minimum (red line) and ``period gap'' (pink rectangle) are identified (see text for details). }   
\end{figure}

We have assembled periods for over half of the CVs in \allsdss, either from the literature or from our analysis of the ZTF data. The main reasons why the remaining systems have no period measurements are either a low inclination (which will reduce the amplitude of any orbital modulation), faintness (i.e. low signal-to-noise ratio of the time-series photometry or spectroscopy), or sparse sampling of the available time-series data. With the possible exception of low \mdot\ short-period CVs (which do not have much orbital modulation from the hot spot and are inherently fainter) these factors are not likely to correlate with the period. The distribution of the SDSS CV periods is therefore assumed to be relatively unbiased, which we compare in Fig.\,\ref{fig:PeriodDist} with that of version 7.24 (December 2015) of the \citet{2003A&A...404..301R} catalogue of CVs (hereafter R\&K). This final version of the R\&K catalogue  lists 1429 CVs (including some from \allsdss); however it focuses on well-studied CVs, many of which were discovered by virtue of their outbursts.  

Interestingly the period gap is clearly evident in the period distribution of CVs from the R\&K catalogue but is not so evident in the \allsdss\, sample. The proportion of SDSS CVs with periods below three hours is also greater than in the R\&K catalogue, presumably due to spectroscopic identification of CVs with infrequent outbursts as reported by \citet{2009MNRAS.397.2170G}, and the fact that the SDSS CV sample reaches, on average, fainter apparent magnitudes than the R\&K sample (19.7 and 18.1 respectively). This is consistent with the results obtained by \citet{2020MNRAS.494.3799P} and \citet{2021MNRAS.504.2420I}. Fig.\,\ref{fig:PeriodDist} highlights that selection effects have to be taken carefully into account when constructing observed CV samples to develop, support, or disprove theories of CV evolution, such as the canonical ``disrupted magnetic braking framework'' \citep{1983ApJ...275..713R,2011ApJS..194...28K} as well as suggestions that the structure of the donor star may have no strong dependence upon the efficiency of magnetic braking \citep{2003ApJ...582..358A,2022MNRAS.517.4916E}.

\begin{figure} 
\includegraphics[width=\columnwidth]{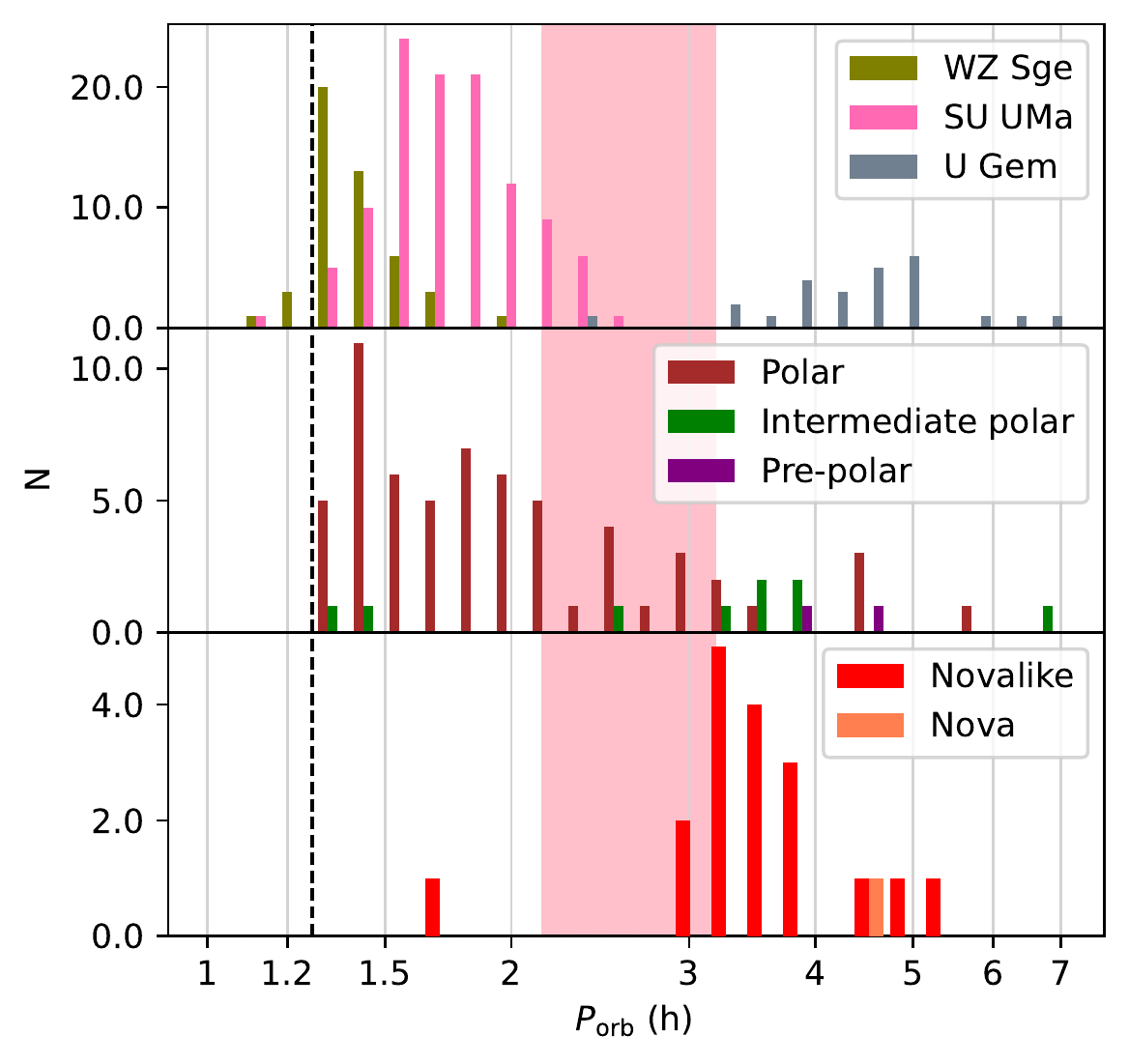}
\caption{\label{fig:PeriodDistbytype}Distribution of the periods of the CV sub-types in \allsdss. CVs evolve from right to left past the ``period gap'' (pink rectangle) to the period minimum (broken black line) and then back out to longer periods. Top panel: U\,Gem CVs typically have periods above the period gap whilst most SU\,UMa CVs have shorter periods below the gap.  Middle panel: Polars tend to have shorter periods whilst intermediate polars and pre-polars tend to be longer.  Bottom panel: Novalike variables congregate above the period gap alongside classical novae.
}   
\end{figure}

The orbital period distribution  broken down by CV-sub-type is shown in  Fig.\,\ref{fig:PeriodDistbytype}. From the top panel it can be seen that U\,Gem CVs typically have periods above the period gap whilst most SU\,UMa CVs are located below the gap. Later in the evolution process low accretion WZ\,Sge CVs approach the period minimum and subsequently evolve back towards longer periods as ``period bouncers''.  The CVs below the period minimum have helium-rich donors which were partially evolved before accretion commenced or formed directly from a detached white dwarf/brown dwarf binary  \citep{1999MNRAS.309.1034K,2007MNRAS.381..827L,2003MNRAS.340.1214P}. 
Polars tend to have shorter periods whilst those of intermediate polars and pre-polars tend to be longer. The ``period gap'' is not so apparent in magnetic CVs, (see for example \citealt{2019MNRAS.488.2881M}) and may be a result of less efficient magnetic braking due to the white dwarf magnetic field \citep{2002MNRAS.335....1W}.  Novalike variables congregate above the period gap alongside classical novae. Many classical novae become novalike variables following their eruptions although there are exceptions where they become dwarf novae in the period between eruptions  \citep{1995ApJ...446..838H,2016Natur.537..649M}.  Short-period novalike variables are very rare;  BK\,Lyn is the most famous example:  \citet{2013MNRAS.434.1902P} speculate that this is due to being the result of a relatively recent ($\sim\,2000$\,yr ago) nova, and it has now morphed into an ER\,UMa dwarf nova. 

\begin{figure} 
\includegraphics[width=\columnwidth]{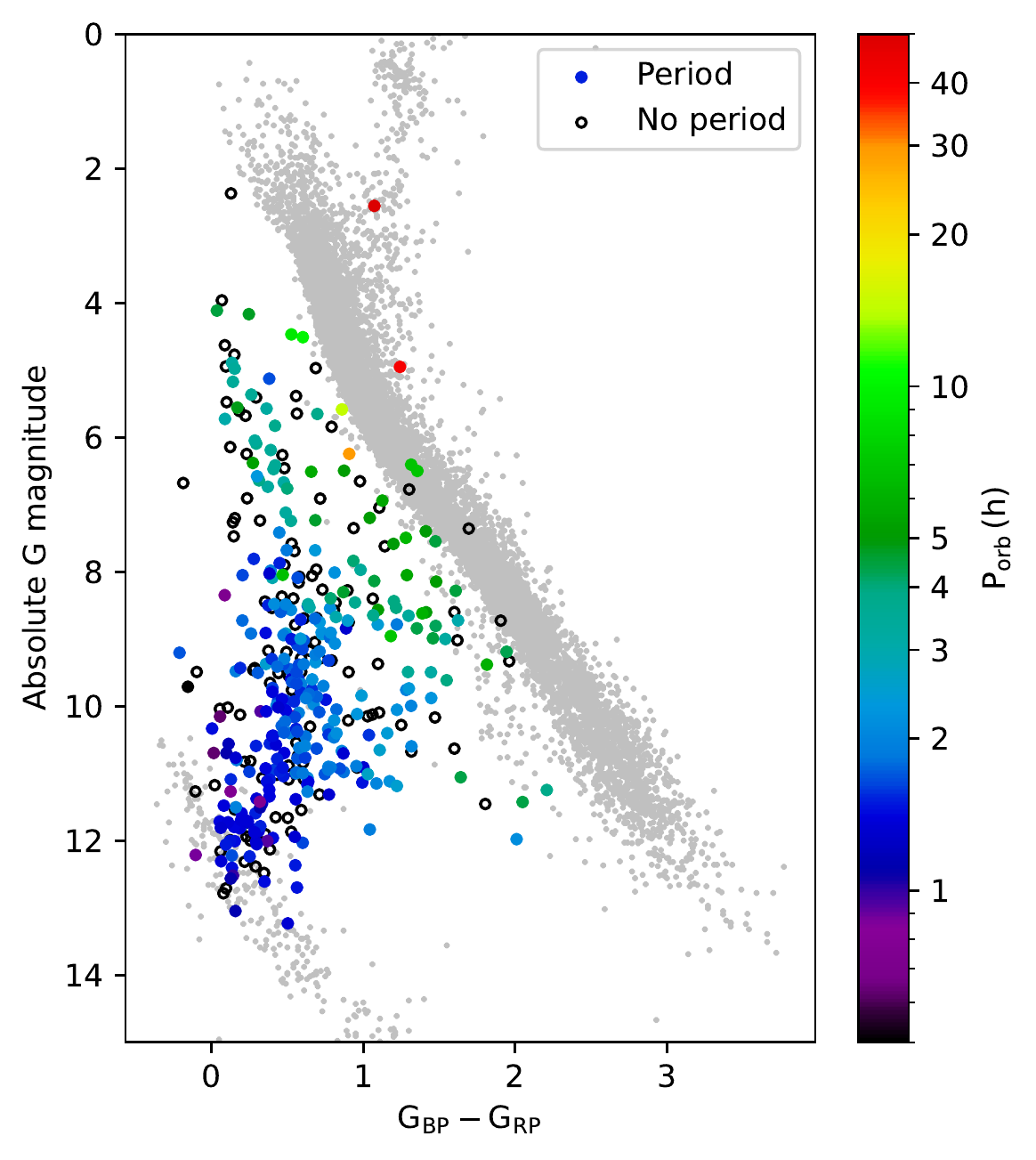}
\caption{\label{fig:HRDiagram} HR diagram of the CVs in \allsdss\, showing the orbital periods where available. CVs without a period are shown as empty black circles. Note that the shortest periods are close to the white dwarf cooling sequence and the longest periods are close to the main sequence.
 }   
\end{figure}

The evolution of orbital periods of CVs across the HR diagram is shown in Fig.\,\ref{fig:HRDiagram}.  CVs originate in or near the main sequence, with their luminosity dominated either by a luminous accretion disc or donor star, and progressively migrate towards the white dwarf cooling sequence as their donor mass decreases and periods shorten. In most WZ\,Sge the contribution of the accretion disc to their optical flux is small, and that of their donor stars entirely negligible.

\subsection{Distribution of CV sub-types}
\begin{figure} 
\includegraphics[width=\columnwidth]{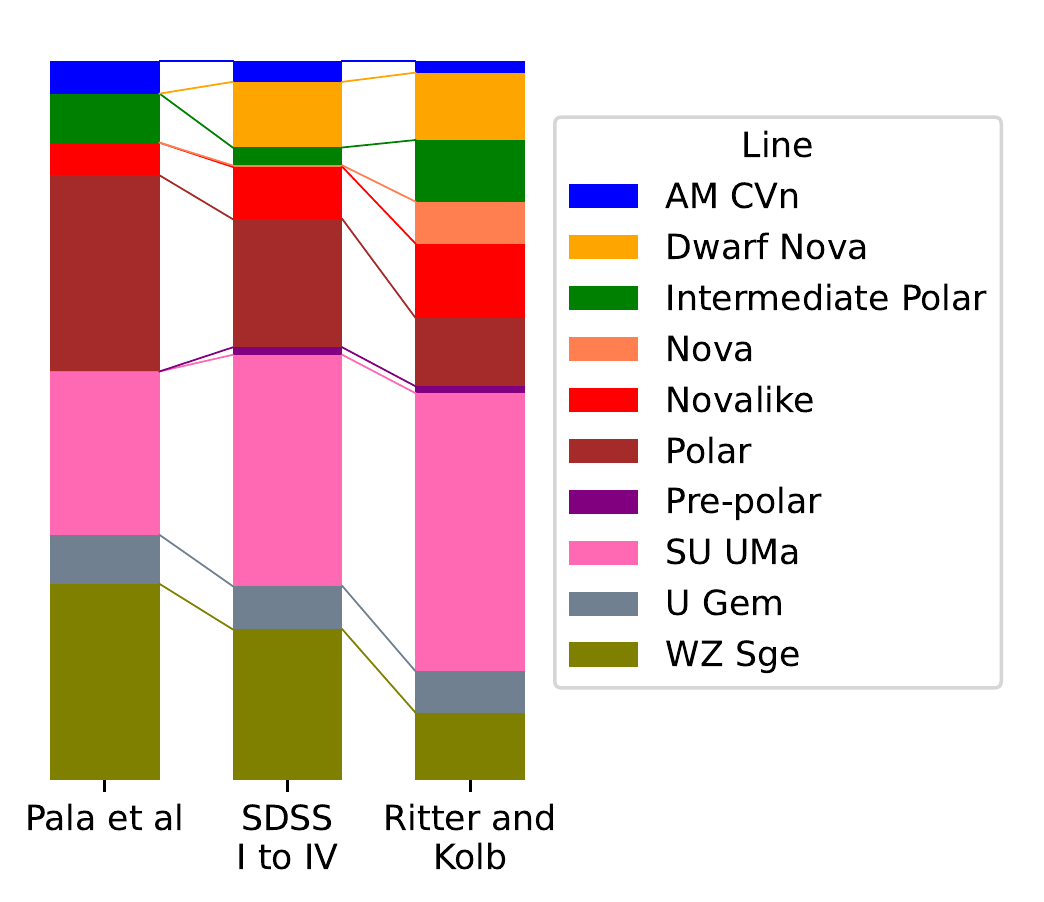}
\caption{\label{fig:Hist_of_types} Comparison of the distribution of sub-types between a volume-limited  150\,pc sample \citep{2020MNRAS.494.3799P}, the \allsdss\ sample and the R\&K catalogue. Objects identified as either CVs or magnetic CVs (30 out of \fxnum{507} in \allsdss) without further classification have been omitted in each case. Selection effects are clearly evident with faint and/or non-outbursting sub-types less common in the R\&K sample. The proportion of magnetic systems is also significantly higher in the 150\,pc sample.
}   
\end{figure}

The distribution of the \allsdss\ CV sub-types is shown in Fig.\,\ref{fig:Hist_of_types}, where we compare it with that of the 150\,pc volume-limited sample \citep{2020MNRAS.494.3799P} and also with that of the R\&K catalogue.

The proportion of magnetic CVs (polars, pre-polars and intermediate polars) in our sample is 20\,per\,cent which is consistent with the 18\,per\,cent found in  \citet{2003A&A...404..301R} but is much lower than the 34\,per\,cent found in \citet{2020MNRAS.494.3799P}. This may be due to the small sample size (42)  in \citet{2020MNRAS.494.3799P} which has a statistical $2\sigma$\, uncertainty of $\pm19$\, per cent. By comparison our sample (\fxnum{507}) has an uncertainty of $\pm0.03$ per cent. Alternatively the difference may be because the absolute luminosity of polars is on average lower than non-magnetic systems causing a selection effect in our  magnitude-limited sample.

The proportion of novae and novalikes is lower in \citet{2020MNRAS.494.3799P} than our sample. This is due to their intrinsic brightness rendering them visible at far greater distances than other CVs and hence there will be a higher proportion in our magnitude-limited sample. The proportion is greater still in \citet{2003A&A...404..301R} due to historic novae identified by their eruptions.

We found a smaller proportion of  WZ\,Sge CVs than \citet{2020MNRAS.494.3799P} most probably because they are inherently faint and only observed at relatively short distances.

\subsection{Variability of CVs}
The photometric variability of CVs is widely used as a selection criterion when searching for CV candidates. However, different CV sub-types display very different types of variability, both in terms of amplitude and time-scales, and hence specific variability selection methods may introduce strong biases towards (or against) individual sub-types. 

\begin{figure} 
\includegraphics[width=0.99\columnwidth]{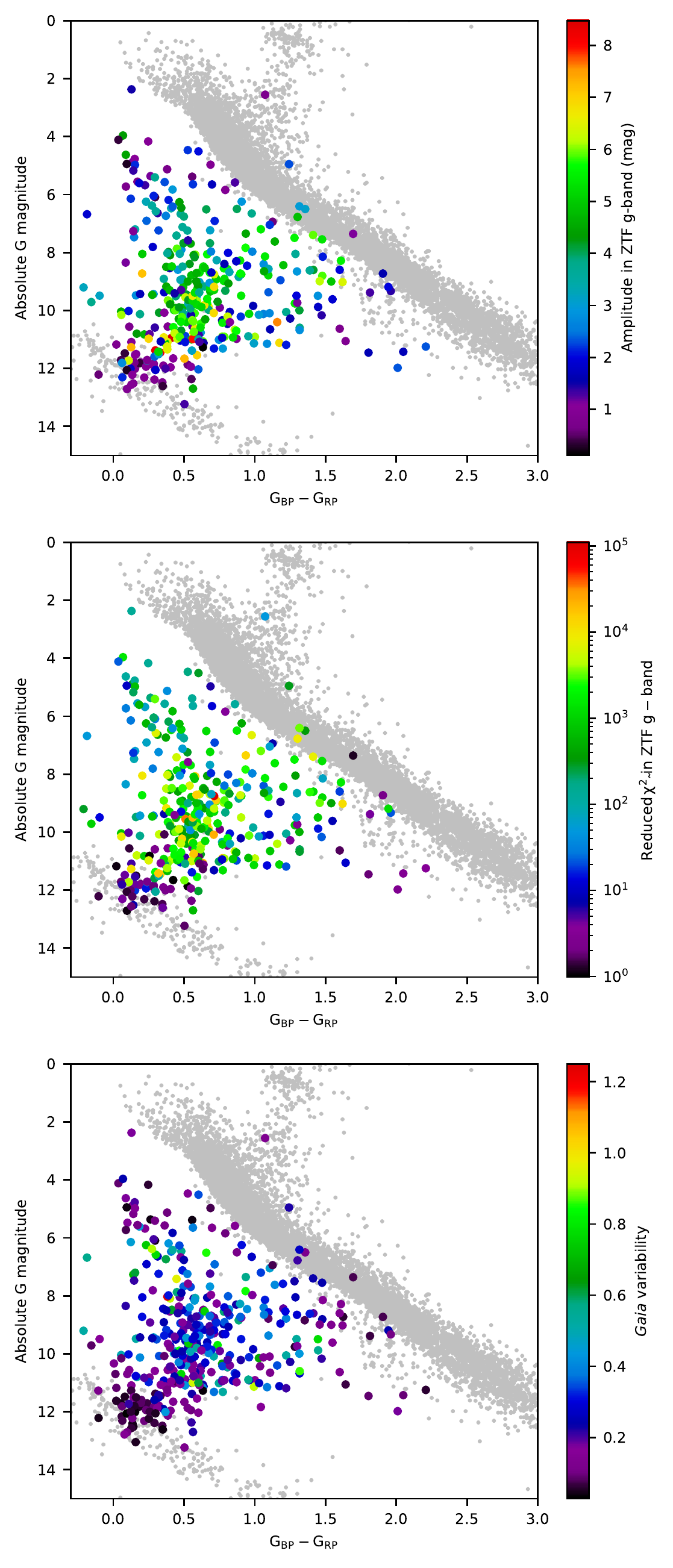}
\caption[]{\label{fig:HRvariability} Variability of CVs related to position in the HR diagram. In each case older CVs which are approaching the white dwarf cooling sequence show less variability.  Top panel: Distribution of maximum variability in the $g$-band ZTF light curve. Middle panel: Reduced chi-squared of the variability of the $g$-band ZTF light curve with respect to the median magnitude. Bottom panel: Variability in \textit{Gaia} $G$-band magnitude (Eq.\,\ref{eq:gvar}). 
}
\end{figure}

To illustrate this point, we computed  three variability criteria for the \allsdss\ CVs: (1) The difference between the maximum and minimum magnitude over the full set of ZTF observations which will be sensitive to outbursts and eclipses, where we only included systems with at least 150 observations. (2) The reduced $\chi^2$ of the ZTF photometry with respect to its median. (3) the \textit{Gaia} $G$-band variability defined as \citep{2021ApJ...912..125G}
\begin{equation}
\label{eq:gvar}
G_\mathrm{var}=\frac{\texttt{phot\_g\_mean\_flux\_error } \times \sqrt{\texttt{phot\_g\_n\_obs}}}{\texttt{phot\_g\_mean\_flux}}
\end{equation}
The distribution of these three types of variability are shown in the \textit{Gaia} HR diagram in Fig.\,\ref{fig:HRvariability}). 

The maximum amplitude of variability in the ZTF light curves (top panel) shows a concentration half-way between the white dwarf cooling track and the main sequence. This reflects the predominant population of SU\,UMa type CVs in this area of the HR diagram (see Fig.\,\ref{fig:CVtypesP}), which have frequent dwarf nova outbursts with amplitudes of $m\simeq3-5$, i.e. it is very likely that ZTF, despite the sparse sampling, will have caught them during an outburst. In contrast, the regions occupied predominantly by novalike variables ($G_\mathrm{BP}-G_\mathrm{RP}\simeq0.25$, $G_\mathrm{abs}\simeq6$) and WZ\,Sge systems ($G_\mathrm{BP}-G_\mathrm{RP}\simeq0.25$, $G_\mathrm{abs}\simeq12$)  show only relatively small variations between the minimum and maximum brightness detected by ZTF. The is reflected by their stable, hot (novalike variables) or cool (WZ\,Sge CVs)  accretion discs. However, a few systems in the region occupied by WZ\,Sge CVs show very large amplitude, $\Delta m\simeq8$, variability, implying that ZTF observed them during one of their rare super-outbursts. The ZTF reduced $\chi^2$ diagram shows overall a similar trend, however, the area occupied by novalike variables exhibits a higher level of variability than in the maximum amplitude diagram, most likely because many of these systems have persistent short-term variability (flickering, \citealt{1992A&A...266..237B}), and on average higher quality ZTF light curves as they are intrinsically bright. Finally, the \textit{Gaia} variability diagram shows again a lack of variability in the area occupied by novalike variables and WZ\,Sge stars, and overall a lower number of strongly variable systems. This is most likely due to the fact that the \textit{Gaia} data spans a shorter baseline ($\simeq2.8$\,yr), and has on average less data points than the ZTF photometry. Future \textit{Gaia} releases will extend the baseline to at least 5\,yr, possibly up to 10\,yr, increasing the diagnostic power of the \textit{Gaia} variability index.

\subsection{Space Density}

Measuring the space density of the various CV sub-types is fundamentally important for validating evolutionary models.

Assuming that the Galaxy is axially symmetrical the space density of a class of objects is a function of the height above the Galactic disc and the radial distance from the Galactic centre. Our analysis is sufficiently local that radial variations will be small and we henceforth ignore this factor. We further assume an exponential vertical distribution with scale height $h$ where $h$ is assumed to be dependent upon the typical age of the CV sub-type. The space density at a height $z$ above the plane is therefore:

\begin{equation} \label{eq:scale1}
\rho\left(z \right)=\rho _{0}\times\exp\left( -\frac{\left | z \right |}{h} \right) 
\end{equation}

Our objective is to estimate the space density of the CV sub-types. Hence if the estimate of the number of a sub-type in a volume $V$ is $N_{\mathrm{obs}}$ the space density is given by:
\begin{equation} \label{eq:scale2}
\rho _{0} =\frac{ N_{\mathrm{obs}}}{\displaystyle\int_V{\exp\left( -\frac{\displaystyle\left | z \right |}{\displaystyle h} \right) }dv}
\end{equation}

\noindent which for a suitably large sample volume would become a constant dependent only on the choice of $h$. This poses four challenges which are addressed in the following sections. (1) some CVs within the SDSS footprint would have been too bright or too faint to be observed by SDSS; (2) an appropriate value of $h$ for each CV sub-type needs to be chosen; (3) SDSS was not complete and did not target a proportion of CVs within its footprint; (4) the SDSS footprint only achieved partial sky coverage and a method of extrapolating the results derived from the SDSS observations is needed. 

\subsubsection{CV brightness}
\allsdss\, spectroscopy did not target any objects with $g\lesssim15$, which has been taken into account in section\,\ref{sec:completeness} below. At the faint end it will be apparent from the HR diagram (Fig.\,\ref{fig:HRDiagram}) that some sub-types have a larger intrinsic luminosity than others and will therefore be observable at greater distances by SDSS. We therefore sought to establish for each sub-type a  limiting distance ($R_\mathrm{lim}$) within which they would have reliably been detected by SDSS. To derive this we first assume that CVs with $g<21$ will be reliably detected from inspection of their SDSS spectrum if targeted; this is a conservative assumption as for example \citet{2017A&A...597A..79P} used $g<22$. In the following, we decided to use the \textit{Gaia} $G$-band magnitude rather than SDSS-$g$, as the multiple observations of \textit{Gaia} more adequately sample the average brightness of an individual system. The two passbands yield very similar apparent magnitudes for a given brightness. We calculated the mean ($\mu$) and standard deviation ($\sigma$) of the absolute \textit{Gaia} $G$ magnitude of each sub-type from the \allsdss\ CVs (ignoring objects where the error in the  \citet{2018AJ....156...58B} distance was greater than $20$\,per\,cent). By definition for any CV of a given sub-type with apparent magnitude $m$, absolute magnitude $M$\, and distance $d$\,pc there is a 99.7\,per\,cent probability that $ M<\mu+ 3 \sigma$ hence 
\begin{align}
m-5\log_{10}(d)+5<&\mu+ 3 \sigma \\
m<&\mu+ 3 \sigma+5\log_{10}(d)-5
\end{align}
Now $d<R_{\mathrm{lim}}$ is the largest value of $d$ corresponding to $m=21$:
\begin{align}
21=&  \mu+ 3 \sigma+5\log_{10}(R_{\mathrm{lim}})-5 \\
5\log_{10}(R_{\mathrm{lim}})=&5+21-\mu- 3 \sigma \\
R_{\mathrm{lim}}=&10^{1+0.2(21-\mu- 3 \sigma)}
\end{align}

\noindent The values of $R_{\mathrm{lim}}$ for each sub-type are shown in Table \ref{tab:limitingdistance}.

\begin{table}
\centering
\caption{Absolute magnitudes and limiting distances of CV sub-types from \allsdss. $N$ is the number of \allsdss\ objects with reliable distances (see text). $\mu,\,\sigma$ are the mean and standard deviation of the absolute \textit{Gaia} $G$ magnitudes.}  \label{tab:limitingdistance}
\begin{tabular}{llccc}
\hline
Type & $N$ & $\mu$ & $\sigma$ & $R_{\mathrm{lim}}$\,(pc) \\  \hline
    U Gem & 16 &   7.83 &  1.13 & 899 \\
SU UMa & 66 &   9.90 &  0.94 & 451 \\
WZ Sge & 33 &  11.77 &  0.39 & 412 \\
Polar & 39 &  10.33 &  1.21 & 256 \\
Intermediate Polar & 7 &   8.59 &  1.02 & 739 \\
Novalike variable & 17 &   5.77 &  1.03 & 2676
\\\hline                                   
\end{tabular}
\end{table}

\subsubsection{Scale heights of the different CV sub-types}
As we shall see the choice of scale height in the model has a significant effect upon the final estimates of the space density. \citet{1984ApJS...54..443P}  used a single scale height of 150\,pc for all sub-types in their analysis as did \citet{1984Ap&SS..99..363D} (who used $125\pm22$\,pc).  \citet{2007MNRAS.374.1495P}  assumed scale heights of 120, 260 and 450\,pc (hereafter the P2007 model) for long-\porb\ systems ($\porb>3$\,h), normal short-\porb\ systems ($\porb<3$\,h) and period bouncers respectively. This was based on the assumption that these systems had increasing ages and therefore, as a result of scattering with other stars, increased space velocities resulting in higher average heights above the Galactic plane \citep{1996MNRAS.282.1454K}. We have used the assumptions of P2007 in our work. Where systems do not have an observed period we assign them to one of the three scale heights in P2007 based on their sub-type. U\,Gem CVs, novalike variables and intermediate polars are assumed to be long-period CVs,  short-period systems include SU\,UMa CVs, polars and WZ\,Sge CVs with the third category reserved for period bouncers. We attempted to analyse the \allsdss\ data in order to support the scale height assumptions in P2007 but in practice our samples were not complete to a sufficient Galactic height.

\subsubsection{Completeness} \label{sec:completeness}

\allsdss\ used complex targeting rules focusing largely on quasars and not CVs (e.g. \citealt{2002AJ....123.2945R,2012ApJS..199....3R,2017MNRAS.469.1065D}). Fortunately CVs have similar colours to quasars and a significant proportion of quasar candidates were observed by SDSS. However, given that CV sub-types occupy different regions of the \textit{Gaia} HR diagram (and in the colour-colour diagrams used by SDSS for their target selection)  the proportion of each sub-type observed by SDSS is not the same.  Whilst in principle an analysis could estimate the degree of completeness for each sub-type, such an endeavour would be very complex. We have adopted an empirical approach by analysing version 7.24 of the R\&K CV catalogue which contains 1429 CVs. We removed the 117 CVs in the R\&K catalogue that were originally discovered by SDSS leaving 1312 systems. The apparent magnitude (``mag1'') of these CVs in their normal state was processed and systems that were too bright ($m<15$) or too faint ($m>21$) were also removed\footnote{Ideally, this analysis would have been done by sub-type but we discovered that the ``type1'' and ``type2''  fields in the R\&K catalogue were unreliable; in particular many WZ\,Sge had been classified as SU\,UMa}. From this we established that 208 CVs fell within the footprint of SDSS.  We then established that 97 of these (46.6\,per\,cent)  were re-discovered by SDSS and we assumed that this proportion applies to all CVs and sub-types within the footprint. We then defined a correction factor $C_{\mathrm{orr}}=2.14$ as the reciprocal of this proportion. 

\citet{2020MNRAS.494.3799P} used a different approach by quantifying the number of \textit{Gaia} objects within the SDSS footprint which fulfilled the quasar colour criteria for SDSS~I and II and did not include any cut in the red colours. This resulted in 5300 objects and \citealt{2020MNRAS.494.3799P} then found the proportion of these objects with an SDSS spectrum (53\,per\,cent) of which $\approx6$ were predicted to be in the sample. Only four were found (three had a spectrum) from which they concluded that the completeness was 71\,per\,cent (i.e. $C_{\mathrm{orr}}=1.4$). It is unsurprising that the completeness of the full \allsdss\ sample is lower than that of the nearby, bright 150\,pc sample.

\subsubsection{SDSS Footprint}\label{sec:SDSSfootprint}

The coordinates of the centres of the plates used in the ``BOSS'', ``eBOSS'' and ``SDSS'' 
surveys, together with their radius (1.49 degrees), define the coverage. However there are two problems to address; firstly the plates overlap and secondly CVs are not uniformly distributed in space but concentrated near the Galactic plane.

\begin{figure} 
\includegraphics[width=\columnwidth]{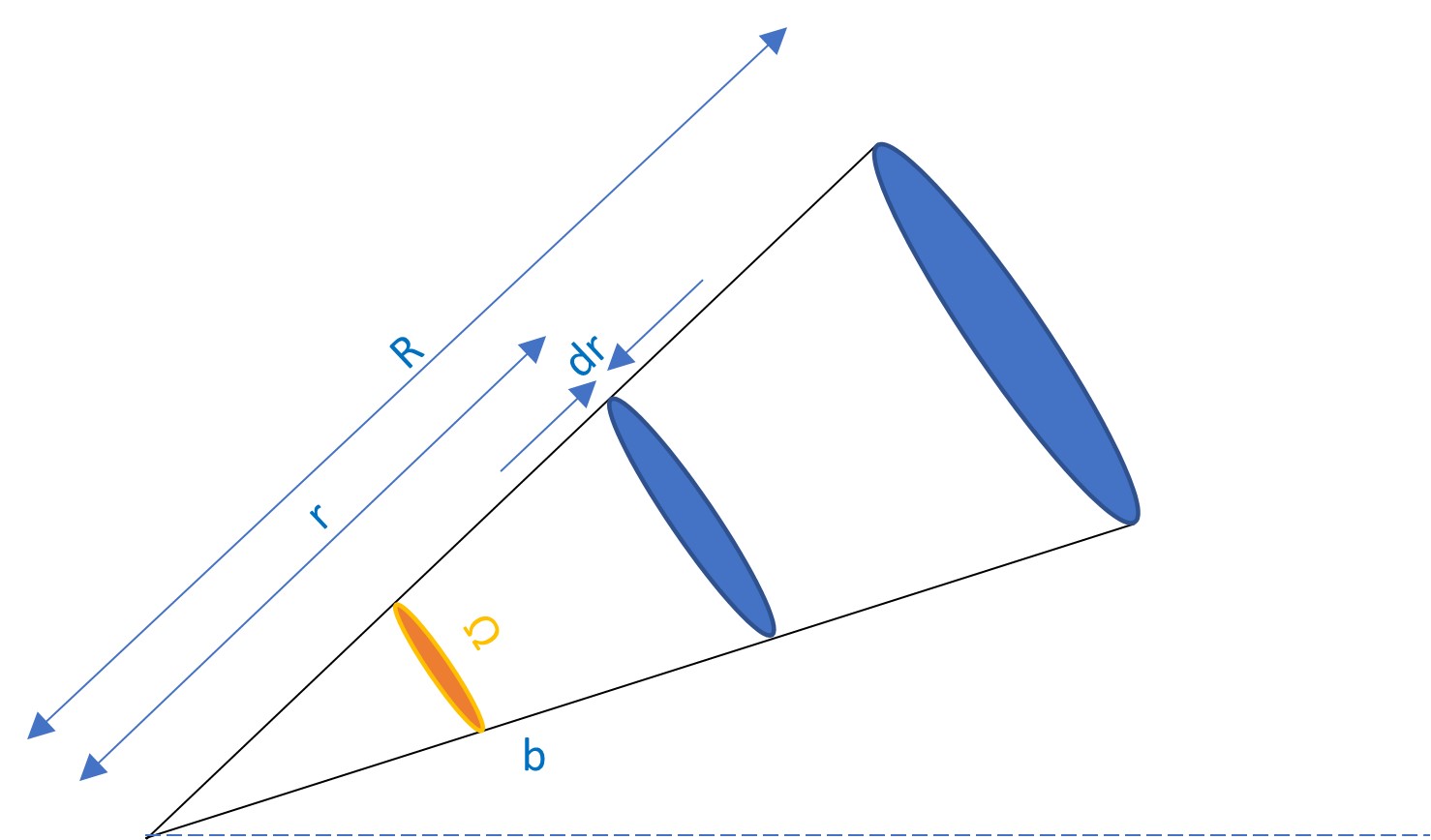}
\caption{\label{fig:diagram1} Diagram depicting  a slice of infinitesimal thickness $dr$ through a cone subtending a solid angle ($\Omega$) at a radial distance (r). The cone represents one HEALPix pixel. The volume of the  slice is  $r^{2}\Omega dr$.
 }   
\end{figure}

For convenience we define from the  denominator of (\ref{eq:scale2})  an effective volume for a sphere radius $R$:
\begin{equation}
V_{\textrm{eff}}\left(h,R \right)=\int_{V(R)}{\exp\left( -\frac{\left | z \right |}{h} \right) }dv
\end{equation}
\noindent To address the overlap problem we model the celestial sphere using $2^8$ \textsc{astropy} HEALPix pixels \citep{2005ApJ...622..759G}. Each HEALPix pixel has the same surface area ($\Omega$ steradians) on the unit sphere and can be represented by a cone (Fig.\,\ref{fig:diagram1}). The effective volume of a sphere radius $R$ is therefore the sum of the volume of the cones:
\begin{equation}
V_{\textrm{eff}}(h,R)=\sum_{\mathrm{All\,pixels}} \int_0^R \exp -\left(\frac{\left | r \sin(b)\right |}{h}\right)( r^{2}\Omega dr)
\end{equation}
The following substitutions are used:
\begin{equation}
A= \frac{\left | R \sin(b))\right |}{h}
\end{equation}
and
\begin{equation}
x=\left(\frac{rA}{R}\right)
\end{equation}
We then compared the coordinates (ra,dec) of each HealPix pixel with the coordinates of the centre of each plate to identify whether that pixel was covered by that plate. We then defined $\mathrm{SDSS_{pix}}$ as the de-duplicated set of HEALPix pixels that lay within the coverage of one or more plates. The effective volume of the SDSS sample is therefore:

\begin{equation}
\begin{split}
V_{\textrm{eff}}(h,R)&=\sum_{\mathrm{SDSS_{pix}}} \int_0^A \exp \left (-x  \right )\left( \left ( \frac{Rx}{A} \right )^{2}\Omega \left ( \frac{Rdx}{A} \right )\right)
\\
&=\Omega R^{3}\sum_{\mathrm{SDSS_{pix}}}\left ( \frac{1}{A^{3}} \right ) \int_0^A x^{2}\exp \left (-x  \right )dx
\end{split}
\end{equation}

\noindent Solving the integral analytically yields
\begin{multline}
V_{\textrm{eff}}(h,R)=\Omega R^{3}\sum_{\mathrm{SDSS_{pix}}}\left ( \frac{1}{A^{3}} \right ) \left ( 2-\left ( A^{2}+2A+2 \right ) \exp\left ( -A \right )\right )
\end{multline}
which can be computed numerically for a given $h$ and $R$. 

\subsubsection{Estimates of space density}

Bringing the preceding sections together the three estimates of the space density (for $h=120, 260, 450$\,pc) are calculated from
\begin{equation}\label{equ:spaced}
\rho _{0} =\frac{N_{\mathrm{obs}}\,\times C_{\mathrm{orr}}}{V_{\mathrm{eff}}(h,R_{\mathrm{lim}})}
\end{equation}
and the results are listed in Table \ref{tab:spacedensity}. 

Next, we need to discuss the  accuracy of these results. From equation (\ref{equ:spaced}) the primary sources of error in $\rho_0$ are in the correction factor and the scale height. The true scale height in each case will very likely lie between the two extreme values that we adopted, i.e. $h=120$\,pc and $h=450$\,pc (Table\, \ref{tab:spacedensity}) and so a reasonable estimate for $\rho_0$ (excepting possibly old WZ\,Sge systems)  is   $\rho_0(260\,\mathrm{pc})$ with the extreme values forming a notional $3\sigma$ error estimate. We have avoided quoting any statistical uncertainties on our space density estimates as the true uncertainties are likely to be dominated by the assumptions on the scale height.

\begin{figure*} 
\includegraphics[width=\textwidth]{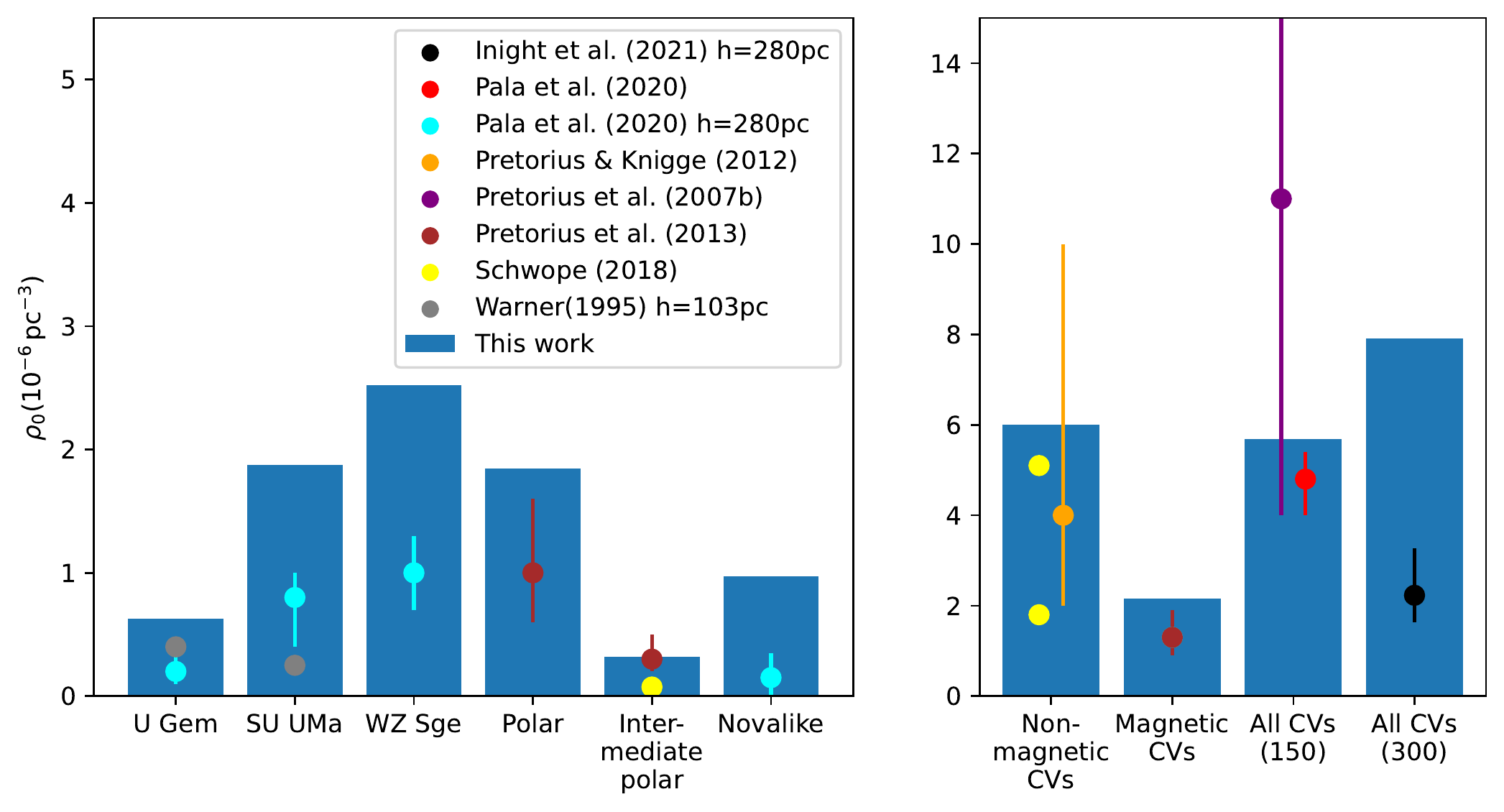}
\caption{\label{fig:space_densities} Comparison of the space density estimates for CVs from this work (bar chart) and published values. All estimates use the P2007 assumptions for scale height $h$ unless otherwise stated. Left panel: Estimates for individual sub-types. Note that \citet{2020MNRAS.494.3799P} are understated as they do not take account of completeness.  Right panel: Composite estimates. \citet{2021MNRAS.504.2420I} specifically excluded selection effects.
 }   
\end{figure*}

It is interesting to compare these estimates with the those based on different approaches used in previous works (Fig.\,\ref{fig:space_densities}) noting that the assumed scale heights have to be comparable.  

Section 9.5 in \cite{1995cvs..book.....W} proposed $\rho_0(103\,\mathrm{pc}) = 0.4 \times 10^{-6}\,\mathrm{pc}^{-3}$ for U\,Gem type systems and  $\rho_0(176\,\mathrm{pc}) = 0.25\times 10^{-6}\,\mathrm{pc}^{-3}$ for SU\,UMa systems. Both values are lower than contemporary estimates presumably as earlier CV discoveries were primarily the result of following up transients and limited to brighter targets.

\citet{2007MNRAS.382.1279P} estimated $\rho_0(120\,\mathrm{pc}) = (15^{+25}_{-8}) \times 10^{-6}\,\mathrm{pc}^{-3}$ for all non-magnetic CVs based on a very small sample of four systems. Our results are towards the lower end of this range.

\citet{2012MNRAS.419.1442P} estimated a density of $\rho_0(260\,\mathrm{pc})= (3.8^{+5.3}_{-1.9}) \times 10^{-6}\,\mathrm{pc}^{-3}$ for non-magnetic CVs which compares with our value of  $\rho_0(260\,\mathrm{pc})  = (4.29) \times 10^{-6}\,\mathrm{pc}^{-3}$ (by adding the individual estimates from Table \ref{tab:spacedensity}).

\citet{2013MNRAS.432..570P} estimated $\rho_0 = (1.3^{+0.6}_{-0.4}) \times 10^{-6}\,\mathrm{pc}^{-3}$ for magnetic CVs, based upon a scale height of $120$\,pc for long period systems and $260$\,pc for short period systems,  which is consistent with our estimates. They also estimated values for polars and intermediate polars which are consistent with this work.  

\citet{2018A&A...619A..62S} estimated $\rho_0(120\,\mathrm{pc})= 0.13 \times 10^{-6}\,\mathrm{pc}^{-3}$ for intermediate polars. This is significantly lower than our estimate.

\citet{2020MNRAS.494.3799P} estimated a composite $\rho_0(280\,\mathrm{pc}) = (4.8^{+0.6}_{-0.8}) \times 10^{-6}\,\mathrm{pc}^{-3}$ for all CVs and our $\rho_0(260\,\mathrm{pc}) = 5.78\times10^{-6}\mathrm{pc}^{-3}$ is consistent with this noting that we use a larger correction factor ($2.1$ compared with $1.4$ in \citet{2020MNRAS.494.3799P}). \citet{2020MNRAS.494.3799P} also provided incomplete (their footnote 9 states that they were unable to estimate the completeness of individual sub-types) estimates for $\rho_0(260\,\mathrm{pc})$   for five  sub-types. These values are lower than our estimates~--~ partly due to (in)completeness but also because we use a larger correction factor ($2.1$ compared with $1.4$ in \citet{2020MNRAS.494.3799P}). Our larger sample size may also have had an effect brought about by the effectiveness of SDSS in discovering short-period systems. 

\citealt{2021MNRAS.504.2420I} estimated $\rho_0 = (2.23^{+1.04}_{-0.6}) \times 10^{-6}\,\mathrm{pc}^{-3}$ for their Gold Sample of 201 systems. The Gold Sample is a reliable subset of the known CVs (at least 305) within 300\,pc and their estimate is therefore a lower limit for the true space density. 

\begin{table*}
\centering
\caption{Analysis of the space density of CV sub-types for different assumptions of scale height. $\rho_0$ (P2007) assumes the scale heights from \citet{2007MNRAS.374.1495P}. $N$ is the number of \allsdss\, objects that are closer than the limiting distance. Note that the values of $\rho_0$ for individual sub-types are slightly understated ($\simeq6$\,per cent)  due to the remaining unclassified CVs and dwarf novae. This does not apply to the ``All CVs'' values which have been calculated for 150\,pc and 300\,pc limiting distances to enable comparison with \citet{2020MNRAS.494.3799P} and \citet{2021MNRAS.504.2420I} respectively.}
\label{tab:spacedensity}
\begin{tabular}{|l|l|l|l|l|l|l|} 
\hline
\multirow{3}{*}{Type} & \multirow{3}{*}{\begin{tabular}[l]{@{}l@{}}Limiting\\ distance\\ (pc)\end{tabular}} & \multirow{3}{*}{$N$} & \multicolumn{4}{c|}{Space Density $\mathrm{(10^{-6} pc^{-3})}$}  \\
                      &                                                                                     &                    & \multicolumn{4}{l|}{}                                            \\ 
\cline{4-7}
                      &                                                                                     &                    & $\rho_0(120)$ & $\rho_0(260)$ & $\rho_0(450)$ & $\rho_0(P2007)$            \\
\hline
    U Gem & 899 & 10 & 0.68 & 0.16 & 0.09 & 0.63 \\
SU UMa & 451 & 30 & 4.21 & 1.71 & 1.19 & 1.88 \\
WZ Sge & 412 & 38 & 6.24 & 2.72 & 1.96 & 2.52 \\
Polar & 256 & 8 & 3.25 & 1.89 & 1.53 & 1.84 \\
Intermediate polar & 739 & 6 & 0.50 & 0.14 & 0.08 & 0.32 \\
Novalike & 2676 & 23 & 1.01 & 0.12 & 0.04 & 0.97 \\
All CVs & 150 & 7 & 8.21 & 5.78 & 5.07 & 5.68 \\
All CVs & 300 & 52 & 13.92 & 7.23 & 5.61 & 7.92  
    \\\hline
\end{tabular}
\end{table*}

\begin{figure} 
\includegraphics[width=\columnwidth]{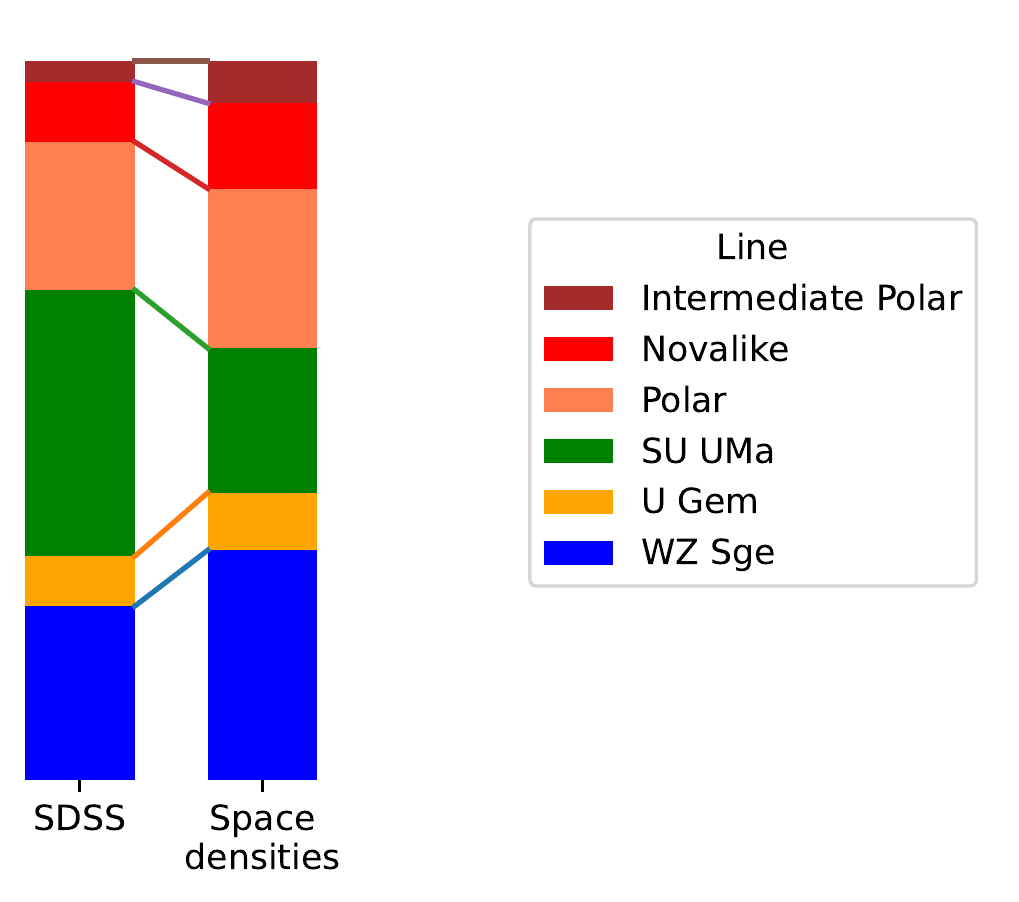}
\caption{\label{fig:comparison} Comparison of the distribution of sub-types between the \allsdss\ sample and our predicted space densities using the P2007 scale height assumptions. This plot ignores sub-types for which we have not estimated a space density.
}   
\end{figure}

In principle, the fractions of sub-types from our space density estimates should reflect the true fractions in the underlying population without selection effects~--~as opposed to the number ratios in the magnitude limited \allsdss\ sample. We therefore compared these fractions with those from the \allsdss\ sample (Fig.\, \ref{fig:comparison}), where the main difference is the larger proportion of WZ\,Sge stars and a lower fraction of SU\,UMa CVs when looking at the space density results. This is in line with the expectation that low-luminosity WZ\,Sge stars should be under-represented in a magnitude limited sample. Whilst demonstrating that the \allsdss\ sample is subject to some selection effects, we also caution not to over-interpret the space density ratios because of the significant effect of scale height assumptions. 

In summary, the long-standing discussion of the CV space density seems to converge to a value of $\simeq5\times10^{-6}\mathrm{pc}^{-3}$. However, the uncertainties in scale heights continue to limit the accuracy of any estimates of space density.  We have calculated space densities for six sub-types and Fig.\,\ref{fig:space_densities} demonstrates that these estimates are consistent with other work.

\subsection{Low accretion rate systems}

Period bouncers are CVs where the mass of the donor has shrunk below the level needed to sustain hydrogen fusion in the core ($\simeq0.061\Msun$) and the donor has evolved to a brown dwarf. A consequence of this is that the orbital separation of the CV increases and \porb\ evolves back from the period minimum (often taken as $\sim76.2$\,min, \citealt{2006MNRAS.373..484K}) towards longer periods. \citet{2011MNRAS.411.2695P} identified a total of 22 candidate period bouncers, \fxnum{eight} of which are in our sample (Table\,\ref{tab:bouncers}). Together with one polar, the \fxnum{13} non-magnetic period-bouncers in Table\,\ref{tab:bouncers} represent \fxnum{14} \,per\,cent of the WZ\,Sge CVs and \fxnum{2.6}\,per\,cent of our complete sample. The two new candidates from this work (J0904+4402 and J2131$-$0039) were identified because of their spectra (white dwarf visible and donor not apparent) and an orbital period significantly larger than the period minimum.
We estimated the space density of period bouncers using \fxnum{14} per cent of the space density of WZ\,Sge from Table \ref{tab:spacedensity} to be $\rho_0(450)=0.27\times\mathrm{10^{-6} pc^{-3}}$. Comparing this with $\rho_0(P2007)=7.92\times\mathrm{10^{-6} pc^{-3}}$ implies that only $3.4$\, per\,cent of CVs are period bouncers which is a long way from the 70\,per\,cent of all CVs predicted by \citet{1993A&A...271..149K}. \citet{2015ApJ...809...80G} modelled CV evolution to predict the  intrinsic proportion of period bouncers as being $38-60$\,per\,cent of all CVs  and the observed proportion (taking account of selection effects) as $2.5-11.2$\,per\,cent. Although period bouncers are intrinsically faint any observational bias is unlikely to fully account for this difference. As long as period bouncers maintain a minimum amount of mass transfer, and associated Balmer emission lines, SDSS spectroscopy is extremely efficient at identifying WZ\,Sge CVs, independent of their orbital period and their outburst frequency. The lack of observed period bouncers remains puzzling, and could be due to the donors in  period bouncers either becoming detached or merging with the white dwarfs.

\begin{table*}
\caption{Period bouncers within the coverage of \allsdss. Candidate systems are shown in italics. The mass ratio of two systems were identified using the method from \citet{2013PASJ...65..115K}; the donor mass is estimated by assuming a maximum white dwarf mass of $1.4 \Msun$ and is therefore definitely low enough to form a brown dwarf. J1212+0136 is an interesting system as it was initially assumed to be a detached system with a white dwarf and low mass companion. It was only recognised as an accreting system from the detection of cyclotron emissions and an X-ray observation which also confirmed its type. } \label{tab:bouncers}
\begin{tabular}{|l|l|l|l|l|}
\hline
SDSS name & CV type & Reference & \porb\ (h) & $M_{\mathrm{donor}} (\Msun)$ \\ \hline
\textit{J0058-0107} & WZ Sge & \citet{2017PASJ...69...75K} & 1.559 & q=0.079 \\
J0804+5103 & WZ Sge & \citet{2021ApJ...918...58A} & 1.433 & 0.042 \\
J0843+2751 & WZ Sge & \citet{1998PASP..110.1290P} & 1.439 & 0.02 \\
\textit{J0904+4402} & WZ Sge: & This work & 1.674 &  \\
J1035+0551 & WZ Sge: & \citet{2006Sci...314.1578L} & 1.368 & 0.06 \\
J1057+2759 & WZ Sge: & \citet{2017MNRAS.467.1024M} & 1.507 & 0.044 \\
\textit{J1212+0136} & Polar & \citet{2008ApJ...674..421F} & 1.474 &  \\
J1216+0520 & WZ Sge: & \citet{2006MNRAS.373..687S} & 1.646 & <.04 \\
J1255+2642 & WZ Sge & \citet{2005PASP..117..427P} & 1.992 & <.05 \\
J1433+1011 & WZ Sge & \citet{2011MNRAS.415.2025S} & 1.302 & 0.0571 \\
J1435+2336 & WZ Sge: & \citet{2022MNRAS.510.6110P} & 1.296 &  \\
J1507+5230 & WZ Sge & \citet{2007MNRAS.381..827L} & 1.11 & 0.056 \\
\textit{J2131-0039} & WZ Sge: & This work & 1.67 &  \\
J2304+0625 & WZ Sge & \citet{2014PASJ...66..116N} & 1.616 & q=.053
\\ \hline
\end{tabular}
\end{table*}

\begin{figure*} 
\includegraphics[scale=0.62]{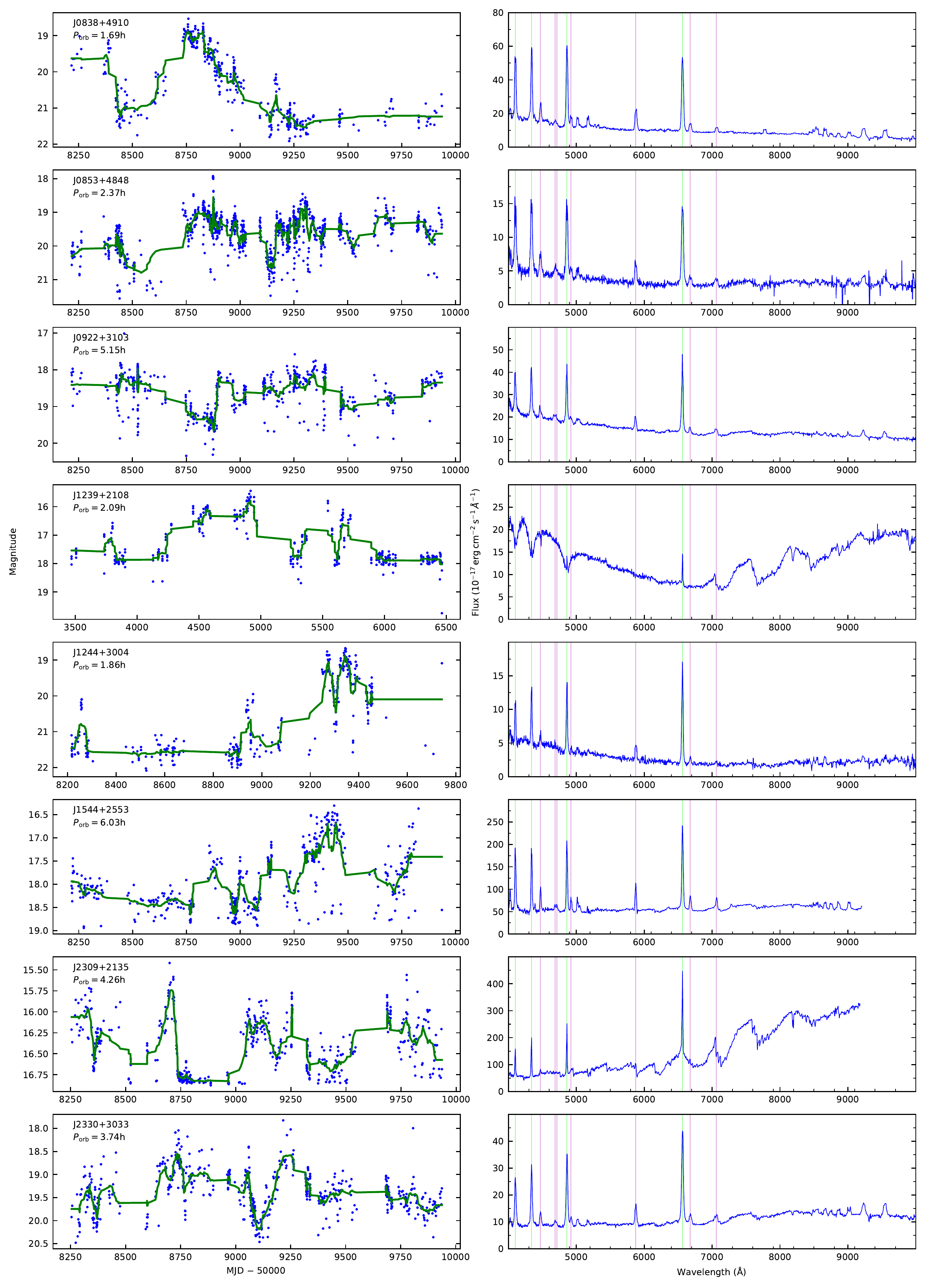}
\caption [.] {\label{fig:unusual} We identified eight CVs with substantial and erratic changes in apparent magnitude over periods of months or years~--~suggesting commensurate changes in accretion rate. The light curves shown are the merged $r$- and $g$-band data from ZTF, with the exception of J1239+2108 for which we show the CRTS data. The green lines show the trend using five point boxcar smoothing. The changes are atypical as the spectra of these systems are inconsistent with either a magnetic CV or novalike variable classification.}
\end{figure*}

\subsection{Unusual state changes}
\label{section:state_changes}
In contrast to dwarf novae, which undergo quasi-regular outbursts because of thermal instabilities in their accretion discs, novalike variables are known to exhibit state changes due to sustained changes in \mdot \citep{1985ApJ...290..707S,1999MNRAS.305..225L,2007MNRAS.377.1747R}. Polars are also known to undergo state changes on a timescale of months to years \citep{2005AJ....130..742K,2004MNRAS.350.1373R,2016MNRAS.463.1342S}, and low states have been observed occasionally in a small number of dwarf novae \citep{1995MNRAS.273..772W,2002A&A...384L...6S}. The cause of the high/low state transitions is still debated, but a likely contender is star spots near the inner Lagrangian point \citep{1994ApJ...427..956L,1998ApJ...499..348K,2000A&A...361..952H}

Serendipitously whilst reviewing the ZTF light curves of the spectroscopically identified CVs we have identified eight CVs (Figure.\,\ref{fig:unusual}) within the \allsdss\ sample that display erratic light curve morphologies, neither displaying disc outbursts, nor clearly identifiable state changes~--~and which have spectra that are inconsistent with being either magnetic CVs or novalike variables. These systems defy the standard CV classification scheme outlined in section\,\ref{section:sub_types}. 
These eight examples of systems  exhibit significant brightness changes on timescales that indicate the cause is a change to \mdot\ rather than instabilities of their discs. Below, we discuss the properties of these systems.

J0838+4910 is below the period gap with $\porb=1.69$\,h. 
\citet{2015AJ....149..128T} rightly concluded that this was an SU\,UMa based on the measured orbital period and the CRTS light curve. The SDSS spectrum is consistent with an SU\,UMa classification. However the ZTF data is contradictory~--~showing no outbursts and two long-lasting faint states. 

J0853+4848 has a spectrum that shows Balmer, \ion{He}{i} and weak \ion{He}{ii}   emission lines. The white dwarf is not visible but the M-type donor is detected in the red part of the spectrum. The system has been brighter since the SDSS spectrum was obtained. There is no CRTS data and the ZTF data shows one small outburst but also state changes of up to two magnitudes. The period is 2.36\,h which is in the period gap. 

J0922+3103 (AR\,Cnc) and  J1544+2553 have periods of $5.15$\,h and $6.03$\,h respectively and spectra which resemble U\,Gem CVs with double-peaked Balmer and helium lines, and clear signatures of their M-dwarf donors. However their light curves more closely resemble those of polars and they exhibit no outbursts. This ambiguity prevents any further classification beyond that of a CV. \citet{2013AJ....146..101P} classified J1544+2553 as a novalike using an automated process but this is not consistent with the strong lines in the spectrum, the detection of the donor and the position in the HR diagram.

J1239+2108 (IR\,Com)  has previously been identified as a CV with erratic brightness changes \citep{1997A&A...325..994R}, and SDSS obtained a spectrum when accretion had virtually ceased for an extended period \citep{2014MNRAS.442L..23M}, revealing both the white dwarf and the donor star.  It is on the lower edge of the period gap with $\porb=2.09$\,h.

J1244+3004 (CSS 080427:124418+300401) was found in a bright state within the CRTS light curve for about ten years, albeit exhibiting outbursts \citep{2015AJ....149..128T}. The ZTF light curve (Fig.\,\ref{fig:unusual}) shows a prolonged drop of $m\simeq2$. It's period of $\porb=1.86$\,h places it below the period gap.

J2309+2135 (V405\,Peg) was identified by \citet{2002A&A...396..895S} as the counterpart of a \textit{ROSAT} X-ray source, and based on the detection of the donor star, and the apparent magnitude, the authors speculated that V405\,Peg may be as close as $\simeq30$\,pc. \citet{2009PASP..121..465T} determined the period to be 4.26\,h, and measured a parallax of $\simeq150\,$\,pc~--~superseded by the \textit{Gaia} parallax of $\simeq173$\,pc. \citet{2014A&A...561A.121S} obtained deep X-ray observations, revealing a relatively low X-ray luminosity, and the authors noted that the system displays characteristics of both magnetic and non-magnetic CVs. The SDSS spectrum very clearly reveals the M-type donor star, with little contribution from either an accretion disc or the white dwarf.  

J2330+3033 has a spectrum showing both Balmer and \ion{He}{i} emission lines, the donor is clearly visible. It is eclipsing and \citet{2017MNRAS.465.4968H} found $\porb=3.74$\,h. The CRTS and ZTF light curves show variability but no outbursts.  

We note that the periods of these systems cluster both near the lower edge of the period gap, and in the $\simeq4-6$\,h range, indicating  widely varying evolutionary stages. The lack of outbursts is indicative of a low \mdot. A low \mdot\ is corroborated by the detection of the donor stars in the short-period systems, as non-magnetic CVs within this period range are typically SU\,UMa systems with accretion disc dominated spectra. The long-period systems coincide with the period range where dwarf novae are usually found (we encourage the reader to inspect section 7.2 of \citealt{2011ApJS..194...28K} for a discussion of the inconsistency between the CV sub-type distribution in the $3-6$\,h period range and the theoretical predictions of \mdot\ and the critical accretion rate for disc instabilities to (not) occur). We speculate that these eight systems may either contain weakly magnetic white dwarfs, with fields that are sufficiently low so that they do not develop the hallmarks of polars or intermediate polars, but strong enough to affect the structure of the disc. However, we also note that a number of intermediate polars exhibit dwarf nova outbursts~--~so the tentative suggestion of weak fields does not explain the absence of  outbursts. An alternative idea is that these systems are undergoing some type of state change in their donor stars, as possibly suggested by their location near the lower edge of the period gap, and in the range where dwarf novae and novalike overlap above the period gap.

\subsection{Tertiary systems}
\begin{table*}
\caption{Tertiary systems identified in \allsdss. $\mu_{\mathrm{RA}}$ and $\mu_{\mathrm{Dec}}$ are the components of proper motion and $a$ is the projected separation.  In each case the calculated binding energy $-E_{\mathrm{bind}}$ (see text for details) is  typical of a bound system as shown in  fig.\,6b of \citet{2007prpl.conf..427B}. 
}
\label{tab:tertiary}
\begin{tabular}{llcccccc}
\hline
System & \begin{tabular}[c]{@{}c@{}}\textit{Gaia} EDR3\\ source\_id\end{tabular} & \begin{tabular}[c]{@{}c@{}}P$_{\mathrm{orb}}$\\(h)\end{tabular} & \begin{tabular}[c]{@{}c@{}}Distance\\(pc)\end{tabular} & \begin{tabular}[c]{@{}c@{}}$\mu_{\mathrm{RA}}$\\ (mas$\,yr^{-1}$)\end{tabular} & \begin{tabular}[c]{@{}c@{}}$\mu_{\mathrm{Dec}}$\\ (mas$\,yr^{-1}$)\end{tabular} & \begin{tabular}[c]{@{}c@{}}$a$\\ (au)\end{tabular} & \begin{tabular}[c]{@{}c@{}}$-E_{\mathrm{bind}}$\\ ($10^{41}\,$erg)\end{tabular} \\ \hline
 J0801+2103 & 670872860843747072 &  & $816\pm106$ & -9.22 & -17.26 & \multirow{2}{*}{$12682\pm2720$} & \multirow{2}{*}{$7\pm1.6$} \\ 
 & 670872860843744768 &  & $1597\pm608$ & -8.97 & -17.20 & & \\ \hline
 J0803+2516 & 681813727390427136 & 2.2 & $1016\pm130$ & -7.63 & -7.52 & \multirow{2}{*}{$4922\pm976$} & \multirow{2}{*}{$22\pm4.3$} \\ 
 & 681813723094550144 &  & $1042\pm72$ & -7.80 & -7.22 & & \\ \hline
 J0840+0455 & 582480170443653376 &  & $1466\pm104$ & -3.92 & -11.11 & \multirow{2}{*}{$862\pm106$} & \multirow{2}{*}{$142\pm17.5$} \\ 
 & 582480166148927744 &  & $1441\pm120$ & -3.88 & -11.00 & & \\ \hline
 J1014+0638 & 3873721404734233344 & 1.4 & $192\pm5$ & -62.27 & -34.29 & \multirow{2}{*}{$1680\pm80$} & \multirow{2}{*}{$56\pm2.7$} \\ 
 & 3873721409029428352 &  & $185\pm1$ & -62.89 & -34.03 & & \\ \hline
 J1131+4322 & 772038105376131456 & 1.5 & $328\pm6$ & -12.78 & 11.70 & \multirow{2}{*}{$1745\pm65$} & \multirow{2}{*}{$54\pm2.0$} \\ 
 & 772038105376626432 &  & $361\pm5$ & -12.66 & 11.94 & & \\ \hline
 J1206+5100 & 1548430890981046272 & 3.8 & $302\pm8$ & 0.46 & -28.23 & \multirow{2}{*}{$2080\pm112$} & \multirow{2}{*}{$39\pm2.1$} \\ 
 & 1548430886684193152 &  & $317\pm49$ & -0.48 & -28.00 & & \\ \hline
 J1702+2235 & 4571234234022916992 & 1.6 & $3237\pm1269$ & 4.42 & -6.84 & \multirow{2}{*}{$1447\pm847$} & \multirow{2}{*}{$63\pm36.9$} \\ 
 & 4571234234022916864 &  & $1605\pm1238$ & 4.40 & -6.74 & & \\ \hline
 J2205+1155 & 2728239161441977216 & 1.4 & $1004\pm451$ & 15.91 & -9.58 & \multirow{2}{*}{$18470\pm11519$} & \multirow{2}{*}{$5\pm2.9$} \\ 
 & 2728239200096378368 &  & $544\pm42$ & 16.67 & -9.42 & &
\\\hline
\end{tabular}
\end{table*}
\citet{2010ApJS..190....1R} suggest that around eight per\,cent of close binaries are part of a triple system and have a distant companion. They are significant as the distant star can influence the evolution of the inner binary through Kozai cycles \citep{2007ApJ...669.1298F}. We searched \textit{Gaia} EDR3 for such systems by identifying candidate stars within 30\,arcsec of each of the CVs in our sample. Potential companions can be expected to have similar proper motions. However proper motions, and hence differences between them, will scale inversely with distance. We therefore adopted the following metric where the denominator is a proxy for distance:

\begin{equation}
\sqrt{\frac{\left(\mathrm{pmra_{cv}}- \mathrm{pmra_{cand}} \right)^{2}+\left(\mathrm{pmdec_{cv}}- \mathrm{pmdec_{cand}} \right)^{2}}{\mathrm{pmra_{cv}}^{2}+\mathrm{pmdec_{cv}}^{2}}}<0.05
\end{equation}
resulting in the eight systems in Table \ref{tab:tertiary}.

The projected separations, $a$, were calculated by multiplying the angular separation by the \citet{2018AJ....156...58B} distance; the errors on the \citet{2018AJ....156...58B} distances are too large to estimate the radial separation.
Estimating the  binding energy is a useful test of whether these systems are gravitationally bound to the CV or not:
\begin{equation}
-E_\mathrm{bind} = G\frac{(M_\mathrm{WD}+M_\mathrm{d})M_3}{a}
\end{equation}
A rough estimate is sufficient for this purpose and so we assume typical values of  $M_\mathrm{WD}=0.83\,\mathrm{M}_\odot$ and a donor of mass $M_\mathrm{d} = 0.14\,\mathrm{M}_\odot$. We estimate $M_3$ from the position of the star in the \textit{Gaia} HR diagram; and find that the CV companion in seven of the tertiaries is on the main sequence; the companion to J1206+5100 is a white dwarf. The calculated values shown in  Table \ref{tab:tertiary}, when compared with fig\,6b of \citet{2007prpl.conf..427B} show clearly that these eight CVs are tertiary systems.  These eight are only 1.6\,per\,cent of our sample and should be considered a lower limit on the fraction of CVs in triple systems. Our search is likely to be incomplete at larger distances where \textit{Gaia} may not resolve close tertiaries or not detect them because they are too faint; in particular tertiaries with very low-mass or cool white dwarf companions would be hard to identify.

\section{Summary} \label{section:summary}
We have reviewed the spectra of CVs observed by \allsdss\, and publish \fxnum{178} for the first time along with \fxnum{59} new periods. We discovered \fxnum{70} completely new CVs and classified or updated \fxnum{262}. These \fxnum{70} new CVs from \sdss\, augment the 282 previously identified in SDSS I and II. This  total is remarkably consistent with the early prediction of 400  from \citet{2002AJ....123..430S}.
We have analysed the results and position the different CV sub-types in evolutionary terms together with revised estimates of space density. The period bouncers and tertiary systems within our sample have been identified and the abundances compared with previous estimates.  

\section*{Acknowledgements}

We thank the anonymous referee for their helpful comments.

We acknowledge with thanks the variable star observations from the AAVSO International Database contributed by observers worldwide and used in this research.

Funding for SDSS-III has been provided by the Alfred P. Sloan Foundation, the Participating Institutions, the National Science Foundation, and the U.S. Department of Energy Office of Science. The SDSS-III web site is http://www.sdss3.org/.

SDSS-III is managed by the Astrophysical Research Consortium for the Participating Institutions of the SDSS-III Collaboration including the University of Arizona, the Brazilian Participation Group, Brookhaven National Laboratory, Carnegie Mellon University, University of Florida, the French Participation Group, the German Participation Group, Harvard University, the Instituto de Astrofisica de Canarias, the Michigan State/Notre Dame/JINA Participation Group, Johns Hopkins University, Lawrence Berkeley National Laboratory, Max Planck Institute for Astrophysics, Max Planck Institute for Extraterrestrial Physics, New Mexico State University, New York University, Ohio State University, Pennsylvania State University, University of Portsmouth, Princeton University, the Spanish Participation Group, University of Tokyo, University of Utah, Vanderbilt University, University of Virginia, University of Washington, and Yale University.

Funding for the Sloan Digital Sky Survey IV has been provided by the Alfred P. Sloan Foundation, the U.S. Department of Energy Office of Science, and the Participating Institutions. SDSS acknowledges support and resources from the Center for High-Performance Computing at the University of Utah. The SDSS web site is www.sdss.org.

SDSS IV is managed by the Astrophysical Research Consortium for the Participating Institutions of the SDSS Collaboration including the Brazilian Participation Group, the Carnegie Institution for Science, Carnegie Mellon University, Center for Astrophysics | Harvard \,\& Smithsonian (CfA), the Chilean Participation Group, the French Participation Group, Instituto de Astrofísica de Canarias, The Johns Hopkins University, Kavli Institute for the Physics and Mathematics of the Universe (IPMU) / University of Tokyo, the Korean Participation Group, Lawrence Berkeley National Laboratory, Leibniz Institut für Astrophysik Potsdam (AIP), Max-Planck-Institut für Astronomie (MPIA Heidelberg), Max-Planck-Institut für Astrophysik (MPA Garching), Max-Planck-Institut für Extraterrestrische Physik (MPE), National Astronomical Observatories of China, New Mexico State University, New York University, University of Notre Dame, Observatório Nacional / MCTI, The Ohio State University, Pennsylvania State University, Shanghai Astronomical Observatory, United Kingdom Participation Group, Universidad Nacional Autónoma de México, University of Arizona, University of Colorado Boulder, University of Oxford, University of Portsmouth, University of Utah, University of Virginia, University of Washington, University of Wisconsin, Vanderbilt University, and Yale University.

Based on observations obtained with the Samuel Oschin Telescope 48-inch and the 60-inch Telescope at the Palomar
Observatory as part of the Zwicky Transient Facility project. ZTF is supported by the National Science Foundation under Grants
No. AST-1440341 and AST-2034437 and a collaboration including current partners Caltech, IPAC, the Weizmann Institute for
Science, the Oskar Klein Center at Stockholm University, the University of Maryland, Deutsches Elektronen-Synchrotron and
Humboldt University, the TANGO Consortium of Taiwan, the University of Wisconsin at Milwaukee, Trinity College Dublin,
Lawrence Livermore National Laboratories, IN2P3, University of Warwick, Ruhr University Bochum, Northwestern University and
former partners the University of Washington, Los Alamos National Laboratories, and Lawrence Berkeley National Laboratories.
Operations are conducted by COO, IPAC, and UW.

Data from the Catalina Real-Time Transient Survey has been used \citep{2009ApJ...696..870D}

This work has made use of data from the European Space Agency (ESA) mission \textit{ Gaia} (\url{https://www.cosmos.esa.int/gaia}), processed by the \textit{Gaia} Data Processing and Analysis Consortium (DPAC, \url{https://www.cosmos.esa.int/web/gaia/dpac/consortium}). Funding for the DPAC has been provided by national institutions, in particular the institutions participating in the \textit{Gaia} Multilateral Agreement.

This paper includes data collected with the TESS mission, obtained from the MAST data archive at the Space Telescope Science Institute (STScI). Funding for the TESS mission is provided by the NASA Explorer Program. STScI is operated by the Association of Universities for Research in Astronomy, Inc., under NASA contract NAS 5–26555.

This work makes use of observations obtained at the MDM Observatory, operated by Dartmouth College, Columbia University, Ohio State University, Ohio University, and the University of Michigan.

Guoshoujing Telescope (the Large Sky Area Multi-Object Fiber Spectroscopic Telescope LAMOST) is a National Major Scientific Project built by the Chinese Academy of Sciences. Funding for the project has been provided by the National Development and Reform Commission. LAMOST is operated and managed by the National Astronomical Observatories, Chinese Academy of Sciences.

Some of the results in this paper have been derived using the healpy and HEALPix package.

BTG was supported by grant ST/T000406/1 from the Science and Technology Facilities Council (STFC). This project has received funding from the European Research Council (ERC) under the European Union’s Horizon 2020 research and innovation programme (Grant agreement No. 101020057). This research was supported in part by the National Science Foundation under Grant No. PHY-1748958.

\section*{Data Availability}
All data used in this article are available from the sources referenced in the text. The Supplementary Data including a machine-readable version of the catalogue is included in the online version of this paper.


\bibliographystyle{mnras}
\bibliography{refs,others}



\newpage

\appendix

\onecolumn

\section {Newly discovered CVs} \label{section:appendix1}
\onecolumn
\begingroup
\renewcommand{\arraystretch}{1.2}
 \begin{longtable}[c]{llllll}
 \caption{\label{tab:newCVs}New CVs from  \sdss. Distances are from \citep{2021AJ....161..147B}.  Apparent magnitudes are from SDSS data release~7 \citep{2009ApJS..182..543A}.  : Tentative values, * Superhump periods, \dag Low accretion SU UMa, \ddag CVs with unusual state changes, \S Period bouncers  \label{long_app}}\\

\hline
SDSS name & Gaia EDR3 source\_id & Period (d) & SDSS g (mag) & Distance (pc) & Variable type \\ \hline
 \hline
 \endfirsthead

\hline
SDSS name & Gaia EDR3 source\_id & Period (d) & SDSS g (mag) & Distance (pc) & Variable type \\ \hline
 \hline
 \endhead
 \hline
 \endfoot
 \hline
 \endlastfoot
J000014.74+255603.1 &  &  & 22.17 &  & CV: \\
J001231.54+280011.1 & 2859847683924640384 &  & 20.98 &  & WZ Sge: \\
J002910.96+324125.8 &  &  & 22.12 &  & DN \\
J004849.86+332832.9 & 361363704879572736 &  & 20.61 & $763^ {+358} _{-230}$ & SU UMa: \\
J005421.90+334815.7 &  &  & 22.12 &  & WZ Sge: \\
J010340.75+200525.9 & 2788762156951156864 & 0.158(1): & 21.45 &  & IP: \\
J012212.21+075546.8 & 2578454444735885056 &  & 20.33 & $343^ {+119} _{-78}$ & WZ Sge: \\
J012402.25+211359.7 & 289388711652054912 & 0.083475(1) & 21.38 & $688^ {+357} _{-241}$ & Polar \\
J013421.83+042200.0 &  &  & 21.41 &  & Polar: \\
J013540.99+340132.9 & 316371177970255232 &  & 22.42 &  & Polar: \\
J014121.21+062007.6 & 2562434457240098432 &  & 20.77 &  & WZ Sge: \\
J014732.85+144443.3 & 2588313902941290112 &  & 22.38 & $907^ {+722} _{-310}$ & Polar \\
J015253.39-004350.4 &  &  & 20.81 &  & SU UMa: \\
J015803.40+282303.6 & 298356019250182016 & 0.0888104(1) & 20.37 & $254^ {+20} _{-16}$ & DN \\
J020712.71-014116.2 & 2494386992562185088 &  & 21.43 &  & Polar: \\
J023046.26-004531.1 &  &  & 21.68 &  & WZ Sge: \\
J024215.05-011054.4 &  &  & 22.3 &  & CV: \\
J072827.34+330741.6 &  &  & 21.94 &  & WZ Sge: \\
J073605.07+182709.8 & 672172522242684800 &  & 21.83 &  & Polar \\
J073648.03+200959.7 &  &  & 22.0 &  & CV \\
J080249.06+363716.4 &  &  & 21.9 &  & WZ Sge: \\
J080625.73+343244.4 &  & 0.0637(1): & 21.46 &  & SU UMa: \\
J081017.41+262238.3 &  &  & 21.31 &  & SU UMa: \\
J081520.83+335634.1 &  &  & 21.6 &  & WZ Sge: \\
J082809.87+222004.2 & 665816516105274496 &  & 20.79 &  & CV: \\
J083404.24+185416.8 & 662643501643372544 &  & 19.23 & $1634^ {+726} _{-438}$ & Polar  \\
J083549.86+292636.9 & 704773999783239808 &  & 19.88 & $1244^ {+657} _{-442}$ & Polar \\
J084228.13+462334.4 &  &  & 22.17 &  & CV \\
J090947.82+110637.0 & 592059145810439808 &  & 20.83 & $601^ {+631} _{-232}$ & WZ Sge: \\
J091107.84+260617.3 &  &  & 21.53 &  & SU UMa: \\
J093130.75+335651.2 & 797435239135183744 & 0.0799(2): & 20.44 & $797^ {+990} _{-248}$ & SU UMa \\
J101335.56+393826.3 & 802734507224071936 &  & 20.78 & $882^ {+741} _{-316}$ & WZ Sge: \\
J101723.18+375343.4 & 754291266149306496 &  & 21.65 &  & SU UMa \\
J104245.15+371819.8 & 751777469035821312 &  & 19.7 & $507^ {+95} _{-91}$ & Polar \\
J104521.19+362433.0 &  &  & 21.44 &  & WZ Sge: \\
J115030.59+404151.8 & 768134186262349568 & 5.902(2) & 21.9 & $1273^ {+549} _{-451}$ & Polar \\
J115419.05+575751.1 & 845441772229861760 &  & 20.62 & $1144^ {+449} _{-352}$ & ER UMa \\
J121015.61+351334.4 & 4029016667559886080 &  & 19.38 & $356^ {+64} _{-58}$ & WZ Sge: \\
J122808.73+454516.3 & 1541836947948009472 & 0.07829(1): & 21.66 & $1599^ {+778} _{-608}$ & Polar \\
J123717.98+365535.2 & 1519559914656417920 &  & 21.22 &  & WZ Sge: \\
J125716.08+582501.5 & 1578799436498250368 &  & 20.13 & $953^ {+109} _{-99}$ & CV \\
J133816.44+371639.8 &  &  & 21.66 &  & WZ Sge: \\
J140430.98+525811.1 &  &  & 21.93 &  & CV \\
J140430.98+552558.5 & 1609460864427043072 &  & 21.48 & $1413^ {+776} _{-636}$ & DN \\
J152603.67+382216.4 &  &  & 21.35 &  & WZ Sge: \\
J155538.12+380250.5 & 1373544697473750272 & 0.104(1) & 20.39 & $486^ {+46} _{-49}$ & Polar \\
J160712.11+392450.6 & 1379871940214896640 & 0.08071(1) & 21.57 &  & Polar \\
J160813.59+301952.2 &  &  & 21.46 &  & WZ Sge: \\
J161853.16+303845.5 & 1319026547281204480 &  & 20.69 & $1822^ {+1404} _{-922}$ & WZ Sge: \\
J163620.62+472442.4 & 1410251583929439744 &  & 20.59 & $708^ {+318} _{-227}$ & WZ Sge: \\
J171016.94+403127.5 &  &  & 22.08 &  & WZ Sge: \\
J211652.28+014144.3 & 2692060074926640000 &  & 20.43 & $2752^ {+2087} _{-1558}$ & SU UMa: \\
J212406.75+010029.8 & 2691114426208274176 &  & 20.84 &  & WZ Sge: \\
J213040.11-001319.1 &  &  & 22.34 &  & WZ Sge: \\
J220818.52+135330.0 &  &  & 21.33 &  & WZ Sge \\
J220935.78-001332.5 & 2678822818216797056 &  & 21.8 &  & WZ Sge: \\
J221333.38+245059.3 &  &  & 21.94 &  & WZ Sge: \\
J221900.24+201831.1 & 1778468682222798080 & 0.140(1) & 19.85 & $821^ {+124} _{-98}$ & DN \\
J223212.79+260725.1 & 1880505014843117696 &  & 22.26 &  & Polar \\
J223352.56+225913.4 & 1875338547143938944 &  & 19.89 & $1294^ {+491} _{-329}$ & Polar \\
J223623.52+323913.1 &  &  & 21.02 &  & WZ Sge: \\
J225613.53+191132.9 &  &  & 22.12 &  & WZ Sge: \\
J225926.51+223902.4 &  &  & 21.81 &  & WZ Sge: \\
J230319.85+274736.0 &  &  & 22.43 &  & DN: \\
J232113.45+251822.3 & 2841125848466425856 &  & 20.96 & $1189^ {+715} _{-482}$ & WZ Sge: \\
J232553.47+011527.3 &  &  & 21.39 &  & WZ Sge: \\
J233705.01+314149.4 & 2871487075361158400 &  & 19.64 & $2574^ {+939} _{-1150}$ & SU UMa: \\
J235131.68+270907.7 & 2854471033409793024 &  & 21.9 & $1089^ {+467} _{-451}$ & DN: \\
J235603.71+001833.3 &  &  & 21.45 &  & WZ Sge: \\
J183131.63+420220.2 & 2111270034246759424 &  & 16.9 & $1490^ {+162} _{-155}$ & AM CVn:

 \end{longtable}
 \endgroup
\noindent


\bsp	
\label{lastpage}
\end{document}